\documentclass[aps,prd,superscriptaddress,amsmath,groupedaddress,twocolumn,nofootinbib]{revtex4}

\usepackage{color,hangcaption,hhline,psfrag,rotating,amssymb}
\usepackage[hang,nooneline]{subfigure}
\usepackage{dcolumn}

\begin{document}
\title{The $\Upsilon$ and $\Upsilon^{\prime}$ Leptonic Widths, $a_{\mu}^b$ and $m_b$ from full lattice QCD }
\author{B. Colquhoun}
\affiliation{SUPA, School of Physics and Astronomy, University of Glasgow, Glasgow, G12 8QQ, UK}
\author{R. J. Dowdall}
\affiliation{DAMTP, University of Cambridge, Wilberforce Road, Cambridge, CB3 0WA, UK}
\author{C. T. H. Davies}
\email[]{christine.davies@glasgow.ac.uk}
\affiliation{SUPA, School of Physics and Astronomy, University of Glasgow, Glasgow, G12 8QQ, UK}
\author{K. Hornbostel}
\affiliation{Southern Methodist University, Dallas, Texas 75275, USA}
\author{G. P. Lepage}
\affiliation{Laboratory of Elementary-Particle Physics, Cornell University, Ithaca, New York 14853, USA}

\collaboration{HPQCD collaboration}
\homepage{http://www.physics.gla.ac.uk/HPQCD}
\noaffiliation

\date{\today}

\begin{abstract}
We determine the decay rate to leptons of the ground-state $\Upsilon$ 
meson and its first radial excitation 
in lattice QCD for the first time. We use radiatively-improved 
NRQCD for 
the $b$ quarks and include $u$, $d$, $s$ and $c$ quarks in the 
sea with $u/d$ masses down to their physical values. 
We find $\Gamma(\Upsilon \rightarrow e^+e^-)$ = 1.19(11) keV 
and $\Gamma(\Upsilon^{\prime} \rightarrow e^+e^-)$ = 0.69(9) keV, both 
in good agreement with experiment. 
The decay constants we obtain 
are included in a summary plot of meson decay constants from 
lattice QCD given in the Conclusions. 
We also test time-moments of the vector current-current 
correlator against values determined from 
the $b$ quark contribution to $\sigma(e^+e^- \rightarrow \mathrm{hadrons})$ 
and calculate the $b$-quark piece of the hadronic vacuum polarisation 
contribution to the anomalous magnetic moment of the muon, $a_{\mu}^b = 0.271(37) \times 10^{-10}$. 
Finally we determine the $b$-quark mass, obtaining in 
the $\overline{MS}$ scheme, 
$\overline{m}_b(\overline{m}_b, n_f=5)$ = 4.196(23) GeV, 
the most accurate result from lattice QCD to date. 

\end{abstract}


\maketitle

\section{Introduction}
\label{sec:intro}
Precision tests of lattice QCD against experiment 
are critical to provide benchmarks against which 
to calibrate the reliability of predictions from 
lattice QCD for masses and matrix 
elements~\cite{ourlatqcd}. 
Weak decay matrix elements calculated in lattice 
QCD, for example, are critical to 
the flavor physics programme~\cite{cdlat11, lunghilat11} 
of over-determining the Cabibbo-Kobayashi-Maskawa 
(CKM) matrix to find signs 
of new physics.  
This is particularly important for mesons containing 
a valence $b$ quark. 
Lattice QCD therefore needs to provide 
a range of results for a variety of hadrons 
containing $b$ quarks to make sure that the 
analysis of systematic errors is sound.  
Focussing on quantities that are well measured 
experimentally enables strong tests to be made. 

Here we give lattice QCD results for the electromagnetic 
annihilation rate for mesons, the $\Upsilon$ and its 
radial excitation the $\Upsilon^{\prime}$, 
containing valence $b$ quarks and antiquarks. The hadronic 
parameter that determines this rate, the matrix 
element of the vector current between the vacuum 
and the $\Upsilon$, is parameterised by a quantity known 
as the decay constant. 
The determination of this intrinsically non-perturbative 
quantity is both a
 test of our lattice QCD approach to $b$ quark 
physics and a test of QCD itself, since other methods
of determining this rate have large systematic errors or 
uncertainties from model-dependence that reduce  
the significance of the comparison with experiment (see, for 
example, the discussion in~\cite{Eichten:1995ch, Pineda:2006ri, qwg10, upsleptpert}). 
The recent success of lattice QCD in determining the leptonic 
width of the $J/\psi$ to 4\%~\cite{psipaper, etmccharm} 
makes clear the power 
of a model-independent nonperturbative approach to such 
calculations and we apply that approach here, for the 
first time doing a complete calculation. For related earlier 
work in Lattice QCD see~\cite{Daviesorigups,bodwin,gray}. 
This calculation also provides the `missing piece' of 
a set of determinations of decay constants for a range 
of mesons by the HPQCD collaboration and others. We provide a plot
which summarises the results. 

Another test against experiment 
that can be done with the same correlation functions
is that of the $b$ quark 
contribution to $\sigma(e^+e^- \rightarrow \mathrm{hadrons})$. 
Again this is an electromagnetic rate and so the comparison 
is free from CKM uncertainties. 
Here we also build on the success of a similar calculation 
of the $c$ quark contribution~\cite{psipaper}. 

Finally we determine the $b$ quark mass using 
the current-current correlator method~\cite{firstcurrcurr, bcmasses}. 
Since we use the NonRelativistic 
QCD (NRQCD) approach to $b$ quarks~\cite{nrqcd} here 
(including now $\mathcal{O}(\alpha_s)$ corrections to nonleading 
terms in the nonrelativistic expansion~\cite{dowdallr1}) 
our result for $m_b$
has very different systematic errors to that from 
the relativistic Highly Improved Staggered Quark (HISQ) 
formalism~\cite{bcmasses}. Since both determinations 
have 0.5\% uncertainties this provides a very 
stringent comparison. 
We can also compare our result to that obtained 
using an NRQCD calculation of $\Upsilon$ binding 
energies coupled with a lattice QCD perturbation theory 
calculation of the NRQCD zero of energy~\cite{mbpert}. 
The agreement of these three very different determinations is 
a good test of our control of systematic errors. 
Accurate determination of the $b$ quark mass is important 
for calculation of the expected Higgs branching fraction
to $b\overline{b}$~\cite{snowmasshiggs,gplhiggs}.

The paper is laid out as follows: Section~\ref{sec:latt} gives 
an overview of the methods used in the lattice calculation and 
then Section~\ref{sec:results} gives each set of results in turn, 
with additional details in Appendices~\ref{appendix:rel} and~\ref{appendix:zv}.
Section~\ref{sec:conclusions} gives our conclusions, including 
a summary of lattice QCD results for meson decay constants. 

\section{Lattice calculation}
\label{sec:latt}
We use ensembles of lattice gluon 
configurations provided by the MILC collaboration~\cite{milchisq} 
at values of the lattice spacing, $a \approx$  0.15 fm, 0.12 fm and 0.09 fm. 
The configurations include the effect of $u$, $d$, $s$ and $c$ 
quarks in the sea using the highly improved staggered quark (HISQ) 
formalism~\cite{hisqdef} and a gluon action improved through 
$\mathcal{O}(\alpha_sa^2)$~\cite{Hart:2008sq}. These then give
significant improvements in the control of systematic 
errors from finite lattice spacing and light 
quark mass effects over earlier configurations. 

We work at two different values of the $u/d$ quark masses (which 
are taken to be degenerate) in the sea. 
One is at one fifth of the $s$ quark mass, the other is the 
physical $u/d$ quark mass ($m_s/27.5$). 
The lattice spacing on these configurations is determined from 
the mass difference between the $\Upsilon^{\prime}$ and 
the $\Upsilon$~\cite{dowdallr1}. 
Table~\ref{tab:params} lists the 
parameters of the ensembles. 

\subsection{NRQCD}
\label{subsec:nrqcd}
On these configurations we calculate $b$ quark propagators 
using the improved NRQCD action developed in~\cite{dowdallr1} and~\cite{nrqcdimp}. 
The NRQCD Hamiltonian we use is given by:
 \begin{eqnarray}
 aH &=& aH_0 + a\delta H; \nonumber \\
 aH_0 &=& - \frac{\Delta^{(2)}}{2 am_b}, \nonumber \\
a\delta H
&=& - c_1 \frac{(\Delta^{(2)})^2}{8( am_b)^3}
            + c_2 \frac{i}{8(am_b)^2}\left(\bf{\nabla}\cdot\tilde{\bf{E}}\right. -
\left.\tilde{\bf{E}}\cdot\bf{\nabla}\right) \nonumber \\
& & - c_3 \frac{1}{8(am_b)^2} \bf{\sigma}\cdot\left(\tilde{\bf{\nabla}}\times\tilde{\bf{E}}\right. -
\left.\tilde{\bf{E}}\times\tilde{\bf{\nabla}}\right) \nonumber \\
 & & - c_4 \frac{1}{2 am_b}\,{\bf{\sigma}}\cdot\tilde{\bf{B}}  
  + c_5 \frac{\Delta^{(4)}}{24 am_b} \nonumber \\
 & & -  c_6 \frac{(\Delta^{(2)})^2}{16n_h(am_b)^2} .
\label{eq:deltaH}
\end{eqnarray}
Here $\nabla$ is the symmetric lattice derivative and $\Delta^{(2)}$ and 
$\Delta^{(4)}$ the lattice discretization of the continuum $\sum_iD_i^2$ and 
$\sum_iD_i^4$ respectively. $am_b$ is the bare $b$ quark mass in units 
of the lattice spacing. The parameter $n_h$ will be discussed below. 
$\bf \tilde{E}$ and $\bf \tilde{B}$ are the chromoelectric 
and chromomagnetic fields calculated from an improved clover term~\cite{gray}.
The $\bf \tilde{B}$ and $\bf \tilde{E}$ are made anti-hermitian 
but not explicitly traceless, to match the perturbative calculations 
done using this action.  

In terms of an expansion in the velocity of the heavy quark, $v$, 
$H_0$ is $\mathcal{O}(v^2)$ and 
$\delta H$ is $\mathcal{O}(v^4)$, including discretisation corrections. 
$H_0$ contains the bare quark 
mass parameter which is nonperturbatively tuned to the correct 
value for the $b$ quark as discussed below. 
The terms in $\delta H$ have coefficients $c_i$ whose values are 
fixed from matching lattice NRQCD to full QCD. This matching takes account 
of high momentum modes that differ between NRQCD and full QCD and 
so it can be done perturbatively, giving the $c_i$ the 
expansion $1 + c^{(1)}_i\alpha_s + \mathcal{O}(\alpha_s^2)$. 
Here we include $\mathcal{O}(\alpha_s)$ corrections to the coefficients 
of the $\mathcal{O}(v^4)$ kinetic terms, $c_1$, $c_5$ and $c_6$, 
and, for some of the ensembles, the chromomagnetic term, $c_4$~\cite{dowdallr1, nrqcdimp}. 
The effect of $\mathcal{O}(\alpha_s)$ corrections to 
other terms (with coefficients $c_2$ and $c_3$) 
at $\mathcal{O}(v^4)$ will be estimated by looking at 
the effect of the $\mathcal{O}(\alpha_s)$ corrections to $c_4$. 
The $\mathcal{O}(\alpha_s)$ coefficients to $c_1$, $c_4$, $c_5$ and $c_6$  
are calculated after tadpole-improvement to 
the gluon field, which means dividing all the links, $U_{\mu}(x)$ 
by a tadpole-parameter, $u_0$, before constructing covariant 
derivatives or 
${\bf E}$ and ${\bf B}$ fields for the Hamiltonian above.  
For $u_0$ we took the mean trace of the gluon field 
in Landau gauge, $u_{0L}$~\cite{dowdallr1}. Tadpole-improvement means that the 
$c_i^{(1)}$ coefficients are typically less than $\mathcal{O}(1)$, 
whereas without tadpole-improvement they can be much larger because of 
the effect of tadpole diagrams in the lattice theory. 
The values used for $U_{0L}$ and $c_i$ on the different ensembles 
are given in Table~\ref{tab:mbc}. 

\begin{table}
\begin{tabular}{llllllll}
\hline
\hline
Set &  $a$/fm & $am_{l}$ & $am_{s}$ & $am_{c}$ & $L_s/a$ & $L_t/a$ & $N_{\mathrm{cfg}}$ \\
\hline
1 & 0.1474(5)(14) & 0.013 & 0.065 & 0.838 & 16 & 48 & 1020 \\
2 & 0.1450(3)(14) & 0.00235 & 0.0647 & 0.831 & 32 & 48 & 1000 \\
\hline
3 & 0.1219(2)(9) & 0.0102 & 0.0509 & 0.635 & 24 & 64 & 1052 \\
4 & 0.1189(2)(9) & 0.00184 & 0.0507 & 0.628 & 48 & 64 & 1000 \\
\hline
5 & 0.0884(3)(5) & 0.0074 & 0.037 & 0.440 & 32 & 96 & 1008 \\
\hline
\hline
\end{tabular}
\caption{ Details of gluon field configurations used in this 
calculation~\cite{milchisq}. $a$ is the lattice spacing, fixed from the mass 
difference between the $\Upsilon^{\prime}$ and $\Upsilon$ in~\cite{dowdallr1}.
The first error is from statistics and the second from NRQCD 
systematics in that determination and from experiment. 
Sets 1 and 2 are `very coarse', sets 3 and 4 are `coarse' and 
set 5 is `fine'. 
$am_l$, $am_s$ and $am_c$ are the light ($u$ and 
$d$ are taken to have the same mass), strange and charm sea quark 
masses.
Sets 1, 3 and 5 have $m_l = 0.2m_s$ and 
sets 2 and 4 have $m_l$ at its physical value. 
$L_s/a$ and $L_t/a$ are the number of lattice sites in 
the spatial and temporal directions respectively and $N_{cfg}$ is 
the number of configurations in the ensemble. We calculate 
propagators from 4 or 16 time sources on each ensemble to increase 
statistics. Correlators are calculated up to a time separation 
between source and sink of 40 on sets 1--4 and 48 on fine set 5.  
}
\label{tab:params}
\end{table}

\begin{table}
\begin{tabular}{llllllll}
\hline
\hline
Set &  $am_b$ & $a\overline{M}_{kin}$ & $a\overline{E}_0$ & $u_{0L}$ & $c_1,c_6$ & $c_5$ & $c_4$ \\
\hline
1 & 3.297 & 7.087(8) & 0.27823(5) & 0.8195 & 1.36 & 1.21 & 1.0 \\
1 & 3.297 & 7.109(10) & 0.25137(6) & 0.8195 & 1.36 & 1.21 & 1.22 \\
1 & 3.42 & 7.303(15) & 0.27669(5) & 0.8195 & 1.36 & 1.21 & 1.0 \\
2 & 3.25 & 6.988(14) & 0.24950(2) & 0.8195 & 1.36 & 1.21 & 1.22 \\
\hline
3 & 2.66 & 5.761(14) & 0.28458(2) & 0.8340 & 1.31 & 1.16 & 1.0 \\
4 & 2.62 & 5.717(9) & 0.25161(2) & 0.8341 & 1.31 & 1.16 & 1.20 \\
\hline
5 & 1.91 & 4.264(11) & 0.27767(2) & 0.8525 & 1.21 & 1.12 & 1.0 \\
\hline
\hline
\end{tabular}
\caption{ Summary of the valence $b$ quark mass and other action parameters 
for the NRQCD action on the different ensembles of Table~\ref{tab:params}. 
The $b$ quark mass in lattice units (column 2) was tuned 
by calculating the spin-average of the `kinetic masses' 
of the $\Upsilon$ and $\eta_b$ as described in the 
text and given in column 3. 
In column 4 we give the corresponding spin-average of the 
ground-state energies, needed to reinstate the `zero of energy' in 
the current-current correlator. 
Results for sets 3 and 5 are from~\cite{dowdallr1}. 
Column 5 gives the parameter $u_{0L}$ used for `tadpole-improving' 
the gluon field~\cite{dowdallr1, DowdallBdecay} and columns 6, 7 and 8 give the 
coefficients of kinetic and chromomagnetic terms used in the 
NRQCD action. $c_1$, $c_5$ and $c_6$ ($c_1$ and $c_6$ have the 
same value) are correct through 
$\mathcal{O}(\alpha_s)$~\cite{dowdallr1}. For $c_4$ we used 
the $\mathcal{O}(\alpha_s)$ corrected value~\cite{dowdallr1, nrqcdimp} 
for sets 2 and 4
but the value 1.0 on sets 1, 3 and 5. 
The top row of parameters for set 1 are our `preferred' ones and 
these are the results that will be plotted in Figures, unless stated otherwise. 
Results for the other values of $c_4$ (row 2) $am_b$ (row 3) allow us 
to judge the effect of changing these parameters. 
}
\label{tab:mbc}
\end{table}

This improved NRQCD action has been used for accurate calculations 
of the $\Upsilon$ spectrum~\cite{dowdallr1, Daldrop, Dowdallhyp} 
and $B$ and $B_s$ meson masses~\cite{rachelBmass} and decay 
constants~\cite{DowdallBdecay}. 

Given the NRQCD action above, 
the heavy quark propagator is readily calculated from 
its lattice time evolution given by:
\begin{eqnarray}
G({\bf x},t+1) &=& 
	\left( 1-\frac{a\delta H}{2}\right)\left(1-\frac{aH_0}{2n_h}\right)^{n_h}U^{\dag}_{t}(x) 
	 \\
        & & \times \left(1-\frac{aH_0}{2n_h}\right)^{n_h}\left(1-\frac{a\delta H}{2}\right) G({\bf x},t) \nonumber
\label{eq:evol}
\end{eqnarray}
with starting condition:
\begin{equation}
G({\bf x},0) = \phi({\bf x})\mathtt{1}.
\label{eq:nrqcdstart}
\end{equation}
Here $1$ is the unit matrix in 
color and (2-component) spin space and $\phi({\bf x})$ is a 
simple function of spatial position, often called a `smearing function'. 
We can use such a function because we fix the gluon field 
configurations to Coulomb gauge. 
At zero spatial momentum the antiquark propagator is the complex conjugate of the quark propagator for a source 
of the kind given in eq.~(\ref{eq:nrqcdstart}). 
The parameter $n_h$ has no physical significance, but is included 
for improved numerical stability of 
high momentum modes that do not contribute to bound states~\cite{nrqcd}. 
Here we use $n_h=4$ throughout. $n_h$ also appears in the final 
term of $\delta H$ (eq.~(\ref{eq:deltaH})) because of the 
correction for the discretisation error in the time derivative~\cite{nrqcd}.  

\subsection{Meson Correlators}
\label{subsec:mescorr}
Quark and antiquark propagators are combined to form meson correlation 
functions by matching up color indices and combining appropriate spin 
indices. We will focus almost entirely on the vector $\Upsilon$ states here, 
created at $\bf{x_1}$ with an interpolating operator 
\begin{equation}
Y^{(\phi)}(\bf{x_1}) = \sum_{\bf{x_2}} \psi^{\dag}(\bf{x_1})\sigma_i \phi(\bf{x_1}-\bf{x_2})\chi^{\dag}(\bf{x_2})
\label{eq:op}
\end{equation}
where $\psi^{\dag}$ creates a 2-component quark, 
$\chi^{\dag}$, an antiquark and 
$\sigma_i$ is the Pauli spin matrix $\sigma_x$, $\sigma_y$ or $\sigma_z$ 
for different $\Upsilon$ polarisations. 
A meson correlation function that uses this operator at the source ($sc$) and 
an equivalent operator to destroy the meson at the sink ($sk$) can then be 
made by combining quark and antiquark propagators at lattice time $t$ into
\begin{eqnarray}
\label{eq:corr}
C(t) &=& \langle 0 | [Y^{(\phi_{sk})}_t]^{\dag} Y^{(\phi_{sc})}_0 | 0 \rangle \\
&=& \sum_{\bf{y_1},\bf{y_2}} \mathrm{Tr}[\sigma_i G^{\dag}_{\delta}({\bf{y_1}},t)\sigma_i \phi_{sk}({\bf{y_1}-\bf{y_2}}) G_{\phi_{sc}}({\bf{y_2}},t)]. \nonumber
\end{eqnarray}
Here $G_{\delta}$ is generated using eq.~(\ref{eq:nrqcdstart}) with 
$\phi(\bf{x})$ set equal to a delta function and $G_{\phi_{sc}}$ with 
$\phi(\bf{x})=\phi_{sc}(\bf{x})$. $\phi_{sk}$ is the sink smearing function 
which need not be the same as at the source. The convolution can 
is implemented using a Fast Fourier Transform. The meson correlation 
function is projected onto zero spatial momentum by the sum over sink spatial 
indices and the trace is over color and spin indices.  
 
Meson correlation functions in principle contain all the states 
of the system consistent with the quantum numbers of the operator used.
Here we can restrict ourselves to bottomonium states because we have not allowed 
any mixing with other sectors; this is expected to have negligible effect for 
$\Upsilon$ mesons in any case. 
By inserting a complete set of states into the first line of eq.~(\ref{eq:corr})
we see that the (Euclidean) time dependence of $C(t)$ is given by:
\begin{equation}
C(t) = \sum_{m=0}^{m_{exp}-1} c(\phi_{sc},m)c^*(\phi_{sk},m)e^{-E_mt}.
\label{eq:corrfit}
\end{equation}
Here $E_m$ denote the (NRQCD) energies of the ladder of $m_{exp}$ 
vector bottomonium states that we include. 
$m=0$ corresponds to the ground-state $\Upsilon$, 
$m=1$ to the $\Upsilon^{\prime}$ etc.  
The components of the amplitude of a given state depend on the overlap 
of the action of the operator $Y$ on the vacuum with that state: 
\begin{equation}
c(\phi,m) = \langle 0 | Y^{(\phi)} | \Upsilon^{(m)} \rangle /\sqrt{2M_{\Upsilon^{(n)}}},
\label{eq:overlap}
\end{equation}
where $M$ denotes the meson mass and we 
use the conventional $2M$ normalisation for states at rest. 
From this it is clear that different choices of $\phi$ allow us to 
change the contributions of different states to the meson correlator.  
At large times all 
correlators are dominated by the ground-state, but at relatively short 
times meson correlators made with smeared propagators can have 
very different time-dependence that allows us to extract the 
properties of excited states. 

We used this technique, a standard one in lattice QCD calculations, 
to determine the masses of the $\Upsilon^{\prime}$ 
and $\Upsilon^{\prime\prime}$ in~\cite{dowdallr1}. 
We used a delta function source and two `hydrogen-wavefunction' 
smearings, adjusting their radius as we changed the lattice spacing. 
This enabled us to make a $5\times 5$ matrix of meson correlators 
by combining different smeared quark and antiquark propagators together 
(i.e. generalising eq.~(\ref{eq:corr}) to the case where both propagators 
have a smeared source). Fitting the elements of this matrix 
simultaneously to eq.~(\ref{eq:corrfit}) enabled 
us to extract the energies, $E_m$. 
The differences $E_1-E_0$ and $E_2-E_0$ correspond to the mass 
differences between the $\Upsilon^{\prime}$ and $\Upsilon^{\prime\prime}$ 
respectively and the $\Upsilon$. We determined the lattice spacing 
in~\cite{dowdallr1} by setting $E_1-E_0$ equal to its experimental 
value.  

\subsection{Tuning Parameters}
\label{subsec:tuning}
The `zero of energy' is missing in NRQCD and so to convert mass 
differences to absolute masses, and hence to tune the quark mass, 
requires a separate calculation of the energy offset. 
This is done by calculating the energy of a meson as a function 
of spatial momentum and determining a `kinetic mass' which 
can be compared to experiment. This is defined by~\cite{gray}: 
\begin{equation}
aM_{kin} = \frac{{\bf p}^2a^2-(a\Delta E)^2}{2a\Delta E}
\label{eq:kinmass}
\end{equation} 
for a meson with spatial momentum ${\bf p}$ and with 
$\Delta E = E({\bf p}) - E(0)$. We determine $E({\bf p})$ 
and $E(0)$ for the ground-state $\Upsilon$ and (pseudoscalar) $\eta_b$ mesons 
(i.e. $m=0$ in eq.~(\ref{eq:corrfit})) using only delta 
function sources for the $b$ quark propagators. We use 
a wall of random numbers drawn from U(1) for these sources, 
patterned with an appropriate Fourier phase, 
since this improves statistical accuracy 
significantly~\cite{oldr1paper}. Although the kinetic mass 
is independent of the momentum ${\bf p}$ to high accuracy~\cite{gray} 
we fix a particular momentum to determine it given by the 
lattice momentum $(1,1,1)2\pi/L_s$. 

Tuning the $b$ quark 
mass means adjusting the value in the action until the kinetic 
mass for a specific meson agrees with experiment, given a 
result for the lattice spacing which is used to convert the 
dimensionless mass $aM_{kin}$ into physical units.  
Since our NRQCD action includes only the leading spin-dependent
terms (along with their radiative corrections) the kinetic 
masses for $\Upsilon$ and $\eta_b$ show a systematic error 
in that they appear in the wrong order with the $\Upsilon$ 
kinetic mass lower than that of the $\eta_b$. As explained in~\cite{dowdallr1}
this is because the difference in binding energy from the chromomagnetic 
term has not been incorporated correctly into the 
kinetic mass, since a relativistic correction to the chromomagnetic 
term is required for this to happen. 
To remove this effect in our quark mass tuning we instead 
tune the spin-average of the $\Upsilon$ and $\eta_b$ 
masses to experiment, defining
\begin{equation}
\overline{M}_{kin} = \frac{3M_{kin,\Upsilon}+M_{kin,\eta_b}}{4}.
\end{equation}
The values we obtain at the valence $b$ quark masses used on 
each ensemble are given in Table~\ref{tab:mbc}. 
The appropriate experimental value to compare this to is the 
spin-average of experimental $\Upsilon$ and $\eta_b$ 
masses ($\overline{M}_{\Upsilon, \eta_b}$)~\cite{pdg}, 
adjusted for the fact that we are working in a world 
without electromagnetism (which pushes up both 
$\Upsilon$ and $\eta_b$ masses by an estimated 1.6 MeV~\cite{gregory}). 
Putting this shift in gives a value for 
$\overline{M}_{\Upsilon, \eta_b}$ of 9.446(2) GeV. 
Here we have taken a 100\% error on this shift and also allowed an 
error (but no shift) for the fact that 
we do not allow our $\eta_b$ 
meson to annihilate to gluons. We earlier estimated 
the absence of gluon annihilation would shift the $\eta_b$ 
mass upwards by approximately 2.4 MeV~\cite{gregory}. Here we simply take 
this (divided by 4 reflecting the $\eta_b$ contribution to 
the spin-average) as an additional uncertainty.  

From Table~\ref{tab:mbc} we see that our $b$ quark mass is typically 
tuned at the level of 1\%, consistent with the accuracy with which 
we have determined the lattice spacing. The statistical accuracy 
on the kinetic mass itself is much better than this.  
We have used two different quark masses on set 1 so that we 
can test the $m_b$ dependence of results. The well-tuned mass 
on this set is $am_b=3.297$ and this is the preferred value for 
our results. $am_b=3.42$ then represents mistuning by 3--4\%. 

In~\cite{dowdallr1} we discussed the impact of improving 
the NRQCD action on kinetic masses and properties of the spectrum. 
For the calculation of this paper we are focussed on 
the amplitudes given in eq.~(\ref{eq:overlap}). In Appendix~\ref{appendix:rel}
we show how the $v^4$ terms in the NRQCD action modify the 
amplitudes to give improved relativistic covariance. 

\subsection{The Vector Current}
\label{subsec:vec}
Our key result here is for 
the overlap or matrix element between the vacuum and an $\Upsilon$ state 
of an operator $Y$ which corresponds to the local vector 
current $J_{V}$ that couples to a photon. The hadronic parameter known 
as the decay constant, $f$, is defined for an $\Upsilon$ at rest, 
or any of its radial excitations, by 
\begin{equation}
\langle 0 | J_{V,i} | \Upsilon^{(m)}_j \rangle = f_{\Upsilon^{(m)}}M_{\Upsilon^{(m)}}\delta_{ij}
\label{eq:fdef}
\end{equation}
where $i$ is the polarisation of the vector current, $j$ is the 
polarisation of the $\Upsilon$ and $M_{\Upsilon}$ is its mass. 
The square of the decay constant is then 
related to the experimentally measurable leptonic width by:
\begin{equation}
\Gamma(\Upsilon^{(m)} \rightarrow e^+e^-) = \frac{4\pi}{3}\alpha_{QED}^2 e_b^2 \frac{f_{\Upsilon^{(m)}}^2}{M_{\Upsilon^{(m)}}}
\label{eq:vdecay}
\end{equation} 
where $e_b$ is the $b$ quark electric charge in units of $e$ (1/3).
The appropriate value for $\alpha_{QED}$ here is that at the 
$b$ quark mass, 
$\alpha_{QED}(m_b)=1/132$~\cite{alpha-em}. Higher-order electromagnetic 
processes are suppressed because the $\Upsilon$ must decay 
to an odd number of photons.  
The experimental values for the leptonic width are accurately known 
for $\Upsilon$, $\Upsilon^{\prime}$ and $\Upsilon^{\prime\prime}$ following 
a dedicated programme by CLEO~\cite{cleo1, cleo2, cleo3}. 

To determine $f_{\Upsilon}$ accurately from lattice QCD we need an 
accurate representation on the lattice, in terms of an 
operator $Y$, of the vector current. 
Since we are using a nonrelativistic formalism we take a nonrelativistic 
expansion of the current including leading and next-to-leading order ($\mathcal{O}(v^2)$) 
corrections. For the vector case the leading operator is given by 
\begin{equation}
J^{(0)}_{V,\mathrm{NRQCD},i} = \chi^{\dag}\sigma_i \psi .
\label{eq:j0}
\end{equation} 
This corresponds to $Y$ of eq.~(\ref{eq:op}) with the 
smearing function $\phi$ set to a delta function. It is therefore 
one of the standard set of operators that we typically include in 
the spectrum calculation. 
There is only one subleading operator to consider at $\mathcal{O}(v^2)$ which 
we can take to be~\cite{gray, Hart:2006ij}: 
\begin{equation}
J^{(1)}_{V,\mathrm{NRQCD},i} = \chi^{\dag}\sigma_i \frac{\hat{\Delta}^{(2)}}{(am_b)^2}\psi.
\label{eq:j1}
\end{equation} 
$\hat{\Delta}^{(2)}$ is a representation of the $\Delta^{(2)}$ operator 
in which we choose (because we are working in Coulomb gauge) 
not to include the gluon links multiplying the shifted 
quark fields. This operator is readily implemented at the source by 
acting with $\hat{\Delta}^{(2)}$ on a delta function source. At the sink 
we simply apply $\hat{\Delta}^{(2)}$ to, say, the quark propagator before 
combining with the antiquark propagator. 
There is a further $\mathcal{O}(v^2)$ operator that has a `D-wave' 
derivative term~\cite{Jones:1998ub, gray, daviesoldups}. However, since the mixing between the S-wave $\Upsilon$ 
states we are considering here and D-wave states is already suppressed 
by powers of $v^2$ and observed to be small~\cite{Daldrop} we can 
safely neglect that term at this order. 

We can then contruct a vector current in NRQCD matched 
order by order in $v^2$ and $\alpha_s$ to the continuum 
vector current whose matrix element appears in eq.~(\ref{eq:fdef}) and
eq.~(\ref{eq:vdecay}). The required matrix 
element can then be determined in our lattice calculation. 
At next-to-leading order in $v^2$ we write
\begin{eqnarray}
\label{eq:currmatch}
J_{V} &=& Z_V J_{V,\mathrm{NRQCD}} \\
&\equiv& Z_V ( J^{(0)}_{V,\mathrm{NRQCD}} + k_1 J^{(1)}_{V,\mathrm{NRQCD}} ). \nonumber
\end{eqnarray}  
where we have dropped the polarisation index for clarity. 
From tree-level matching of NRQCD and continuum vector currents 
$k_1 = 1/6$. Calculations in lattice perturbation theory show substantial 
corrections are possible at $\mathcal{O}(\alpha_s)$~\cite{Hart:2006ij}.
Here we determine both $k_1$ and the overall normalisation, $Z_V$, 
nonperturbatively on the lattice by comparing to continuum 
QCD perturbation theory for time-moments of the vector current-current 
correlator. This is described in detail in Appendix~\ref{appendix:zv} 
where the values of $Z_V$ and $k_1$ obtained are given. 
Since we use current-current correlator methods both to determine 
$Z_V/k_1$ and to determine the $b$ quark mass, we give here some 
of the notation and key equations that we use. 

\subsection{Moments}
\label{subsec:moments}
Vector bottomonium correlators $C_{V,\mathrm{NRQCD}}(t)$ are constructed using
$J_{V,\mathrm{NRQCD}}$ of eq.~(\ref{eq:currmatch}) at source and sink. 
The time-moments are then defined by:
\begin{eqnarray}
\label{eq:timemomv}
&& G_n^{V,\mathrm{NRQCD}} = \\
&& 2 \sum_{t} (t/a)^n {C}_{V,\mathrm{NRQCD}}(t) \exp(-[\overline{M}_{kin}-\overline{E}_0]t) \nonumber
\end{eqnarray}
where the factor of 2 is needed to relate the moments to continuum 
values since the nonrelativistic quark propagators only propagate 
forwards in time (see eq.~(\ref{eq:evol})) and not in both directions.  
The moment number $n=4, 6, 8, \ldots$. 
The exponential factor gives the NRQCD meson
correlator the correct time-dependence by restoring the `zero of energy' 
missing from the Hamiltonian. We use the values of $\overline{M}_{kin}$ 
and the spin-averaged ground-state energy, $\overline{E}_0$, obtained from 
the results of tuning the $b$ quark mass discussed above and given 
in Table~\ref{tab:mbc}. 

The NRQCD current-current correlator is related by the $Z_V$ renormalisation
factor for the current to that of the continuum current-current correlator, 
up to discretisation and relativistic corrections: 
\begin{equation}
G_n^{V} = Z_V^2 G_n^{V,\mathrm{NRQCD}} .
\label{eq:gandz}
\end{equation}
Continuum time-moments can be derived~\cite{firstcurrcurr} 
from $q^2$-derivative moments of 
the heavy-quark vacuum polarisation function that are 
calculable in continuum QCD perturbation theory~\cite{qcdpt1, qcdpt2, qcdpt3, qcdpt4, qcdpt5}. 
\begin{equation}
G_n^V = \frac{g_n^V(\alpha_s, \mu/m_b)}{[a\overline{m}_b(\mu)]^{n-2}}
\label{eq:gnpert}
\end{equation}
where $g_n^V$ is known through $\mathcal{O}(\alpha_s^3)$ either 
completely or approximately~\cite{qcdpt5} up to $n=22$. 
If we work in the $\overline{MS}$ scheme then 
$\overline{m}_b$ is the $b$ quark mass in that scheme at the scale $\mu$. 
Because the continuum perturbation theory has been obtained to such high 
order (next-to-next-to-next-to-leading order) we make use of this in our 
lattice calculation, rather than using lower order lattice QCD perturbation theory. 

\begin{table}
\begin{tabular}{llll}
\hline
\hline
$n$ &  $r_n^{(1)}$ & $r_n^{(2)}$ & $r_n^{(3)}$ \\
\hline
4 & 0.7623 & 0.2750 & -0.2347 \\
6 & 0.7727 & 0.7190 & -0.1865 \\
8 & 0.6102 & 0.7990 & -0.1398 \\
10 & 0.3500 & 0.7170 & -0.2420 \\
12 & 0.0248 & 0.5907 & -0.4147 \\
14 & -0.3475 & 0.5018 & -0.5806 \\
16 & -0.7563 & 0.5096 & -0.6972 \\
18 & -1.1935 & 0.6618 & -0.7592 \\
20 & -1.6550 & 0.9958 & -0.7894 \\
22 & -2.1360 & 1.5433 & -0.8546 \\
\hline
\hline
\end{tabular}
\caption{ Coefficients of the perturbative series 
$r_n^V = 1 + \sum_i r_n^{(i)}\alpha_s(\mu)$ for 
$\mu=\overline{m}_b(\mu)$. Results are taken 
from~\cite{qcdpt1, qcdpt2, qcdpt3, qcdpt4, qcdpt5} 
for the case with $n_l=4$ light quarks in the sea ($u$, $d$, 
$s$ and $c$) and no heavy ($b$) quarks ($n_h=0$), except 
for $r_n^{(3)}$, which uses the $n_l=4, n_h=1$ case 
from~\cite{qcdpt5}. 
}
\label{tab:rn}
\end{table}

To reduce discretisation errors and sensitivity to tuning 
of the lattice $b$ quark mass we make use of ratios 
of $G_n^{V}$ to the result obtained in the free case, i.e. 
by setting the gluon field, $U_{\mu}(x)$, to the unit matrix in 
color space and using tree-level values for all of the 
coefficients in the NRQCD Hamiltonian. Then  
\begin{equation}
G_n^{V,U=1} = 2 \sum_{t} (t/a)^n {C}_{V,\mathrm{NRQCD},U=1}(t) \exp(-2m_bt).
\label{eq:momu1}
\end{equation}
where now the zero of energy offset is simply twice the $b$ 
quark mass in the NRQCD action for that ensemble. 
Then 
\begin{eqnarray}
\label{eq:rndef}
R_n^V &\equiv& G_n^{V}/G_n^{V,U=1}  \\
 &=& r_n^V(\alpha_{\overline{MS}},\mu/m_b) \left[\frac{m_b}{\overline{m}_b(\mu)}\right]^{n-2} .\nonumber
\end{eqnarray}
$r_n^V$ is a perturbative expansion starting with 1, being 
the ratio of the expansion for $G_n^{V}$ to the leading, zeroth order, 
coefficient and $m_b/\overline{m}_b$ is the ratio of the lattice 
NRQCD quark mass and the mass in the $\overline{MS}$ scheme, i.e. 
the inverse of the mass renormalisation factor for lattice 
NRQCD\footnote{Note that, although we follow 
the same procedure as in~\cite{firstcurrcurr} our definitions of 
$R_n$ and $r_n$ given here are not the same.}. 

Table~\ref{tab:rn} gives the perturbative coefficients for 
the perturbative series for $r_n^V$ up to and including 
that for $\alpha_s^3$ (i.e. next-to-next-to-next-to-leading 
order). The analytic calculations for the coefficients 
are done for the vacuum polarisation function of a 
heavy quark loop with $n_l$ light quark loops and $n_h$ quark 
loops with the same mass as the heavy quark. 
Since we are working with $u$, $d$, $s$ and $c$ quarks in the sea but 
no $b$ quarks we use, where possible, $n_l=4$ and 
$n_h=0$. However this treats the $c$ quark mass as zero with 
potential errors from this of $\mathcal{O}(\alpha^2_s(m_c/m_b)^2$, 
i.e. $0.05\alpha^2_s$. The $\alpha_s^3$ coefficients are 
taken from the numerical approximate results in~\cite{qcdpt5} 
for the case $n_h=1$, $n_l=4$. From comparing exact formula 
for the first few moments~\cite{qcdpt1, qcdpt2, qcdpt3, qcdpt4}, results for different 
$n_l$ values and also the double-quark loop results of~\cite{Czakon:2007qi}  
it is clear that $n_h$ values differing by 1 make negligible difference at this 
order. 
We evaluate the series using $\alpha_s$ in the $\overline{MS}$ scheme 
at the scale $\mu$ equal to $\overline{m}_b(\overline{m}_b)$, thus 
avoiding additional logarithms of $\mu/m_b$. 
We take $\alpha_{\overline{MS}}(n_f=5, M_Z) = 
0.1185(6)$ and $\overline{m}_b(\overline{m}_b)$ = 4.18(3)~\cite{pdg}, 
giving  
$\alpha_{\overline{MS}}(n_f=4, \overline{m}_b(\overline{m}_b))$ = 0.2268(24).  
We take a truncation uncertainty in $r_n^V$ of $0.25\alpha_s^3$, which 
covers the perturbative error from treating $m_c$ as zero, the approximations 
at $\mathcal(O)(\alpha_s^3)$ and unknown coefficients of size 1 at 
$\mathcal{O}(\alpha_s^4)$. Almost all of the 
coefficients in Table~\ref{tab:rn} for orders below 
$\alpha_s^4$ are less than 1.  

Here we use $R_n$ to determine the NRQCD-continuum 
current matching parameter $Z_V$ which appears 
in $G_n^{V}$ in eq.~(\ref{eq:gandz}). For this we need to 
take ratios of powers of different moments to cancel factors 
of the quark mass, as described in 
Appendix~\ref{appendix:zv}. 

\subsection{The Quark Mass}
\label{subsec:mass}
We also determine the quark 
mass in the $\overline{MS}$ scheme from $R_n$ and for this we need to 
cancel factors of $Z_V$. Since $Z_V$ appears to the same 
power in each time-moment the ratio of successive time moments 
is independent of $Z_V$. To extract $\overline{m}_b$ in terms of 
a physical quantity we multiply by the ratio of the spin-average 
of the $\Upsilon$ and $\eta_b$ kinetic masses to twice the lattice 
NRQCD $b$ quark mass. Then  
\begin{equation}
\left[\frac{R_nr_{n-2}}{R_{n-2}r_n}\right]^{1/2}\frac{\overline{M}_{kin}}{2m_b} = \frac{\overline{M}_{\Upsilon,\eta_b}}{2\overline{m}_b(\mu)}.
\label{eq:masseq}
\end{equation}
for $n \ge 6$. From the right-hand side we can determine $\overline{m}_b$ 
using the experimental spin-average of the $\Upsilon$ and $\eta_b$ masses.  

In this way we are able to extract a great deal of 
physics from the correlator $C_{V,\mathrm{NRQCD}}(t)$ and the results will be 
given in Section~\ref{sec:results}. 

Having defined the NRQCD current from perturbation theory 
for the correlator time-moments we can fit $C_{V,\mathrm{NRQCD}}(t)$ 
as a function of $t$ in combination 
with the matrix of smeared correlators described above 
and determine the 
matrix elements of $J_{V}$ that give the 
$\Upsilon$ and $\Upsilon^{\prime}$ decay constants. 
Hence we can determine their leptonic widths from eq.~(\ref{eq:vdecay}). 
Simultaneously time-moments of $C_{V,\mathrm{NRQCD}}(t)$ 
can be directly compared 
to inverse $s$-moments of the 
$b$-quark contribution to $\sigma(e^+e^- \rightarrow \mathrm{hadrons})$.
In Section~\ref{sec:results} we give the lattice 
results for the moments from eq.~(\ref{eq:timemomv}) and compare 
to the values extracted 
from experimental results on
$e^+e^- \rightarrow \mathrm{hadrons}$~\cite{kuhnmc07}. 
A final application is that of the accurate determination of the 
$b$ quark mass in the $\overline{MS}$ scheme using eq.~(\ref{eq:masseq}). 

\section{Results}
\label{sec:results}

\subsection{$\Upsilon$ Leptonic Width}
\label{subsec:leptwidth}
For the leptonic width of the $\Upsilon$ we need to 
determine the matrix element given in eq.~(\ref{eq:overlap}) 
for the ground state, $m=0$. Since the ground-state dominates 
the correlation function at large values of $t$ this 
can be done from a meson correlation function in which 
we simply use the local current of eq.~(\ref{eq:currmatch}) 
at source and sink. By calculating separately the pieces 
corresponding to $J^{(0)}_{V,\mathrm{NRQCD}}$ and 
$J^{(1)}_{V,\mathrm{NRQCD}}$ we can generate correlation 
functions for different values of the current correction coefficient  
$k_1$ and investigate the effect of that on the determination of 
the overall renormalisation factor, $Z_V$, using continuum 
perturbation theory for the current-current correlator. This 
is described in Appendix~\ref{appendix:zv} and the results 
we obtain for $k_1$ and $Z_V$ are given in Table~\ref{tab:j1Z}. 

\begin{table}
\begin{tabular}{llllll}
\hline
\hline
Set &  $am_b$ & $c_4$ & $c(J^{(0)}_{V},0)$ & $c(J^{(1)}_{V},0)$ & $a^{3/2}f_{\Upsilon}\sqrt{M_{\Upsilon}}$ \\
\hline
1 & 3.297 & 1.0 & 0.9422(22) & -0.2439(6) & 1.334(4)(33) \\
1 & 3.297 & 1.22 & 0.9194(24) & -0.2355(7) & 1.346(4)(34) \\
1 & 3.42 & 1.0 & 0.9695(23) & -0.2373(6) & 1.376(4)(40) \\
2 & 3.25 & 1.22 & 0.9087(21) & -0.2371(6) & 1.304(3)(35) \\
\hline
3 & 2.66 & 1.0 & 0.7153(17) & -0.2360(6) & 0.929(2)(26) \\
4 & 2.62 & 1.20 & 0.6821(18) & -0.2268(6) & 0.914(3)(23) \\
\hline
5 & 1.91 & 1.0 & 0.4523(8) & -0.2109(4) & 0.604(1)(11) \\
\hline
\hline
\end{tabular}
\caption{ 
Columns 4 and 5 give the 
ground-state ($\Upsilon$) amplitudes for operators 
corresponding to the leading ($J^{(0)}_V$ abbreviating 
$J^{(0)}_{V,\mathrm{NRQCD}}$) and 
next-to-leading ($J^{(1)}_{V,\mathrm{NRQCD}}$) pieces of the 
NRQCD vector current for $b\overline{b}$ annihilation 
({\it{before}} multiplication by $Z_V$). 
Errors are statistical/fitting errors. 
Column 6 gives the corresponding values for 
the decay constant parameter $f_{\Upsilon}\sqrt{M_{\Upsilon}}$ 
in lattice units. The first error is statistical 
and the second from the $Z_V$ factor used to 
normalise the current.  
}
\label{tab:ampres}
\end{table}

\begin{figure}
\begin{center}
\includegraphics[width=\hsize]{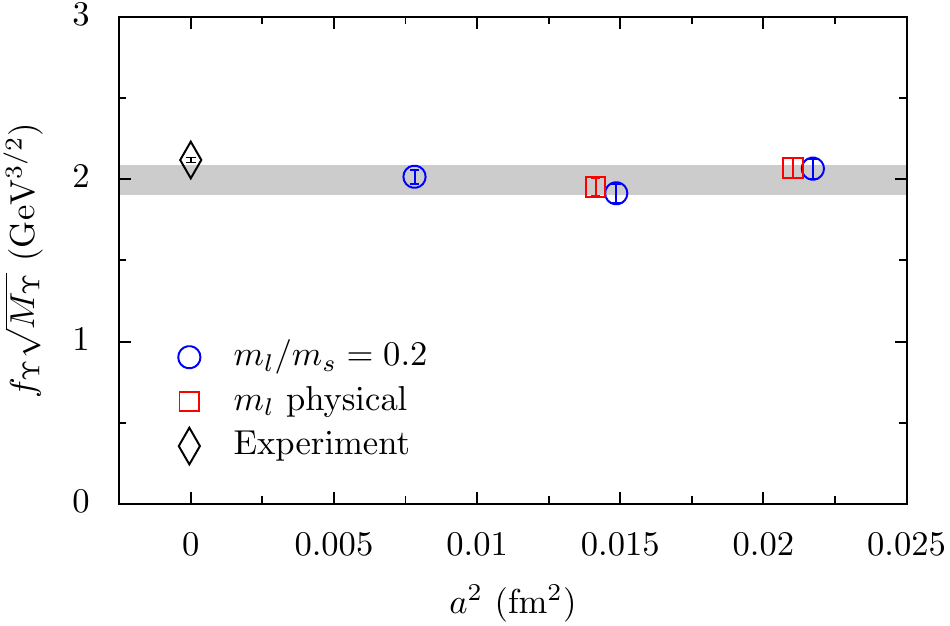}
\end{center}
\caption{The hadronic parameter $f_{\Upsilon}\sqrt{M_{\Upsilon}}$ 
from our lattice calculation plotted against the square of the 
lattice spacing. Open blue circules give results from 
sets 1, 3 and 5 and open red squares, sets 2 and 4. 
The grey band gives the physical value resulting from a fit 
to lattice-spacing and sea-quark mass dependence combined with 
other sources of systematic error as described in the 
text. The width of the grey band is then twice our total error. 
The black diamond gives the result derived from the experimental leptonic 
width using eq.~(\ref{eq:vdecay}). 
}
\label{fig:frootm}
\end{figure}

We fit the $2\times2$ matrix of correlation  
functions that correspond to $J^{(0)}_{V,\mathrm{NRQCD}}$ 
or $J^{(1)}_{V,\mathrm{NRQCD}}$ at 
source and sink to the form given in 
eq.~(\ref{eq:corrfit}) to extract the ground-state amplitude for 
each current, $c(J^{(j)}_{V,\mathrm{NRQCD}},0)$. 
We use a standard Bayesian fitting approach~\cite{gplbayes} 
constraining the amplitudes with priors of width between 
3 and 5
times the ground state amplitude, and energy differences between 
excited states with prior $600 \pm 300$ MeV. 
The amplitudes are given, for each set of gluon configurations, in 
Table~\ref{tab:ampres}. Statistical errors are small in this 
case. Note that the matrix elements of $J^{(1)}$ are negative, 
as expected from the form it takes (eq.~(\ref{eq:j1})).  

We can combine these amplitudes with the values for $k_1$ 
and $Z_V$ from Table~\ref{tab:j1Z} to form the 
amplitude for $J_{V,i}$. In fact what we do is to make up 
correlators that correspond to using operator $J_{V,i}$ at 
source and sink and fit that as above to obtain the ground-state 
amplitude. The two methods give the same result. 
Multiplying these amplitudes by $\sqrt{2}$ and $Z_V$  
gives $f_{\Upsilon}\sqrt{M_{\Upsilon}}$ 
in lattice units using eqs.~(\ref{eq:overlap}) and~(\ref{eq:fdef}). 
These values are given in the rightmost column in Table~\ref{tab:ampres}. 
The error on the amplitude is dominated by that 
from $Z_V$ and, in determining this error, we pay 
attention to the correlation between the uncertainty in $k_1$ and 
that in $Z_V$ as given in Table~\ref{tab:j1Z} 
\footnote{Note that the normalisation of the amplitudes that we are using 
here is that appropriate to that of the decay constant. A normalisation 
that is frequently used instead in NRQCD calculations~\cite{gray} is 
that appropriate to determining a wavefunction. The difference 
between the two normalisations for the amplitude is $\sqrt{6}$.}. 

$f\sqrt{M}$ is the hadronic parameter that is the direct output from 
our lattice QCD calculation and from which we must determine a 
physical result to be compared with experiment. 
The results for $f\sqrt{M}$ 
(converted to physical units 
using the lattice spacing values in Table~\ref{tab:params}) 
are plotted against the square of the lattice 
spacing in Figure~\ref{fig:frootm}. We see relatively 
little dependence on either the lattice spacing or 
the sea quark masses. Table~\ref{tab:ampres} also 
shows that changing 
the coefficient $c_4$ in the 
NRQCD action has insignificant effect. 
Changing the $b$ quark mass from 3.297 (well-tuned, and plotted
on Figure~\ref{fig:frootm}) on 
set 1 to 3.42 (badly-tuned) has a visible effect and we can 
use this to estimate tuning uncertainties. 

To obtain a physical result from our lattice values we 
must fit them as a function of lattice spacing and 
of sea light quark mass. Our results on sets 2 and 4 correspond 
to a physical value of the $u/d$ sea mass but in order 
to incorporate fully any lattice spacing dependence we need also 
to include sets 1, 3 and 5 in the fit. 

For the fits we use the method developed in~\cite{dowdallr1}
allowing for both `standard' discretisation errors that 
come from the gluon or light quark actions but also 
higher order discretisation errors in the NRQCD action that 
may have $am_b$-dependent coefficients. 
Adding these terms 
in to our fit allows them to contribute to the error on the 
physical result. Since we will use this 
fit for other quantities we simply denote the hadronic parameter 
which is the subject of the fit by $h$, here $f_{\Upsilon}\sqrt{M_{\Upsilon}}$. 
We use the form: 
\begin{eqnarray}
\label{eq:hfit}
h(a,m_{sea}) &=& h_{\mathrm{phys}} [ 1 + b_l \delta m_{sea}/(10m_s) + \\
&& \sum_{j=1,3} c_j(a\Lambda)^{2j} + \nonumber \\ 
&& \sum_{j=1,2} (a\Lambda)^{2j} [c_{jb}\delta x_m + c_{jbb} (\delta x_m)^2 ] \nonumber
\end{eqnarray}
The second term in square brackets accounts for the sea quark 
mass dependence using a simple linear dependence expected at 
leading order. Since the sea mass dependence is very small this 
is sufficient. 
$\delta m_{sea}$ is the difference between the sum of twice the light and 
strange sea quark masses and its physical value. The physical 
values of the $s$ quark mass (for the lattice spacing 
values in Table~\ref{tab:params}) are given in~\cite{dowdallr1}
and we take the ratio of physical $s$ to light quark mass as 27.5~\cite{pdg}. 
The factor of $10m_s$ in the denominator is a convenient way (cancelling 
the mass renormalisation) to 
introduce the chiral scale of 1 GeV. 
The third term accounts for standard discretisation errors, 
using a scale of $\Lambda$ where we set 
$\Lambda =$ 500 MeV. The terms containing $\delta x_m$ allow for 
discretisation effects with dependence on the $b$ quark mass in the NRQCD action 
by modelling this with a linear and quadratic term. $\delta x_m$ is 
chosen to vary from -0.5 to 0.5 across our range of masses by 
taking $\delta x_m = (am_b-2.7)/1.5$.
We use eq.~(\ref{eq:hfit}) within a Bayesian fitting approach~\cite{gplbayes}
taking priors on the coefficients of the fit as 0.0(1.0) except 
for $c_1$ which we take as 0.0(0.5) since tree-level $a^2$ errors 
are absent from our action and so we expet this term to be at most 
$\mathcal{O}(\alpha_s)$. We take a prior width of 50\% on $h_{\mathrm{phys}}$.  

The physical value for the leptonic width that we obtain from the fit
($\chi^2=0.57$ for 5 degrees of freedom)  
is 1.995(90) ${\mathrm{GeV}}^{3/2}$. To this we must add systematic 
errors corresponding to:
\begin{itemize} 
\item missing higher order current corrections. These are of $\mathcal{O}(v^4)$ 
in a relativistic expansion and so this can be estimated at 1\% for 
the $\Upsilon$, where $v^2 \approx$ 10\%. 
\item uncertainty in tuning the $b$ quark mass. This is at most 1\% 
from Table~\ref{tab:params} and~\ref{tab:mbc} and is mainly a 
consequence of the 
uncertainty in the determination of the lattice spacing giving 
the physical value for $M_{kin}$. 
From Table~\ref{tab:ampres}, comparing results from $am_b$ = 3.297 
and 3.42, we see that this leads to a possible 1\% uncertainty in 
the decay constant. 
\item electromagnetic effects (missing from our calculation). 
Electromagnetic effects in the $\Upsilon$ 
and $\eta_b$ masses have alresdy been accounted for but at 0.02\% are 
negligible. Effects of the decay constant arising from the additional 
electromagnetic attraction of quark and antiquark can be estimated from 
a potential model, to give 0.2\%~\cite{fdsupdate}. 
\item missing $b$ quarks in the sea. The effect of $b$ quarks in the sea 
induces a short-distance potential~\cite{fdsupdate} between heavy quarks similar to the 
hyperfine potential which causes differences between $f_{\Upsilon}$ and 
$f_{\eta_b}$. Since these differences are small~\cite{McNeile:2012qf} the 
effect is negligible. 
\end{itemize}

This gives a final physical result of 1.995(94) ${\mathrm{GeV}}^{3/2}$ with 
error budget given in Table~\ref{tab:errorbudget}. Errors are
dominated by those from the lattice spacing dependence and $Z_V$. 
Dividing by the square root of the experimental $\Upsilon$ mass 
gives a decay constant result with a 5\% uncertainty:
\begin{equation}
f_{\Upsilon} = 0.649(31) \mathrm{GeV} .
\label{eq:fres}
\end{equation}
In Section~\ref{sec:conclusions} we will include this value in a summary 
plot of decay constants from across the meson spectrum. 

We can use the experimental value of the $\Upsilon$ leptonic width, 
1.340(18) keV to determine a value of $f_{\Upsilon}\sqrt{M_{\Upsilon}}$ 
of 2.119(14) ${\mathrm GeV}^{3/2}$ (and a value for $f_{\Upsilon}$ of 
0.689(5) GeV) using eq.~(\ref{eq:vdecay}). 
The value for $f_{\Upsilon}\sqrt{M_{\Upsilon}}$ is marked on 
the plot in Figure~\ref{fig:frootm} for comparison to our results. 
The agreement is good, within 1.5$\sigma$. The value for 
$f_{\Upsilon}$ will be compared to our results in Figure~\ref{fig:decaysum} 
in the Conclusions. 
Alternatively we can compute a leptonic width from our 
result for $f_{\Upsilon}\sqrt{M_{\Upsilon}}$ using 
eq.~(\ref{eq:vdecay}), along with the experimental value for 
the $\Upsilon$ mass and $\alpha_{QED}$. We obtain $\Gamma(\Upsilon \rightarrow e^+e^-) =$ 1.19(11) keV, again in good agreement with 
the experimental result. 

\begin{table}
\begin{tabular}{lrr}
\hline
\hline
Error &  $f_{\Upsilon}\sqrt{M_{\Upsilon}}$ & $\overline{m}_b(10 \mathrm{GeV})$ \\
\hline
Statistics & 0.3 & 0.0  \\
$Z_V/k_1$ & 2.5 &  0.3 \\
perturbation theory/$\alpha_s$ & - & 0.3 \\
uncertainty in $a$ & 1.6 & 0.0 \\ 
lattice spacing dependence & 3.4 & 0.4 \\
sea-quark mass dependence & 1.0 & 0.0 \\
$b$-quark mass tuning & 1.0 & 0.0 \\
NRQCD systematics & 1.0 & 0.3 \\
electromagnetism\ $\eta_b$ annihilation & 0.0 & 0.0\\
\hline 
total & 4.8 & 0.7 \\
\hline
\hline
\end{tabular}
\caption{ 
Error budget for the quantities determined in this paper. 
Errors are given as a percentage of the final answer. 
For $f_{\Upsilon}\sqrt{M_{\Upsilon}}$ the perturbation theory errors are 
included in the errors from $Z_V/k_1$ and not separated. 
Errors from the lattice spacing dependence are determined 
from the fit and include NRQCD uncertainties. 
Errors smaller than 0.1\% are denoted by 0.0. 
}
\label{tab:errorbudget}
\end{table}
 
\subsection{$\Upsilon^{\prime}$ Leptonic Width}
\label{subsec:upsprime}
To determine the $\Upsilon^{\prime}$ leptonic width we can 
make use of the ratio of amplitudes with that of the $\Upsilon$ to cancel 
$Z_V$ and reduce the uncertainty from that source. We also expect 
lattice spacing and tuning uncertainties to cancel to a large extent. 
The ratio of the amplitudes for $J_{V,i}$ in the ground and 
first-excited states gives:
\begin{equation}
A = \frac{\langle 0 | J_{V,i} | \Upsilon^{(1)} \rangle}{\langle 0 | J_{V,i} | \Upsilon^{(0)} \rangle} 
= \frac{f_{\Upsilon^{\prime}}}{f_{\Upsilon}}\sqrt{\frac{M_{\Upsilon^{\prime}}}{M_{\Upsilon}}} .
\label{eq:amprat}
\end{equation}
To determine the properties of excited states accurately it is important to use 
smeared sources, 
as described in Section~\ref{subsec:mescorr} 
and used in~\cite{dowdallr1} to obtain excited state masses. 
Here we combine results from a local source (corresponding to $J^{(0)}_{V,\mathrm{NRQCD}}$) 
and sink operator $J^{(1)}_{V,\mathrm{NRQCD}}$ with the matrix of correlators used 
in~\cite{dowdallr1}. We use the $3\times 3$ matrix of 
smearings called $l$, $g$ and $e$ in~\cite{dowdallr1}. 
The `l' smearing is the
local operator corresponding 
to $J^{(0)}_{V,\mathrm{NRQCD}}$ so the $ll$ correlator 
already has this operator at source and sink. 
The other correlators 
in the matrix ($lg$, $ge$, $gg$ etc~\cite{dowdallr1}) add information 
about the excited states.  
From fits to all of the correlators we can then 
extract matrix elements for $J^{(0)}_{V,\mathrm{NRQCD}}$ and 
$J^{(1)}_{V,\mathrm{NRQCD}}$ in both the ground-state and excited states. 
We rapidly lose statistical accuracy, however, and so restrict ourselves 
here to the ground and first excited state. 
We use 9-exponential fits of the form given in eq.~(\ref{eq:corrfit}) 
with standard priors on energies and amplitudes ($600\pm 300$ MeV 
on excited state mass splittings and an amplitude prior width 
corresponding to 3--5 times the ground state local amplitude).  

Table~\ref{tab:ampex} gives results for the matrix elements for 
$\Upsilon$ and $\Upsilon^{\prime}$ for sets 1, 3 and 5. 
The results for the $\Upsilon$ agree with those from Table~\ref{tab:ampres} 
but are more accurate because of the additional information being 
used here. For the $\Upsilon^{\prime}$ we see that the matrix 
element for $J^{(0)}_{V,\mathrm{NRQCD}}$ is smaller in magnitude than that for the 
$\Upsilon$ and the matrix element for $J^{(1)}_{V,\mathrm{NRQCD}}$ is bigger in 
magnitude. The table also gives the ratio, $A$, above, obtained by 
combining the results using the value of $k_1$, along with its uncertainty, 
obtained in Appendix~\ref{appendix:zv}. Our set 1 results are for 
our mistuned (by 4\%) $b$ quark mass but we expect this to make 
little difference to the ratio. 

\begin{table}
\begin{tabular}{llll}
\hline
\hline
 &  1  &  3 & 5 \\
\hline
$am_b$ & 3.42 & 2.66 & 1.91  \\
$c(J^{(0)},0)$ & 0.9720(2) & 0.7160(1) & 0.4523(1)\\
$c(J^{(1)},0)$ & -0.2376(1) & -0.2362(1) & -0.2109(1) \\
$c(J^{(0)},1)$ & 0.791(8) & 0.570(8) &  0.360(2)\\
$c(J^{(1)},1)$ & -0.277(4) & -0.261(5)  & -0.225(1)\\
$A$ & 0.854(16) & 0.813(14) & 0.774(7) \\
\hline
\hline
\end{tabular}
\caption{ 
Amplitudes for the operators corresponding to the 
leading $(J^{(0)}_{V,\mathrm{NRQCD}})$ (abbreviated to $J^{(0)}$) 
and next-to-leading $(J^{(1)}_{V,\mathrm{NRQCD}})$ 
pieces of the NRQCD vector current for both 
the $\Upsilon$ $(0)$ and $\Upsilon^{\prime}$ $(1)$ 
mesons. $A$ is the ratio given in eq.~(\ref{eq:amprat}).
The error on $A$ includes the error from the 
uncertainty in $k_1$.
Results are sets 1 (with $am_b=3.42$), 3 and 5. 
}
\label{tab:ampex}
\end{table}

\begin{figure}
\begin{center}
\includegraphics[width=\hsize]{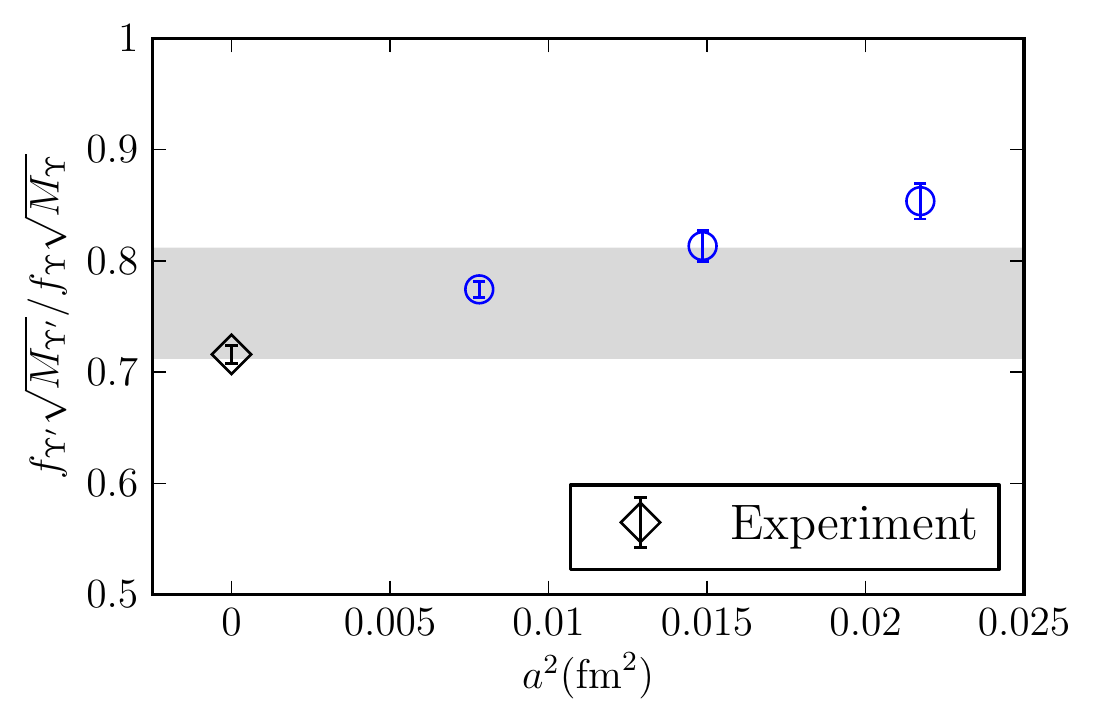}
\end{center}
\caption{The ratio of hadronic parameters $f\sqrt{M}$ 
for $\Upsilon^{\prime}$ to $\Upsilon$ 
plotted against the square of the 
lattice spacing. 
The grey band gives the physical value resulting from a fit 
to lattice-spacing combined with 
other sources of systematic error as described in the 
text. The width of the grey band is then twice our total error. 
The black burst gives the result derived from the experimental leptonic 
widths using eq.~(\ref{eq:vdecay}). 
}
\label{fig:frat}
\end{figure}

Figure~\ref{fig:frat} shows the results for $A$ plotted against the 
square of the lattice spacing. 
We fit the ratio as a function of lattice spacing 
using the fit form given in eq.~(\ref{eq:hfit}). 
We take the 
same set of priors as those described earlier for the 
decay constant except that we increase the prior on the 
conventional $a^2$ dependence to 1.0 since strong $a$-dependence 
is seen (the fit chooses a slope of 0.8(7) for this term). 
We allow 
for light sea quark mass dependence as before, although we might 
expect these effects to also cancel to a large extent. Our 
results do not have a lot of information about sea-quark mass dependence since they 
all come from ensembles with similar light sea quark masses in 
units of the $s$ quark mass. This fit parameter then simply contributes 
3.5\% to the error on the ratio.  

The physical value for $A$ obtained from the fit is 0.762(50), with 
the uncertainty dominated by the $a$-dependence, sea-quark mass dependence
and statistics. 
To this we should add an additional systematic error of 
1\% for missing $v^4$ terms in the NRQCD vector current, 
giving 0.762(51). We do not expect any other sources of systematic 
error to be significant, for example tuning errors will largely 
cancel. 
The experimental value for the ratio of $f\sqrt{M}$ for $\Upsilon^{\prime}$ 
and $\Upsilon$ obtained from their masses and decay widths 
to $e^+e^-$ via eq.~(\ref{eq:vdecay}) is 0.716(8), marked with a black 
diamond on Figure~\ref{fig:frat}. Our result is in good agreement with 
that determined from experiment, but a lot less accurate. 

Using the value from Section~\ref{subsec:leptwidth} for the 
decay constant of the $\Upsilon$ and the experimental ratio 
of masses, we obtain 
\begin{equation}
f_{\Upsilon^{\prime}} = 0.481(39) \mathrm{GeV}. 
\label{eq:fupsprime}
\end{equation}
This will be shown on our summary plot in Section~\ref{sec:conclusions}, 
where it can be compared to the experimental result of 0.479(4) GeV 
determined from the leptonic width. 

We can also use our physical value for the ratio $A$, the 
experimental leptonic width for the $\Upsilon$ and the experimental 
mass ratio to determine a result for the leptonic width of 
the $\Upsilon^{\prime}$. We obtain 0.69(9) keV, again in good 
agreement with the experimental result of 0.612(11) keV~\cite{pdg}.  

\subsection{$R_{e^+e^-}$}
\label{subsec:R}

\begin{figure}
\begin{center}
\includegraphics[width=\hsize]{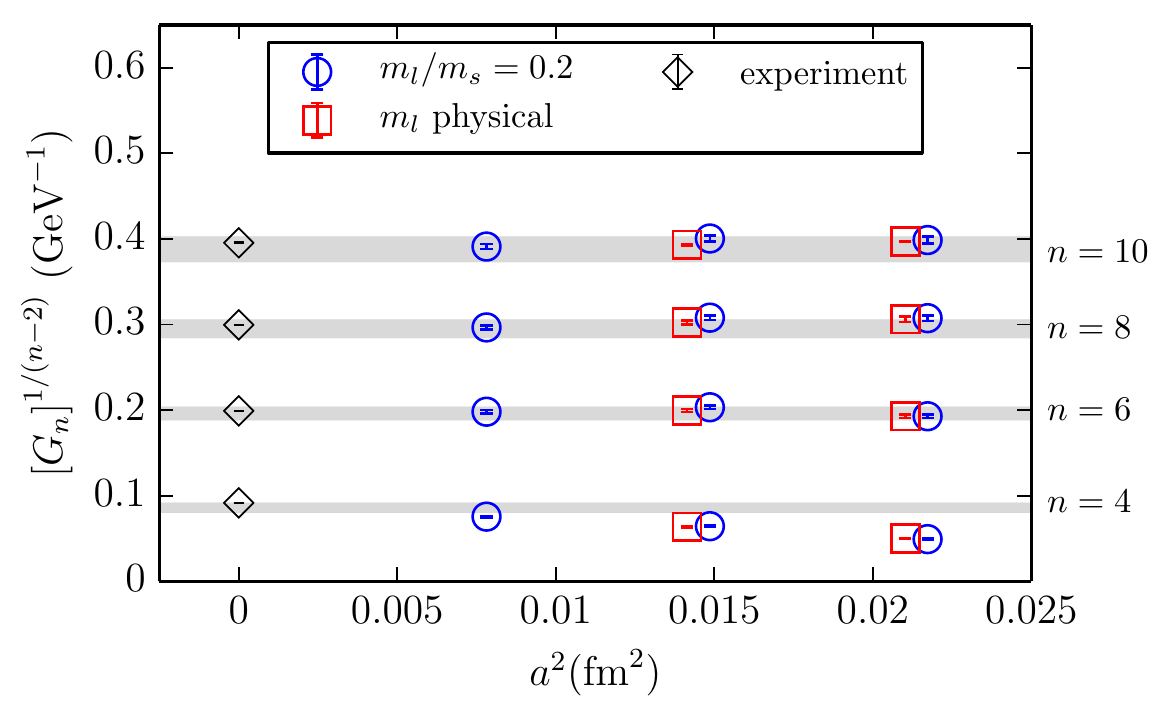}
\end{center}
\caption{
The $1/(n-2)$th root of the $n$th time-moment of 
the vector current-current correlator for (bottom to 
top) $n=$ 4, 6, 8 and 10, plotted 
against the square of the lattice spacing. 
Blue open circles are for sets 1, 3 and 5 and red open 
squares for 2 and 4. The errors on the points include 
the uncertainty from $Z_V$, $k_1$ and the lattice spacing. 
The grey band shows our physical 
result with its full error including that from fitting 
the points and additional systematic errors. 
The black diamonds are results determined from 
the experimental data for $R_{e^+e^-}$
}
\label{fig:moms}
\end{figure}

\begin{table}
\begin{tabular}{lllllll}
\hline
\hline
Set &  $am_b$ & $c_4$ & $n=4$ & $n=6$ & $n=8$ & $n=10$\\
\hline
1 & 3.297 & 1.0 & 0.0492(6) & 0.193(2) & 0.307(3) & 0.399(4) \\
1 & 3.297 & 1.22 & 0.0487(6) & 0.192(2) & 0.306(3) & 0.398(4) \\
1 & 3.42 & 1.0 & 0.0453(6) & 0.185(2) &  0.299(3) & 0.387(4) \\
2 & 3.25 & 1.22 & 0.0500(7) & 0.193(2) & 0.306(3) & 0.397(4) \\
\hline
3 & 2.66 & 1.0 & 0.0643(8) & 0.203(2) & 0.308(3) & 0.401(4) \\
4 & 2.62 & 1.20 & 0.0635(7) & 0.200(2) & 0.302(2) & 0.393(3) \\
\hline
5 & 1.91 & 1.0 & 0.0755(9) & 0.198(2) & 0.297(2) & 0.391(3) \\
\hline
\hline
\end{tabular}
\caption{ 
Values for $(G_n)^{1/(n-2)}$ in $\mathrm{GeV}^{1/2}$ for 
$n=4$, 6, 8 and 10 for 
each ensemble and set of parameters that we use. 
Errors are from statistics, $Z_V$, $k_1$ and the determination of the 
lattice spacing. 
}
\label{tab:moms}
\end{table}

Given a correctly normalised vector current operator, as described in 
the previous section and Appendix~\ref{appendix:zv}, we can return to give 
values for the time-moments from eqs.~(\ref{eq:timemomv}) and~(\ref{eq:gandz}). 
Results are given for 
$(G^V_n)^{1/(n-2)}$ for $n=4$ to 10 
in Table~\ref{tab:moms} on each of our ensembles. 
The power $1/(n-2)$ is taken to 
reduce all the moments to the same dimension. 
Figure~\ref{fig:moms} shows the results 
plotted against the square of the lattice spacing. 
The errors on the points come from uncertainty in $Z_V$ and $k_1$ 
(taking account of their correlation) and in the lattice spacing. 

We fit the results for each moment as a function of lattice 
according to the standard fit in eq.~(\ref{eq:hfit}) using 
the priors given there, except for the case $n=4$ where we 
increase the width of the prior on the $a^2$ term to 3.0. 
Very strong $a$-dependence is seen for that case in 
Figure~\ref{fig:moms}, consistent with the fact that this moment 
has a big contribution from relatively large spatial momenta. 

To the fitted values we must add systematic errors from:
\begin{itemize}
\item NRQCD systematics. Our NRQCD vector current is missing relativistic 
corrections at $\mathcal{O}(v^4)$. In Appendix~\ref{appendix:zv} 
we estimate that the important spatial momenta for moment $n$ 
correspond to $v^2 \approx 1/n$. We can test this expectation 
by studying 
the effect of the relativistic corrections we include in $J^{(1)}_{V,\mathrm{NRQCD}}$ 
(i.e. at $\mathcal{O}(v^2)$). We find shifts for the different 
moments compared to the leading order result on fine set 5 lattices 
of $n=4:25\%$, $n=6:22\%$, $n=8:18\%$, $n=10:15\%$. This agrees 
reasonably well, but is a bit larger, as $n$ increases, than 
the naive expectation of $1/n$.  
To determine the systematic error from missing $v^4$ terms, we 
therefore take the square of the result we see at $\mathcal{O}(v^2)$, 
giving an uncertainty in the moment of 6\% for $n=4$, 4\% for $n=6$, 
3\% for $n=8$ and 2\% for $n=10$. 
In the $1/(n-2)$th root of the 
moment, the quantity determined here, the systematic error becomes:
3\% for $n=4$, 1\% for $n=6$, 0.5\% for $n=8$ and 0.4\% for $n=10$. 
\item $b$ quark mass tuning. The results in Table~\ref{tab:moms} 
for set 1 show that mistuning the $b$ quark mass has a visible effect, 
with an increasing lattice value for $am_b$ giving a smaller value 
for the moment.  
This is most evident for the 4th moment. 
Mass tuning relies on the lattice spacing determination and the tuning error arises 
from the uncertainty in the lattice spacing (since $\overline{M}_{kin}$ is 
determined more accurately than $a$). Here the change in the quark 
mass counteracts the change in lattice spacing so that tuning uncertainties 
are relatively small. We take 1.5\% for the 4th moment and 
0.5\% for the others. 
\item electromagnetism. The effect of electromagnetism in experiment (e.g. photons in the 
final state) missing from our calculation were estimated for the charm case in~\cite{psipaper}. 
The uncertainties were very small there, and will be negligible here because of 
the smaller electric charge of the $b$ quark. 
\end{itemize}

Including these systematic uncertainties along with those from the fit above gives 
the physical results from our calculation: 
\begin{eqnarray}
\left(G_4^V\right)^{1/2} &=& 0.086(5)(3) \mathrm{GeV}^{-1} \nonumber \\
\left(G_6^V\right)^{1/4} &=& 0.196(8)(2) \mathrm{GeV}^{-1} \nonumber \\
\left(G_8^V\right)^{1/6} &=& 0.295(11)(2) \mathrm{GeV}^{-1} \nonumber \\
\left(G_{10}^V\right)^{1/8} &=& 0.388(15)(2) \mathrm{GeV}^{-1} . 
\end{eqnarray}
The first error is from the fit, taking into account lattice spacing 
dependence, and the second error is from systematic errors estimated 
above. 

The results agree well with the values extracted for 
the $q^2$ derivative moments, $\mathcal{M}_k$ ($n=2k+2$), 
of the $b$ quark vacuum 
polarization using experimental values for 
$R_{e^+e^-} = \sigma(e^+e^- \rightarrow \mathrm{hadrons})/\sigma_{pt}$~\cite{kuhn09update}. 
These values, 
appropriately normalised for the comparison to ours, are:
\begin{eqnarray}
(M^{\mathrm{exp}}_1 4!/(12\pi^2 e_b^2))^{1/2} &=& 0.0915(3) \, \mathrm{GeV}^{-1} \nonumber \\
(M^{\mathrm{exp}}_2 6!/(12\pi^2e_b^2))^{1/4} &=& 0.1991(5) \, \mathrm{GeV}^{-1} \nonumber \\
(M^{\mathrm{exp}}_3 8!/(12\pi^2e_b^2))^{1/6} &=& 0.2996(5) \, \mathrm{GeV}^{-1} \nonumber \\
(M^{\mathrm{exp}}_{4} 10!/(12\pi^2e_b^2))^{1/8} &=& 0.3955(6) \, \mathrm{GeV}^{-1}. 
\label{eq:rnexp}
\end{eqnarray}
These are shown as the black diamonds in Figure~\ref{fig:moms}. 
Our results from lattice NRQCD have significantly larger errors 
than those derived from experiment. As discussed above this is primarily 
because of NRQCD systematic errors for these low moments. 
Nevertheless this provides a good 
test of QCD that is complementary to our tests using the leptonic width 
in Sections~\ref{subsec:leptwidth} and~\ref{subsec:upsprime}. 

One application of our results is to the determination of the effect 
on the anomalous magnetic moment, $a_{\mu}=(g_{\mu}-2)/2$, of the 
$\mu$ lepton from coupling to a 
$b$ quark loop i.e. that part of the `hadronic vacuum polarisation' 
(HVP) contribution that comes from $b$ quarks. We use the method developed 
in~\cite{Chakraborty:2014mwa} which converts the moments determined above 
to $q^2$-derivatives of the hadronic vacuum polarisation and thereby 
determines, via Pad\'{e} approximants, 
the $q^2$-dependence of the integrand required for the contribution to $a_{\mu}$. 
We obtain $a_{\mu}^b = 0.271(37)\times 10^{-10}$ from our lattice results.
This can be compared with the result using our approach but substituting 
the values for the moments extracted from experiment as given 
in eq.~(\ref{eq:rnexp}) of $0.307(2)\times 10^{-10}$ or that 
from using QCD perturbation theory~\cite{Bodenstein:2011qy} 
of $0.29(1) \times 10^{-10}$. 

Our error is sizeable and dominated by NRQCD 
systematics. This is because the small $q^2$ region dominates the 
integral for the contribution to $a_{\mu}$ and the integrand there is 
given almost entirely by the fourth time-moment, which is the one we can 
determine least well using NRQCD. The $b$-quark piece of the HVP contribution  
to $a_{\mu}$ is very small, however, compared to the total hadronic 
vacuum polarisation contribution which is $\approx 700 \times 10^{-10}$. 
Its error is therefore not critical 
to the issue of reducing the theoretical uncertainty in the Standard Model 
result for $a_{\mu}$. It is nevertheless important to have results 
for this quantity from lattice QCD as a cross-check of other methods. 
Results using a relatvistic formalism for the $b$ quark should 
give smaller errors in future for this quantity. See~\cite{Davies:2013dem} for preliminary 
results using the HISQ formalism~\cite{hisqdef} for the $b$ quarks. 

\subsection{Mass of the $b$ quark}
\label{subsec:mb}

We can also use our calculation of the time-moments of the 
vector current-current correlator to determine the mass of 
the $b$ quark. The continuum expression for the moments in eq.~(\ref{eq:gnpert})
contains a perturbative series divided by powers of
 the $b$-quark mass in the $\overline{MS}$ 
scheme. To obtain the continuum moments from the lattice 
moments requires multiplication by 
the current renormalisation factor $Z_V$ (as in Section~\ref{subsec:R})
and this introduces significant uncertainties in using the 
moments directly. We can cancel $Z_V$, however, in ratios of successive moments 
(in which the mass does not cancel) 
and this gives a much more accurate and robust method, because at the 
same time we can reduce other systematic errors. 
We also multiply by the ratio of the spin-average of $\Upsilon$ 
and $\eta_b$ kinetic masses to twice the lattice $b$ quark mass. 
This cancels factors of the lattice $b$ quark mass and allows 
us to extract the $b$ quark mass in the $\overline{MS}$ 
scheme as a ratio to the spin-average of experimental $\Upsilon$ 
and $\eta_b$ masses. The relevant equations are given in eqs.~(\ref{eq:rndef}) 
and~(\ref{eq:masseq}), yielding. 
\begin{equation}
\overline{m}_b(\mu) = \frac{\overline{M}_{\Upsilon,\eta_b}}{2}\left[ \frac{R_{n-2}r_n}{R_nr_{n-2}} \right]^{1/2} \frac{2m_b}{\overline{M}_{kin}} 
\label{eq:massex}
\end{equation}

\begin{figure}
\begin{center}
\includegraphics[width=\hsize]{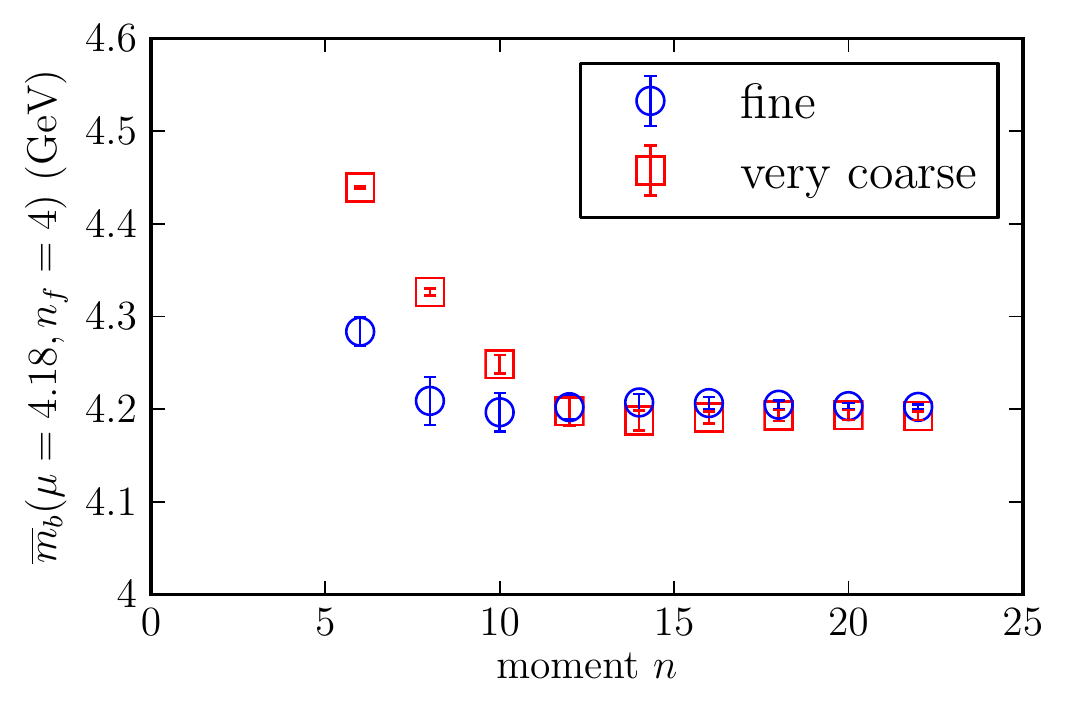}
\end{center}
\caption{The $b$ quark mass in the $\overline{MS}$ scheme 
determined from our calculation of time-moments of the 
vector current-current correlator as a function of the 
moment number, $n$, in eq.~(\ref{eq:massex}). 
Blue open circles are for the fine set 5 lattices and 
the red open circles for the very coarse set 1. 
The errors on the points are dominated by the 
uncertainty in the value of $k_1$, the current 
correction coefficient. 
}
\label{fig:massn}
\end{figure}

\begin{table}
\begin{tabular}{llllll}
\hline
\hline
Set &  $am_b$ & $c_4$ & $n=14$ & $n=18$ & $n=22$ \\
\hline
1 & 3.297 & 1.0 & 4.187(11) & 4.193(6) & 4.192(5) \\
1 & 3.297 & 1.22 & 4.188(11) & 4.194(5) & 4.193(6) \\
1 & 3.42 & 1.0 & 4.189(13) & 4.197(6) &  4.193(5) \\
2 & 3.25 & 1.22 & 4.192(11) & 4.197(6) & 4.196(5) \\
\hline
3 & 2.66 & 1.0 & 4.209(10) & 4.210(7) & 4.208(4) \\
4 & 2.62 & 1.20 & 4.210(10) & 4.214(7) & 4.211(4) \\
\hline
5 & 1.91 & 1.0 & 4.207(9) & 4.204(5) & 4.202(3) \\
\hline
\hline
\end{tabular}
\caption{ 
Values for the $b$ quark mass in GeV in the $\overline{MS}$ scheme, 
determined from eq.~(\ref{eq:massex}) for $n=14$, 18 and 22 
on each set of configurations that we use. The errors 
are those from the uncertainty in $k_1$; statistical errors 
are very small here. 
}
\label{tab:mb}
\end{table}

Table~\ref{tab:mb} gives our results from eq.~(\ref{eq:massex}) for 
$n = 16$, 18 and 20 on all sets and Fig.~\ref{fig:massn} shows results 
from sets 1 and 5 as a function of $n$. 
We expect to see $\overline{m}_b$ 
reach a plateau as $n$ increases when internal spatial momenta 
in the current-current correlator become small enough for our NRQCD 
vector current to be a good approximation to the continuum 
vector current and hence to the continuum perturbation theory. 
In a similar way to that for $Z_V$ (see Appendix~\ref{appendix:zv}) we 
see that this happens down to moment numbers as low as $n=8$ in eq.~(\ref{eq:massex}) 
on the fine lattices, but needs somewhat higher moment numbers 
on the coarser lattices. The results for the two, very different, 
lattice spacing values agree where they have both reached a plateau.  

\begin{figure}
\begin{center}
\includegraphics[width=\hsize]{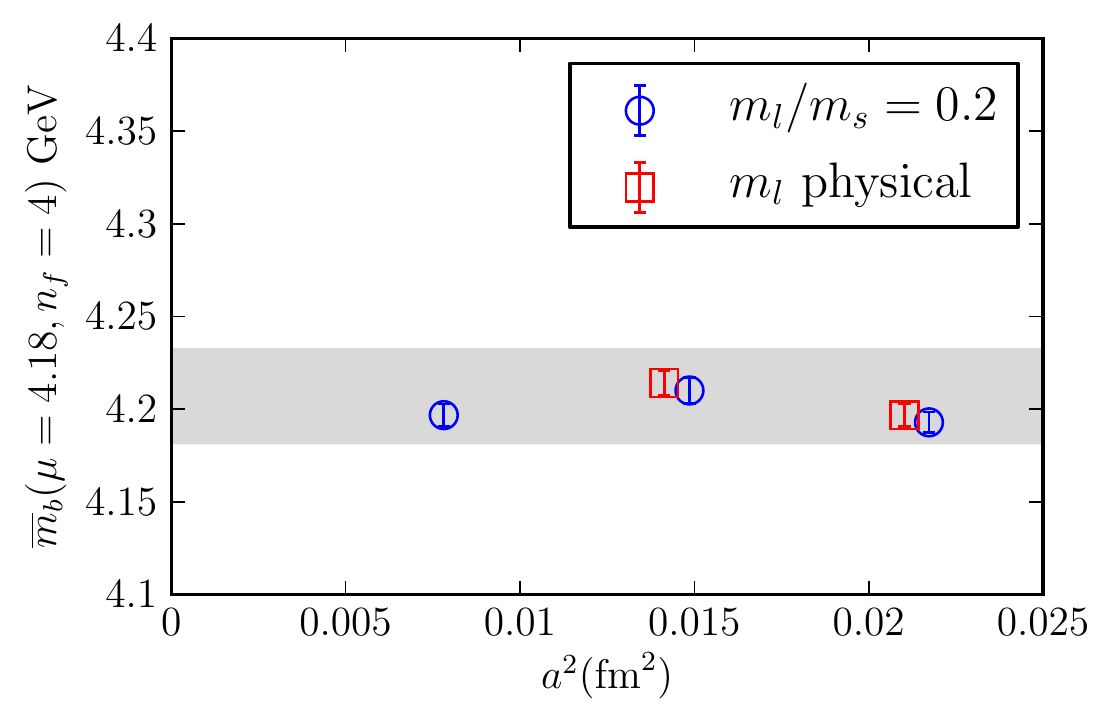}
\end{center}
\caption{The $b$ quark mass in the $\overline{MS}$ scheme 
determined from our calculation of time-moments of the 
vector current-current correlator using 
eq.~(\ref{eq:massex}) with $n=18$. 
Blue open circles are from sets 1, 3 and 5 and 
red open squares from sets 2 and 4. 
The errors on the points include uncorrelated 
errors only and are dominated by the 
uncertainty in the value of $k_1$, the current 
correction coefficient. 
The grey band is the physical value we obtain with its 
total error, including the error from lattice spacing 
and quark mass dependence obtained from a fit to the points 
as well as additional systematic errors described in the text. 
}
\label{fig:massa2}
\end{figure}

We consequently take results from $n=18$ for our central value 
and plot these as a function of $a^2$ in Fig.~\ref{fig:massa2}.  
There is very little dependence on sea-quark mass or lattice 
spacing. Indeed, as Table~\ref{tab:mb} also shows, there is very little 
dependence on the $c_4$ coefficient in the NRQCD action or on the 
lattice $b$ quark mass (since this dependence is largely cancelled 
by $\overline{M}_{kin}$). The errors on the masses are dominated by that from 
the uncertainty in the value of $k_1$; statistical errors are 
negligible here. As expected, the error from changing $k_1$ falls 
as $n$ increases and the moments become more nonrelativistic.   

To determine a physical value for the mass, we again fit the 
results as a function of lattice spacing and sea-quark mass, allowing 
for $am_b$-dependent NRQCD errors. We use eq.~(\ref{eq:hfit}) for 
the appropriate hadronic parameter, which here 
is $\overline{m}_b-\overline{M}_{\Upsilon,\eta_b}/2$. This is the 
`binding energy' of the meson which is the consequence of the 
QCD interactions that we include in our lattice calculation. 
The physical value for $\overline{m}_b(\mu=4.18 \,\mathrm{GeV}, n_f=4)$ that 
we obtain from our fit is 4.207(21) GeV. The result from fitting values 
from $n=14$ or 22 are the same within a fraction of 1$\sigma$.  
To the error on the physical value we must add systematic errors 
(which are correlated between the points on Fig.~\ref{fig:massa2} 
and therefore not included there)
from:
\begin{itemize}
\item continuum perturbation theory. The perturbative coefficients 
in our reduced perturbation theory are well-behaved, as shown 
in Table~\ref{tab:rn}. For $m_b$ we use the square root of 
the ratio of the perturbative series for successive moments, 
reducing further the size of the coefficients multiplying 
powers of $\alpha_s$ that can appear. 
We take an error on $\overline{m}_b$ of $0.25\alpha_s^3/2$ (the 
factor of 2 for the square root) which is 
0.15\% (7 MeV). This covers uncertainties 
from missing $\alpha_s^4$ terms as well as uncertainty in the 
$\alpha_s^3$ coefficients~\cite{qcdpt5} and small uncertainties at 
lower order from mass effects as discussed in Section~\ref{subsec:moments}. 
A test of this error is simply to miss out the $\alpha_s^3$ coefficients 
from our perturbation theory. This increases the value of $m_b$ we obtain
almost uniformly by 5 MeV, so a 7 MeV error on including the $\alpha_s^3$ 
coefficients is conservative. 
\item value of $\alpha_s$. Changing the value of $\alpha_s(m_b)$ by 
$1\sigma$ in our perturbative formulae 
changes the value of $m_b$ we obtain by 3 MeV (in the opposite direction 
to the change in $\alpha_s$). 
\item NRQCD systematics. Our NRQCD action is improved almost 
completely through $\alpha_sv^4$, but we are missing $v^4$ terms in the 
vector current. Following Appendix~\ref{appendix:zv} we estimate 
the effect of this at $v^4 \approx (1/n)^2$. 
For $n=18$ this gives 0.3\% (13 MeV). We can test this estimate by determining 
masses from using the leading-order current alone (i.e. missing $v^2$ 
corrections). We find a shift (downwards) of 30 MeV   
on very coarse lattices and 8 MeV on fine lattices. So an uncertainty of 13 MeV 
is conservative for missing higher order $v^4$ terms in the current. 
\item $b$ quark mass tuning. This is negligible, as is clear from the entries
seen in Table~\ref{tab:mb} for set 1 at different masses. 
\end{itemize}
Electromagnetic effects appear in the value of the spin-average of $\Upsilon$ 
and $\eta_b$ masses that we use for tuning. This has negligible impact 
(1 MeV) on the result for $m_b$. The effect of missing $b$ quarks in 
the sea will be accounted for using perturbation theory below. 

Adding the errors above in quadrature gives 
$\overline{m}_b(\mu=4.18 \,\mathrm{GeV}, n_f=4)$ = 4.207(26) GeV. To 
compare results at the conventional point we must convert 
this to an $n_f=5$ quark mass at its own scale and we do this 
using perturbation theory~\cite{Chetyrkin:1997un}. We obtain 
\begin{equation}
\overline{m}_b(\overline{m}_b,n_f=5) = 4.196(23) \, \mathrm{GeV},
\label{eq:massval}
\end{equation}
with the error squeezed down by the evolution of the mass to its 
own scale, but we include an error from uncertainties in this 
evolution. Evolving to 10 GeV gives a 
value $\overline{m}_b(10 {\mathrm{GeV}},n_f=5)$ = 3.650(25) GeV. 
The error budget for $m_b$ at the scale 10 GeV is given 
in Table~\ref{tab:errorbudget}.

\begin{figure*}
\begin{center}
\includegraphics[width=0.8\hsize]{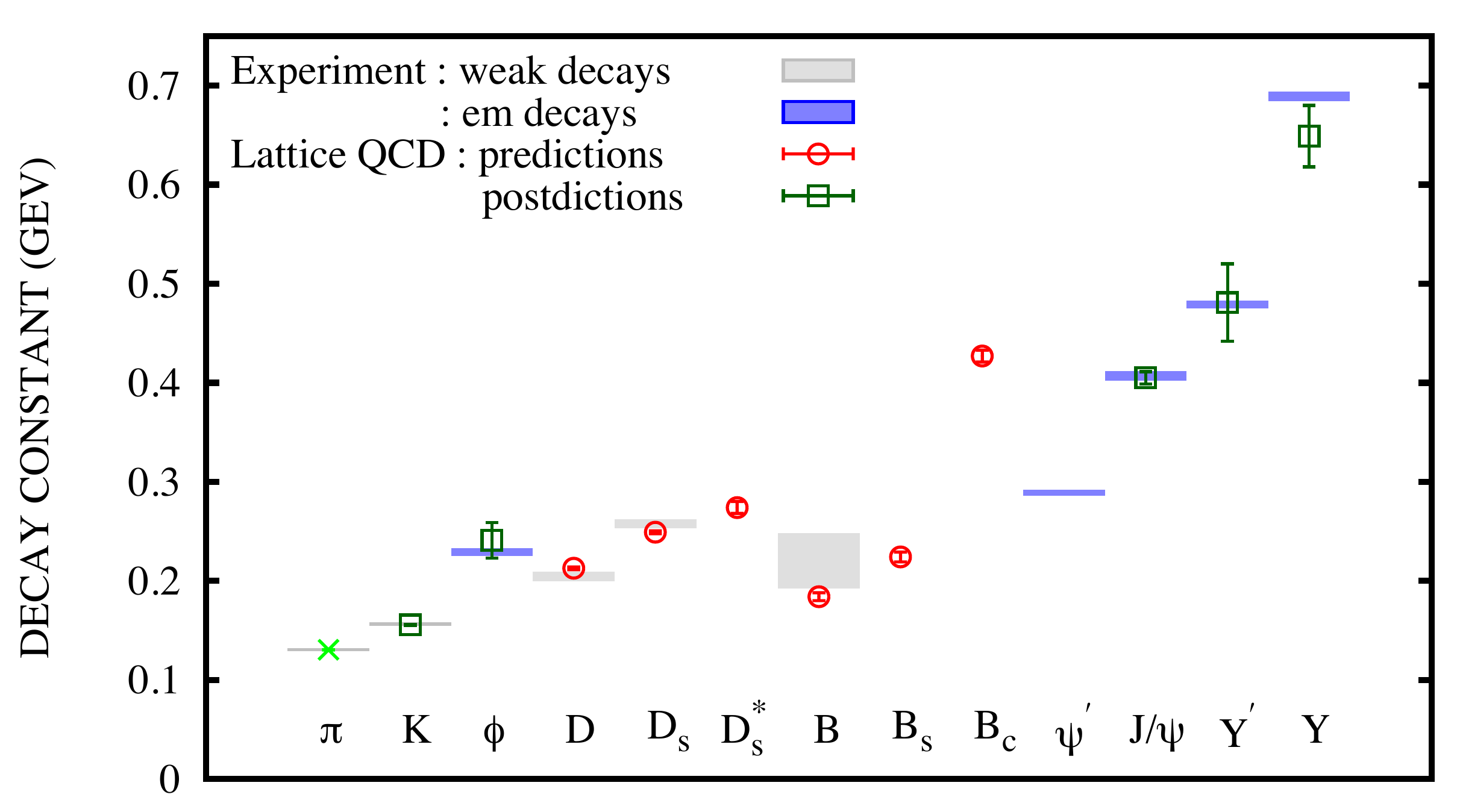}
\end{center}
\caption{ A summary of values for decay constants 
of mesons that are narrow and so well-characterised in experiment. 
The experimental results are taken from appropriate average 
weak or electromagnetic annihilation rates 
in the Particle Data Tables~\cite{pdg} and, for weak 
decays, using average values of the appropriate CKM element. 
For $f_K$, $f_D$ and $f_{D_s}$ experimentally determined values 
are taken from the decay constant review. For the $B^+$ we use 
the average branching fraction~\cite{pdg} obtained by 
Belle and BaBar~\cite{belle1, belle2, babar1, babar2}, along 
with a value for $V_{ub}$ from a unitarity fit to the CKM matrix~\cite{pdg} 
to obtain $f_{B^+}=0.220(28)$ GeV. For the lattice QCD results we 
use world's best values. They are divided into predictions, 
in which lattice calculations originally predated an experimental result, 
and postdictions, in which good experimental values existed before 
lattice results. 
The lattice result for $f_{\pi}^+$ 
is marked with a cross to indicate that it has been used to set the scale for 
some analyses (although not here). 
$\phi$, $J/\psi$,  $D_s^*$, $\eta_c$, $\eta_b$ and $B_c$ results come 
from~\cite{Donald:2013pea,psipaper,dsstar, fdsupdate, McNeile:2012qf} 
using $n_f=3$ configurations. 
For $K^+$ we use our results on the $n_f=4$ configurations used 
here~\cite{Dowdallfkpi} and for $D$ and $D_s$ we use recent results 
from the MILC/Fermilab Lattice collaborations on these 
configurations~\cite{milc14}, 
updating our earlier results on $n_f=3$ 
configurations~\cite{fdsorig, fdsupdate}. For $B$ and $B_s$ we 
use our results on the $n_f=4$ configurations using NRQCD $b$ 
quarks as here~\cite{DowdallBdecay}. 
Finally, the decay constants for $\Upsilon$ 
and $\Upsilon^{\prime}$ come from this paper. 
}
\label{fig:decaysum}
\end{figure*}

\begin{figure}
\begin{center}
\includegraphics[width=\hsize]{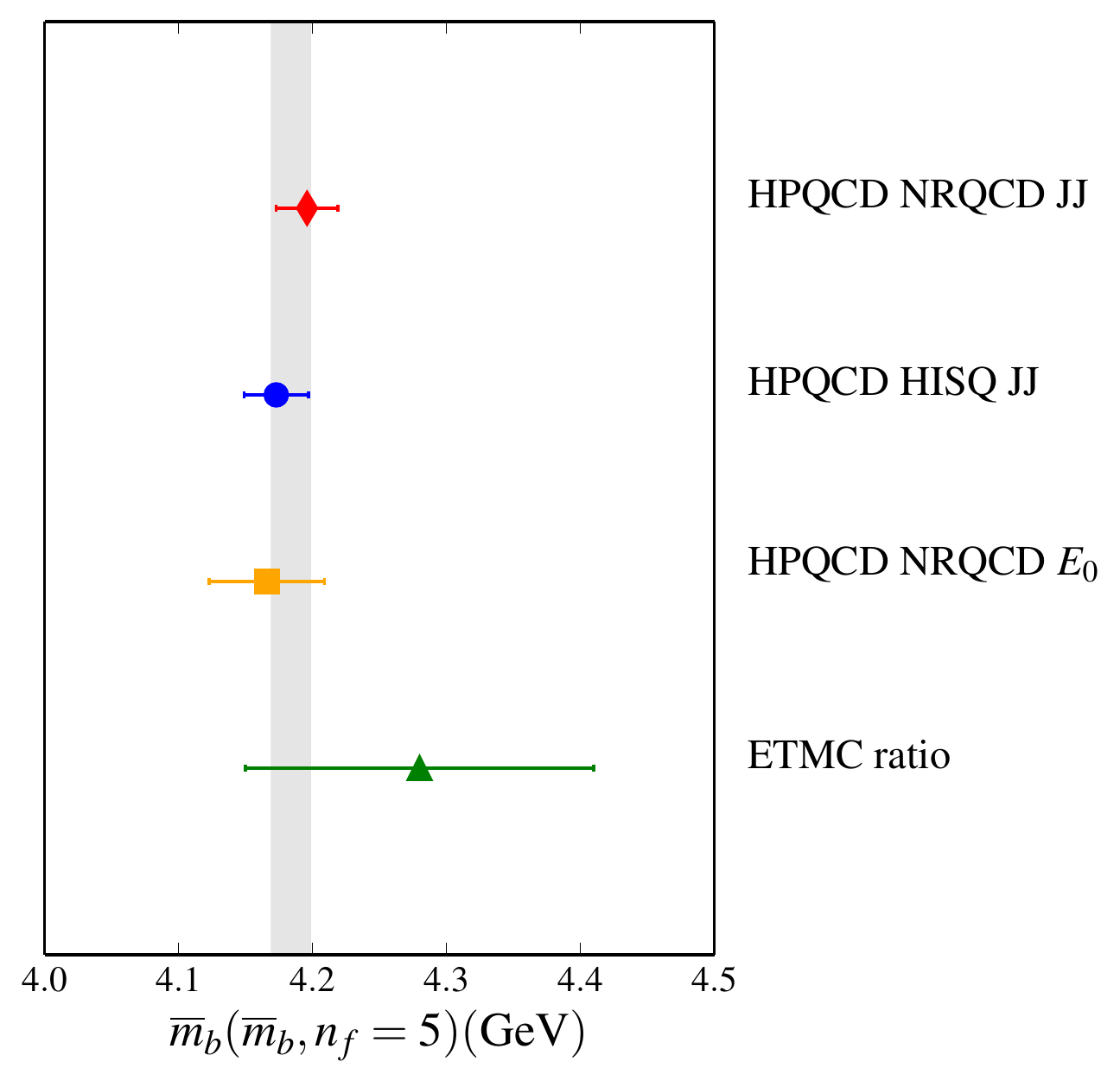}
\end{center}
\caption{Lattice QCD results for $m_b$ in the 
$\overline{MS}$ scheme with 5 flavours and evaluated 
at its own scale. Results are from calculations that 
include either 3 or 4 flavours of sea quarks and so 
can be perturbatively corrected to 5 flavours.    
All 4 results use different methods, indicated on 
the right. The top result is from this paper, the second 
from~\cite{hisqnew}, the third from~\cite{mbpert} and the fourth 
from~\cite{Carrasco:2013naa}, adjusted perturbatively to $n_f=5$.
The grey band gives the weighted average of the lattice results: 
4.184(15) GeV. 
}
\label{fig:mbsum}
\end{figure}

\section{Conclusions}
\label{sec:conclusions}
We have presented here the first complete nonperturbative 
calculation of the leptonic widths of the 
$\Upsilon$ and $\Upsilon^{\prime}$ using full lattice 
QCD including $u, d, s$ and $c$ quarks in the sea. 
These are hard calculations to do in lattice QCD 
because they require an accurate matching of the lattice 
QCD vector current to the continuum vector current 
and because they are short-distance quantities, sensitive 
to discretisation errors. 
We use a matching method which is nonperturbative on the 
lattice, making use of high-order continuum QCD perturbation 
theory. 
We obtain 5\% uncertainty on the $\Upsilon$ decay constant and 
8\% on that of the $\Upsilon^{\prime}$ (7\% uncertainty on the 
ratio of the two). 

Our results are:
\begin{eqnarray}
\label{eq:fres2}
f_{\Upsilon} &=& 0.649(31) \, \mathrm{GeV} \nonumber \\
f_{\Upsilon^{\prime}} &=& 0.481(39) \, \mathrm{GeV}  
\end{eqnarray}
giving leptonic decay widths: 
\begin{eqnarray}
\label{eq:gamres}
\Gamma({\Upsilon}\rightarrow e^+e^-) &=& 1.19(11) \, \mathrm{keV} \nonumber \\
\Gamma({\Upsilon^{\prime}}\rightarrow e^+e^-) &=& 0.69(9) \, \mathrm{keV}  
\end{eqnarray}
in good agreement with experiment. 

In Figure~\ref{fig:decaysum} we summarise lattice QCD results 
for decay constants of well-characeterised mesons with comparison 
to experimental results determined from weak (for charged pseudoscalars) 
or electromagnetic (for neutral vectors) decays. This is an update 
of the summary given in~\cite{McNeile:2012qf}. To determine decay 
constants from experimental results for weak decays we use average 
values for the appropriate Cabibbo-Kobayashi-Maskawa (CKM) matrix elements from 
the Particle Data Tables~\cite{pdg}. The lattice results given 
in the Figure all come from simulations that include $u, d, s$ quarks 
in the sea or (as here) $u, d, s$ and $c$ quarks in the sea. The figure 
demonstrates the ability of lattice QCD to cover a wide range of physics 
results working simply with the QCD Lagrangian and the input parameters 
for QCD. The agreement with experiment is good (within 2$\sigma$), where experimental 
results are available. 
Lattice QCD is also able to make predictions (given 
by violet circles in the Figure - some of these now also have experimental 
results) and confidence in the 
reliability of these is enhanced by the fact that they sit within this 
wider picture.  

The results for $\Upsilon$ and $\Upsilon^{\prime}$ from this paper are 
in fact the least accurate.  To improve these results in future requires 
further study of the current correction operators in the nonrelativistic 
expansion of the vector current. Different representations of this current 
will have different matrix elements and different `mixing-down' 
behaviour with the leading-order current which may reduce 
uncertainties in $Z_V$ and discretisation effects. Higher-order 
current corrections should also be considered. Finer lattices, with 
a spacing of $a \approx$ 0.06 fm are also available~\cite{milchisq} and calculations 
on these would have reduced discretisation errors, for example 
in the ratio of $\Upsilon^{\prime}$ to $\Upsilon$ decay constants.  

Since our method uses time-moments of the vector heavyonium  
correlation function we are also able to compare results 
directly to values for low moment number derived from 
experiment for $\sigma(e^+e^- \rightarrow \mathrm{hadrons})$ 
in the $b$ quark region. We find good agreement, although 
NRQCD systematic errors are large for these moments. Improved 
results will come from the use of relativistic formalisms such 
as HISQ. Our results can be converted into the first lattice 
result for the $b$ quark contribution to the anomalous 
magnetic moment of the muon, $a_{\mu}^b = 0.271(37)\times 10^{-10}$.  

Finally we give a new determination of the $b$ quark mass 
from matching ratios of time-moments of the vector current-current 
correlator to continuum QCD perturbation theory through NNNLO. 
Our result is 
\begin{equation}
\overline{m}_b(\overline{m}_b, n_f=5) = 4.196(23) \, \mathrm{GeV}
\end{equation}
with an error that puts this result among the best lattice QCD 
determinations of this Standard Model parameter. 

Figure~\ref{fig:mbsum} gives a summary plot of lattice QCD results 
for $m_b$. We compare values obtained on configurations that 
include either $u, d, s$ or $u, d, s, c$ quarks in the sea and 
can then be converted into a value for 
$\overline{m}_b(\overline{m}_b, n_f=5)$ by adding in the $c, b$ 
or $b$ quarks respectively using perturbation theory. 
The four results use different methods. The top result is from the 
work described here. The second~\cite{hisqnew} uses the 
relativistic HISQ formalism for 
the $b$ quark and pseudoscalar current-current correlators that 
are absolutely normalised on the lattice. Low moments ($n=4-10$) are 
compared to continuum QCD perturbation theory for a range of 
masses up to the $b$ quark mass on the finest lattices. Results are 
combined from $n_f=3$ and $n_f=4$ calculations, updating~\cite{bcmasses}. 
The third~\cite{mbpert} calculates the quark mass from the binding energy for 
$\Upsilon$ and $B_s$ mesons using the NRQCD formalism on $n_f=3$ 
gluon field configurations. This combines the nonperturbative 
lattice calculation with continuum QCD perturbation 
theory for the mass renormalisation and lattice QCD perturbation 
theory for the heavy quark self-energy, both through 
$\mathcal{O}(\alpha_s^2)$, in a method developed in~\cite{Davies:1994pz}. 
The fourth result uses the twisted mass formalism for a range of quark 
masses from $c$ to $b$ on configurations that include $n_f=4$ sea quarks. 
A ratio is taken of heavy-light meson masses to quark masses, for successively larger 
masses in a procedure that has a well-defined static (infinite quark mass) 
limit. This allows interpolation to the $b$ quark mass. The result 
given of 4.29(13) GeV is for $n_f=4$. Perturbative adjustment to $n_f=5$ 
gives 4.28(13) GeV and that is the value plotted in Figure~\ref{fig:mbsum}.

The results all have very different systematic errors. 
Even the two results that use current-current correlator methods are 
working in a very different range of moment number requiring different 
methods (i.e. a direct extraction vs using a ratio of moments) with different mesons 
and a different quark action. There is therefore no obvious correlation 
between the results and we can take a weighted average to obtain 
4.184(15) GeV, plotted as the grey band in Figure~\ref{fig:mbsum}. 
This result is very compatible with, but twice as accurate as, the current 
evaluation in the Particle Data Tables~\cite{pdg}. 
The value also agrees well with determinations from continuum methods, 
for example using $R_{e^+e^-}$ results in the $b$ region~\cite{kuhn09update}. 

The method we have given here is applicable to other lattice formalisms 
for heavy quarks, for example that of the Fermilab 
Lattice Collaboration~\cite{fermilab}. Further determinations of $m_b$ 
from other formalisms would be useful in the long-term goal of reducing 
uncertainties in Standard Model parameters needed for precision 
characterisation of the Higgs boson. 

{\bf{Acknowledgements}} We are grateful to the MILC collaboration 
for the use of 
their configurations and to R. Horgan for useful 
discussions. 
Computing was done on the Darwin supercomputer at the University 
of Cambridge as part of STFC's DiRAC facility. 
We are grateful to the Darwin support staff for assistance. 
Funding for this work came from STFC, 
the Royal Society, the Wolfson Foundation and NSF. 

\appendix

\section{Determination of $Z_V$}
\label{appendix:zv}

The perturbative analysis of heavy-heavy current-current correlators 
is well developed in continuum QCD perturbation 
theory~\cite{qcdpt1, qcdpt2, qcdpt3, qcdpt4, qcdpt5} and here 
we make use of that to normalise the lattice NRQCD vector current 
for $b\overline{b}$ annihilation 
that we use to determine the $\Upsilon$ leptonic width. 
The method is a variation of that used for the $J/\psi$ leptonic width 
in~\cite{psipaper}. In that case we were working with a relativistic 
discretisation of the QCD quark action on the lattice. Since here 
we have a nonrelativistic discretisation there are some differences 
in the approach that we lay out in this section\footnote{Note also that, 
in a nonrelativistic formalism, the annihilation and scattering currents 
do not have the same renormalisation factor}. 

Time-moments of current-current correlators, 
being ultraviolet-finite quantities, 
can be calculated in lattice QCD and extrapolated to the continuum 
limit to give a continuum result that can be compared 
to experiment~\cite{psipaper}. The current used in 
the correlator must be matched to the continuum current, however. 
When the Highly Improved Staggered Quark 
discretisation~\cite{hisqdef} 
is used, for example,
the local pseudoscalar density is absolutely 
normalised~\cite{firstcurrcurr, bcmasses} but the 
vector current normalisation has to be fixed.
For heavy quarks this can be done using the continuum 
QCD perturbation theory for the vector current-current correlator moments. 
The multiplicative renormalisation factor $Z_V$ is simply determined 
by matching the lattice result at a given lattice spacing 
for a specific moment to the perturbative result. 
We can choose which moment to use, since differences in $Z_V$ 
that arise from a different choice are discretisation effects 
that must disappear 
in the continuum limit, along with other discretisation errors 
that result from working at a non-zero lattice spacing. 
The low moments, 4--10, are known through $\mathcal{O}(\alpha_s^3)$ 
so are clearly to be preferred over higher ones. 
It is convenient to use ratios of vector 
to pseudoscalar current-current correlator moments since then
factors of the quark mass cancel~\cite{psipaper}. 

When a nonrelativistic discretisation of the QCD quark action 
is used, neither the pseudoscalar nor the vector currents is absolutely 
normalised and the lattice current is only determined to a given 
order in a relativistic expansion. Hence the match to continuum 
QCD perturbation theory has both discretisation errors and relativistic 
errors, which are mixed by the higher dimension operators used to 
implement corrections, and so we cannot simply take a value 
of $Z$ from the match for a specific moment.  

\begin{figure}
\begin{center}
\includegraphics[width=0.9\hsize]{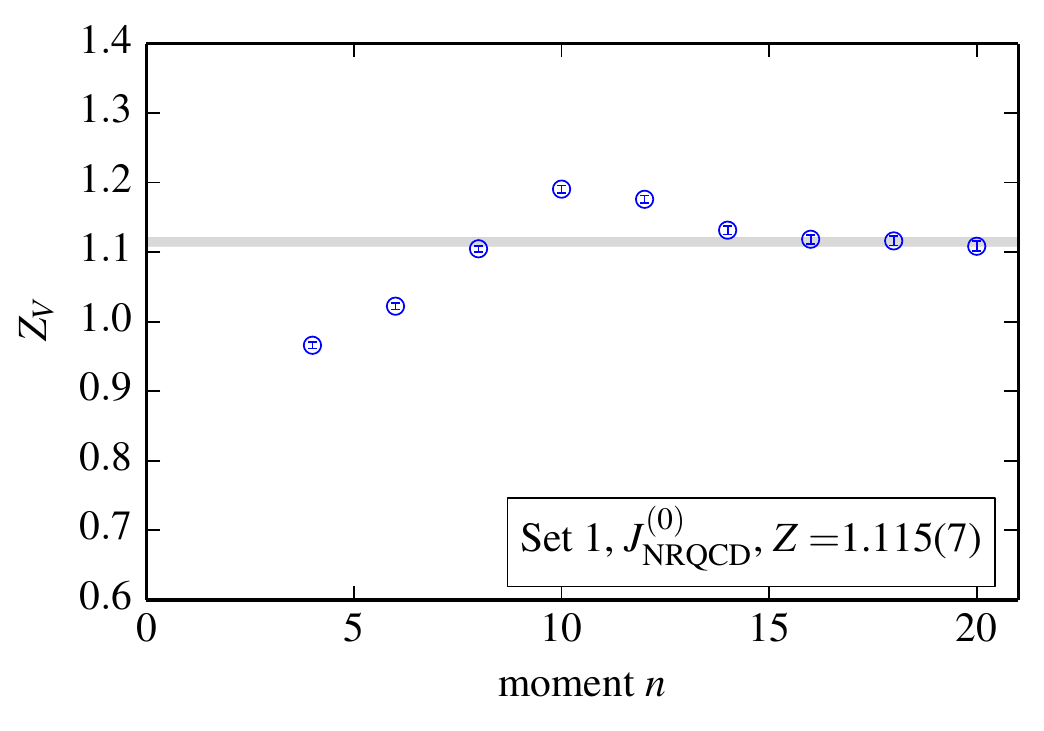}
\includegraphics[width=0.9\hsize]{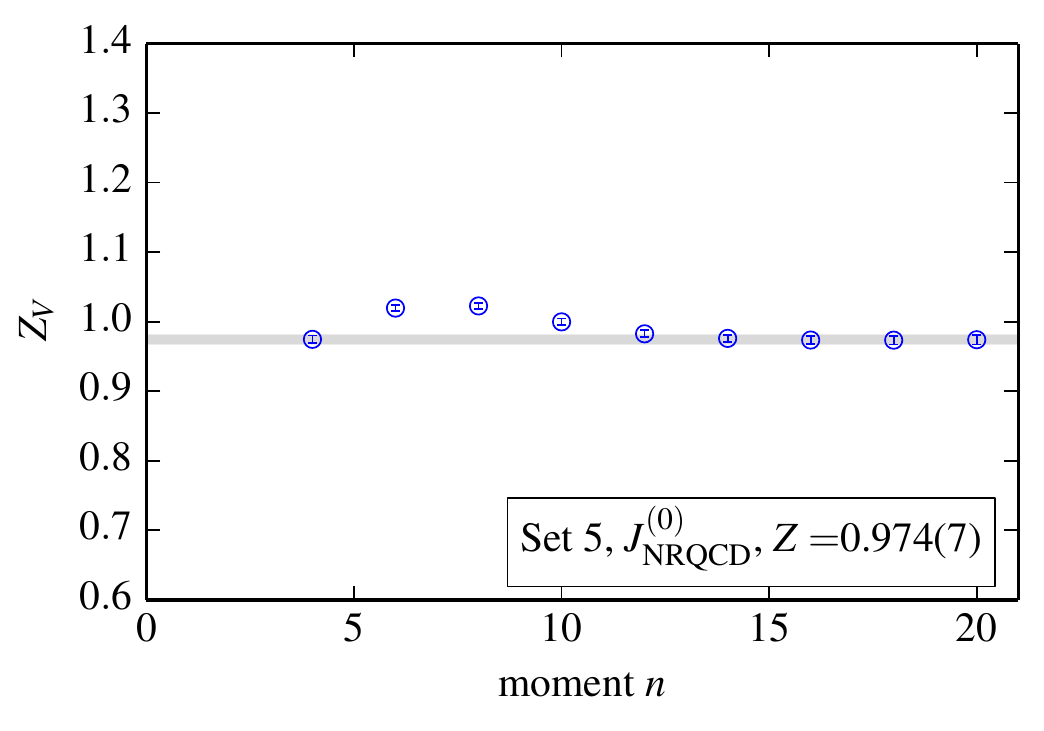}
\end{center}
\caption{Renormalisation factor $Z_V$ for the leading 
term in the NRQCD vector current, $J^{(0)}_{V,i,\mathrm{NRQCD}}$, 
determined from continuum QCD perturbation theory 
for the current-current correlator using 
eq.~(\ref{eq:zeq2}) 
and plotted as a function 
of moment number, $n$. 
The top plot shows results on the very coarse lattices, set 1 
(for $am_b=3.297$, the preferred value), 
and the lower plot shows results on the fine lattices, set 5. 
The error on the points includes uncertainty 
in the continuum perturbation theory. 
The grey bands give the results of a fit to a constant for range 
16--20 for set 1 and 14--20 for set 5. 
}
\label{fig:treeZ}
\end{figure}

\begin{figure}[h]
\begin{center}
\includegraphics[width=0.9\hsize]{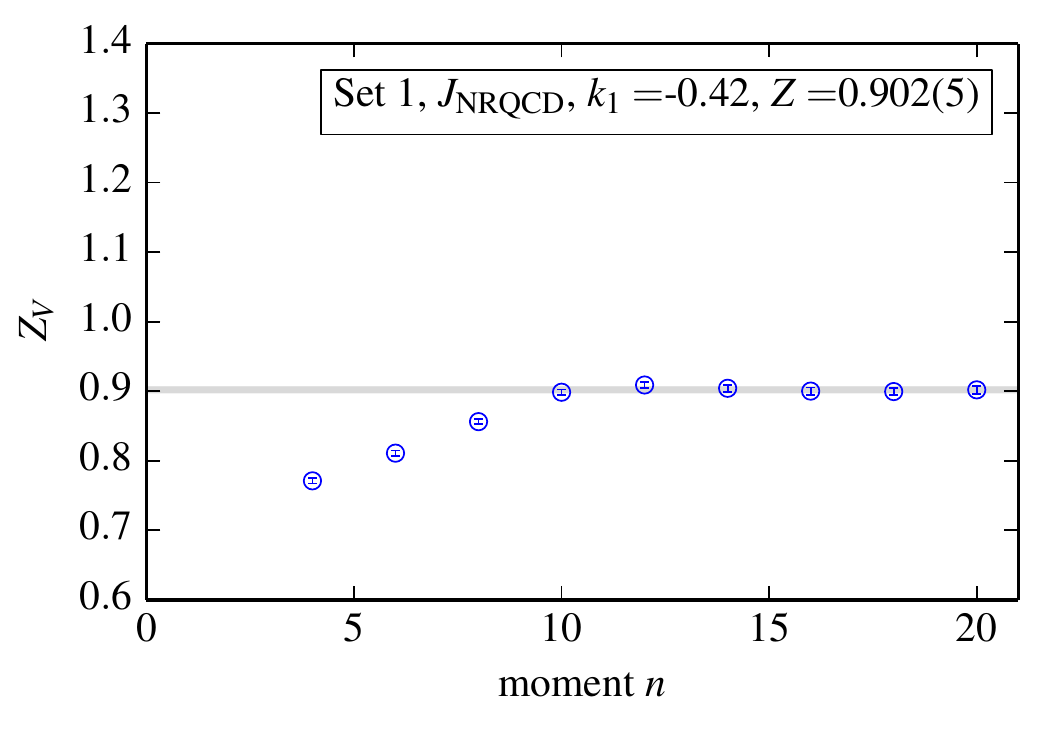}
\includegraphics[width=0.9\hsize]{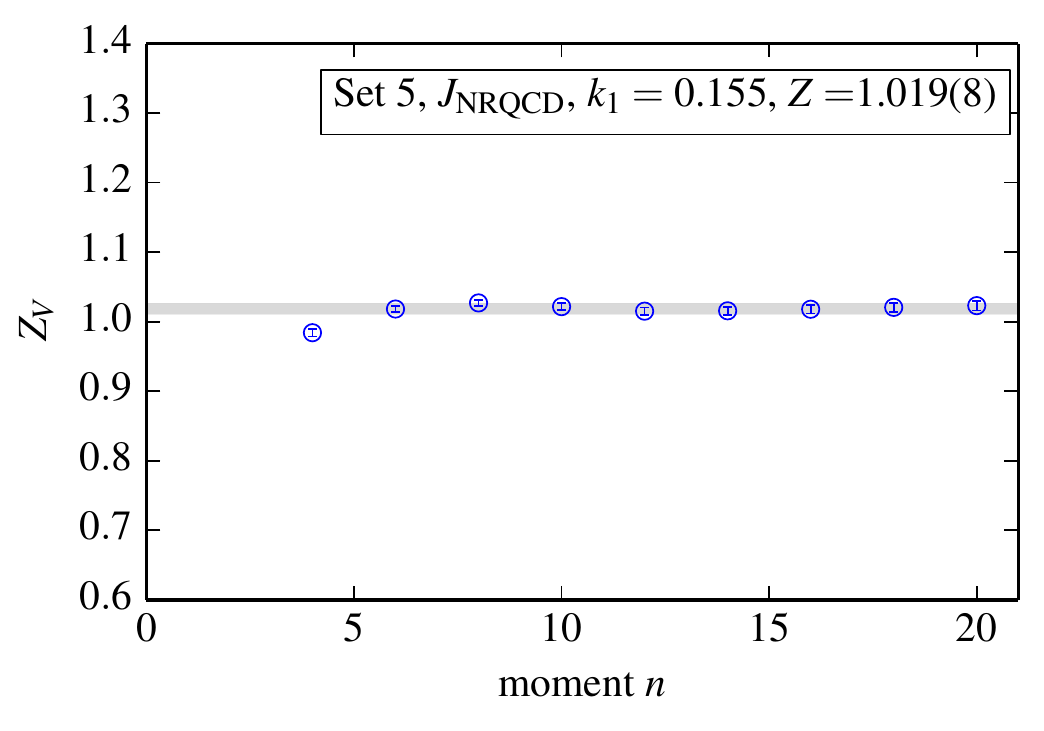}
\end{center}
\caption{Renormalisation factor $Z_V$ for the NRQCD vector 
current including leading and next-to-leading terms,
$J^{(0)}_{V,i,\mathrm{NRQCD}} + k_1 J^{(1)}_{V,i,\mathrm{NRQCD}}$, 
plotted as a function 
of moment number, $n$. 
From top to bottom results are from very coarse set 1
and fine set 5. 
The error on the points includes uncertainty
in the continuum perturbation theory. 
The grey bands give the results of a fit to a constant in each case. 
}
\label{fig:j1Z}
\end{figure}

In determining the normalisation of the current 
we can, however, make use of the fact 
that time-moments with low moment number emphasise very short times in the 
current-current correlator 
and are therefore sensitive to much higher 
internal spatial momenta within the quark-antiquark pair
(the overall momentum of the pair is zero)
than higher moments are~\cite{bcmasses}. Thus, as the moment 
number changes, the sensitivity to relativistic corrections 
changes. This is easily seen 
in an analysis of the free case. 
At leading relativistic order, for vector or pseudoscalar moments,  
multiplying the free quark and antiquark propagators together we have
\begin{eqnarray}
G_n &=& 4\int d^4x\, t^n \int \frac{dE_1 d^3p_1}{(2\pi)^4} \frac{dE_2 d^3p_2}{(2\pi)^4} \\
&& \frac{e^{-2mt}e^{i(E_1+E_2)t}e^{i({\bf p_1 + p_2})\cdot {\bf x}}}{(iE_1+p_1^2/2m)(iE_2+p_2^2/2m)} \nonumber
\label{eq:freeg1}
\end{eqnarray}
where the quarks have mass $m$. Integrating over ${\bf x}$ and ${\bf p}$
gives
\begin{equation}
G_n = 4\int dt \, t^n \Theta(t) e^{-2mt} \int \frac{d^3p}{(2\pi)^3} 
e^{(-p^2/m)t} .
\label{eq:freeg2}
\end{equation}
Performing the integral over $t$ allows us to study the contribution 
to the integral as a function of $v^2$, the square of the heavy 
quark velocity (in units of $c^2$) in the quark-antiquark pair and the expansion 
parameter in the nonrelativistic expansion. In 
\begin{equation}
G_n = \frac{n!}{\pi^22^{n+1}m^{n-2}} \int \frac{(v^2)^{1/2} d(v^2)}{(1+v^2/2)^{n+1}} 
\label{eq:freeg3}
\end{equation}
the integrand peaks at $v^2 = 1/(n+1/2)$, falling as expected 
with increasing $n$.
 
We therefore expect that the comparison of continuum QCD 
perturbation theory to the NRQCD correlator moments will in general 
be poor at very small moment number ($n=4,\ldots$), 
where the internal velocity
within the quark-antiquark pair can be large
and NRQCD, as an expansion in $v^2$,  will have sizeable 
systematic errors. The comparison will improve 
as the moment number increases ($n \ge 6$) and the internal 
momentum falls to 
nonrelativistic values. 
The improvement will 
be visible as the development of a plateau region in a 
plot of the renormalisation 
constant $Z_V$ as a function of moment number. This 
will happen at a moment number where discretisation and 
relativistic corrections missing from the NRQCD calculation 
have become small compared to the unknown higher order 
terms in $\alpha_s$ in the continuum perturbation theory 
expansion for $Z_V$.  

As discussed in Section~\ref{subsec:vec} we calculate 
NRQCD vector current-current correlators using a local 
NRQCD current at source and sink, $J_{V,\mathrm{NRQCD}}$. 
Allowing for a renormalisation of this current to match 
the continuum vector current we have 
\begin{equation}
J_{V,i} = Z_V J_{V,\mathrm{NRQCD},i}.
\end{equation}
Time moments of the vector correlator calculated from $J_{V,\mathrm{NRQCD}}$, 
$C_{V,\mathrm{NRQCD}}(t)$, are then related to those from 
the continuum current, $G_n^V$, by
\begin{eqnarray}
G_n^{V} &\equiv& Z_V^2 C_n^{V} \\
&=& 2 Z_V^2 \sum_{t} (t/a)^n {C}_{V,\mathrm{NRQCD}}(t) \exp(-[\overline{M}_{kin}-\overline{E}_0]t) \nonumber
\label{eq:timemomv2}
\end{eqnarray}
up to discretisation and relativistic correction terms, 
where we reproduce eq.~(\ref{eq:timemomv}) from Section~\ref{subsec:moments}. 
$G_n^{V}$ is given by a perturbative expansion 
\begin{equation}
G_n^V = \frac{g_n^V(\alpha_s, \mu/m_b)}{(a\overline{m}_b(\mu))^{n-2}}
\end{equation}
where $g_n^V$ is known through $\mathcal{O}(\alpha_s^3)$
~\cite{qcdpt1, qcdpt2, qcdpt3, qcdpt4, qcdpt5}. 
$\overline{m}_b$ is the $b$ quark mass in the $\overline{MS}$ scheme. 

Dividing by the tree-level value for the correlator 
moments reduces both relativistic 
and discretisation errors. 
$Z_V$ is then given by
\begin{equation}
Z_V^2 =  \frac{C_n^{V,U=1}}{C_n^V} r_n^V \left( \frac{m_b}{\overline{m}_b} \right)^{n-2}
\label{eq:zeq1}
\end{equation}
where $C_n^{V,U=1}$ is the appropriate time-moment of 
the free NRQCD correlator 
(with the coefficients in the NRQCD action 
of eq.~(\ref{eq:deltaH}) set to their tree-level values of 1),
$m_b$ is the quark mass in the NRQCD Hamiltonian and $r_n^V$ is the 
perturbative series $g_n^V$ divided by its leading, $\mathcal{O}(\alpha_s^0)$ 
result and for which coefficients are given in Table~\ref{tab:rn}.  
Taking appropriate powers to cancel factors of the quark mass means that 
we can extract $Z_V$ from a ratio using different moments:
\begin{equation}
Z_V = X_n^{(n^{\prime}-2)/(2(n^{\prime}-n))} /X_{n^{\prime}}^{(n-2)/(2(n^{\prime}-n))} 
\label{eq:zeq2}
\end{equation}
with 
\begin{equation}
X_n = \frac{C_n^{V,U=1}}{C_n^V}r_n^V  .
\label{eq:xdef}
\end{equation}
We will simply use $n^{\prime}=n+2$. We evaluate $r_n^V$ using 
$\mu=m_b$ and take 
$\alpha_{\overline{MS}}(n_f=4, m_b)$ = 0.2268(24)~\cite{pdg}. 

Figure~\ref{fig:treeZ} shows an example of this approach in 
the case where the NRQCD current operator used is the leading 
term in the relativistic expansion of the current, 
$J^{(0)}_{V,\mathrm{NRQCD},i} \equiv \chi^{\dag} \sigma_i \psi$ 
(eq.~(\ref{eq:j0})). We see a plateau in $Z_V$ 
for moments between 14 and 20 for very coarse lattices and 
12 to 20 for fine lattices. The errors on the points 
include the truncation errors from the continuum perturbation 
theory, taken as $0.25\alpha_s^3$ to include uncertainty at this 
order~\cite{qcdpt5}, unknown terms at higher orders and possible 
missing mass-effects at lower orders. 
The statistical errors from the calculation of the NRQCD correlators 
are very small but the statistical error in the determination 
of the kinetic mass that appears 
in the time-moments (eq.~(\ref{eq:timemomv}) is significant here. 
Nonperturbative contributions to the moments from 
the gluon condensate divided by the fourth power of the quark 
mass, discussed at length for charmonium correlators 
in~\cite{firstcurrcurr, bcmasses, hisqnew}, are negligible here because of 
the size of the $b$ quark mass. 

The $Z$ values given in Figure~\ref{fig:treeZ} are obtained from fitting 
the results to a constant over the range of moment number. 
We use 14--20 on the fine lattices ($\chi^2/\mathrm{dof}=0.05$) 
and 16--20 on the very coarse ($\chi^2/\mathrm{dof}=0.6$). 

Our NRQCD Hamiltonian is completely improved through 
$\mathcal{O}(v^4)$ (eq.~(\ref{eq:deltaH})), which is next-to-leading 
order in the relativistic expansion. We also include almost 
all corrections at $\mathcal{O}(\alpha_s v^4)$. We therefore expect that 
the behaviour of $Z_V$ for moment numbers at the low end of 
the plateau in Figure~\ref{fig:treeZ} can be improved 
by the addition of next-to-leading order 
relativistic corrections 
to the current, since the current $J^{(0)}_{V,\mathrm{NRQCD}}$ 
is the only source of 
errors at this order. 
In fact we can use this to determine the coefficient of 
the current correction term nonperturbatively. 

We take, as in eq.~(\ref{eq:currmatch}), 
\begin{equation}
J_{V,i} = Z_V ( J^{(0)}_{V,\mathrm{NRQCD},i} + k_1 J^{(1)}_{V,\mathrm{NRQCD},i} )
\label{eq:currmatch2}
\end{equation}  
and use the behaviour of $Z_V$ to determine $k_1$. 
Since there is only one relativistic (and discretisation) 
current correction operator 
at this order, this is straightforward to do. 

\begin{table}
\begin{tabular}{lllll}
\hline
\hline
Set &  $am_b$ & $c_4$ & $k_1$ & $Z_V$ \\
\hline
1 & 3.297 & 1.0 & -0.42(16) & 0.902(5)(58)\\
1 & 3.297 & 1.22 & -0.29(15) & 0.963(5)(60)\\
1 & 3.42 & 1.0  & -0.52(20) & 0.890(5)(67)\\
2 & 3.25 & 1.22 & -0.36(16) & 0.926(5)(62)\\
\hline
3 & 2.66 & 1.0 & -0.18(10) & 0.865(6)(59)\\
4 & 2.62 & 1.20 & -0.11(9) & 0.913(6)(50) \\
\hline
5 & 1.91 & 1.0 & 0.155(35) & 1.019(8)(35)\\
\hline
\hline
\end{tabular}
\caption{ 
Values for $k_1$ and $Z_V$ of eq.~(\ref{eq:currmatch2}) obtained 
from our current matching procedure for the NRQCD vector current 
(see text for details). 
Two errors are given for $Z_V$. The 
first is from the fit for $Z_V$ at the central value 
of $k_1$ and is dominated by that from 
truncation errors in the continuum perturbation theory and statistical 
errors in the determination of the meson kinetic mass. 
The second comes from the uncertainty in $k_1$ and is correlated 
with that uncertainty, so that $Z_V$ increases as $k_1$ increases. 
}
\label{tab:j1Z}
\end{table}

Figure~\ref{fig:j1Z} shows the behaviour of $Z_V$ for our 
preferred values of $k_1$ on the very coarse and 
fine lattices. Note that the $C_n^{V,U=1}$ needed 
for the ratio in eq.~(\ref{eq:xdef}) is calculated with the 
tree-level value of $k_1$ i.e. 1/6. By adjusting $k_1$ we are able to 
achieve a plateau in $Z_V$ down to $n=10$ or below in 
all cases. 
To achieve a plateau in $Z$ to lower 
$n$ values would require current corrections at higher 
order in $v^2$ (and $a^2$). 
We take the central value of $k_1$ as the 
point of minimum $\chi^2/\mathrm{dof}$ (0.8 for set 1, 0.2 for 
set 3 and 0.4 for set 5) and the uncertainty on $k_1$ 
as the range that gives $\Delta \chi^2/\mathrm{dof} = 1$. 

The corresponding values of $k_1$ and $Z_V$ obtained 
are given in Table~\ref{tab:j1Z}. The $k_1$ values 
agree well with those from matching the 
NRQCD vector current using $\mathcal{O}(\alpha_s)$ 
lattice QCD perturbation 
theory~\cite{Hart:2006ij} (although note that this perturbation 
theory is not directly applicable to our calculation). There it was found that, 
at a quark mass value $am_b$ close to that we use 
on the fine lattices, the $\mathcal{O}(\alpha_s)$ corrections 
to $k_1$ were very small, leaving it at its tree-level value 
of 1/6. At larger values of $am_b$, corresponding to our 
coarser lattices, the $\mathcal{O}(\alpha_s)$ corrections 
to $k_1$ became large and negative, dominating 
the tree-level result and leading to a change in 
sign for $k_1$. 

The value for $Z_V$ depends on $k_1$ because radiative corrections 
to the current corrections can generate the leading-order 
current in a process known as `mixing down'~\cite{Hart:2006ij}. 
Thus the $Z_V$ values given in Table~\ref{tab:j1Z} have 
two errors. 
The first comes from the fit result at a fixed value of 
$k_1$ and the second comes from the uncertainty in $k_1$ and is correlated 
with that uncertainty. 
The uncertainty in $k_1$ is estimated as described above and 
is sizeable. However the associated shift in $Z_V$ has the effect 
of counteracting this change when determining the decay constant 
(as expected for a mixing-down effect)
so that the uncertainty in that quantity is significantly smaller than 
the errors in Table~\ref{tab:j1Z} 
naively imply (see Section~\ref{subsec:leptwidth}). 
We also see, from that Table that, as expected for a renormalisation 
constant, the values for 
$k_1$ and $Z_V$ do not change significantly between sets 
at approximately the same value of the lattice spacing. 

The NRQCD current that we obtain with the values of $k_1$ 
and $Z_V$ given here still has systematic errors from missing 
current corrections at $\mathcal{O}(v^4)$. These errors 
will be accounted for in the error budget for the different 
quantities that we determine using this current in Section~\ref{sec:results}.  

\section{Tests of relativistic covariance of NRQCD correlators}
\label{appendix:rel}

As part of testing the NRQCD framework, it is important to 
check for consistency against relativistic behaviour as 
NRQCD is improved. Here we provide tests of amplitudes that 
complement earlier results~\cite{dowdallr1} on meson 
energies. The tests involve studies of the amplitudes 
of the leading order NRQCD vector current ($J^{(0)}_{V,\mathrm{NRQCD}}$) 
for mesons of non-zero spatial momentum. 
Since the rest of this paper looks at an improved 
NRQCD current and mesons with zero spatial momentum, the 
results here are not directly relevant to the rest of 
our results. They are nevertheless useful as part of the 
NRQCD `bigger picture' and so we include them here as an Appendix.  

As relativistic corrections are added to the NRQCD action it 
starts to behave, not surprisingly, more like a relativistic action.
In~\cite{dowdallr1} (see also~\cite{Bernard:2010fr}) the behaviour 
of the meson energy as a function of spatial momentum was discussed, in 
particular the r\^{o}le of $v^4$ corrections to the action in feeding 
the meson binding energy into the kinetic mass of eq.~(\ref{eq:kinmass}), 
so that the dispersion relation for energy as a function of 
momentum is correct.  

\begin{figure}
\begin{center}
\includegraphics[width=0.9\hsize]{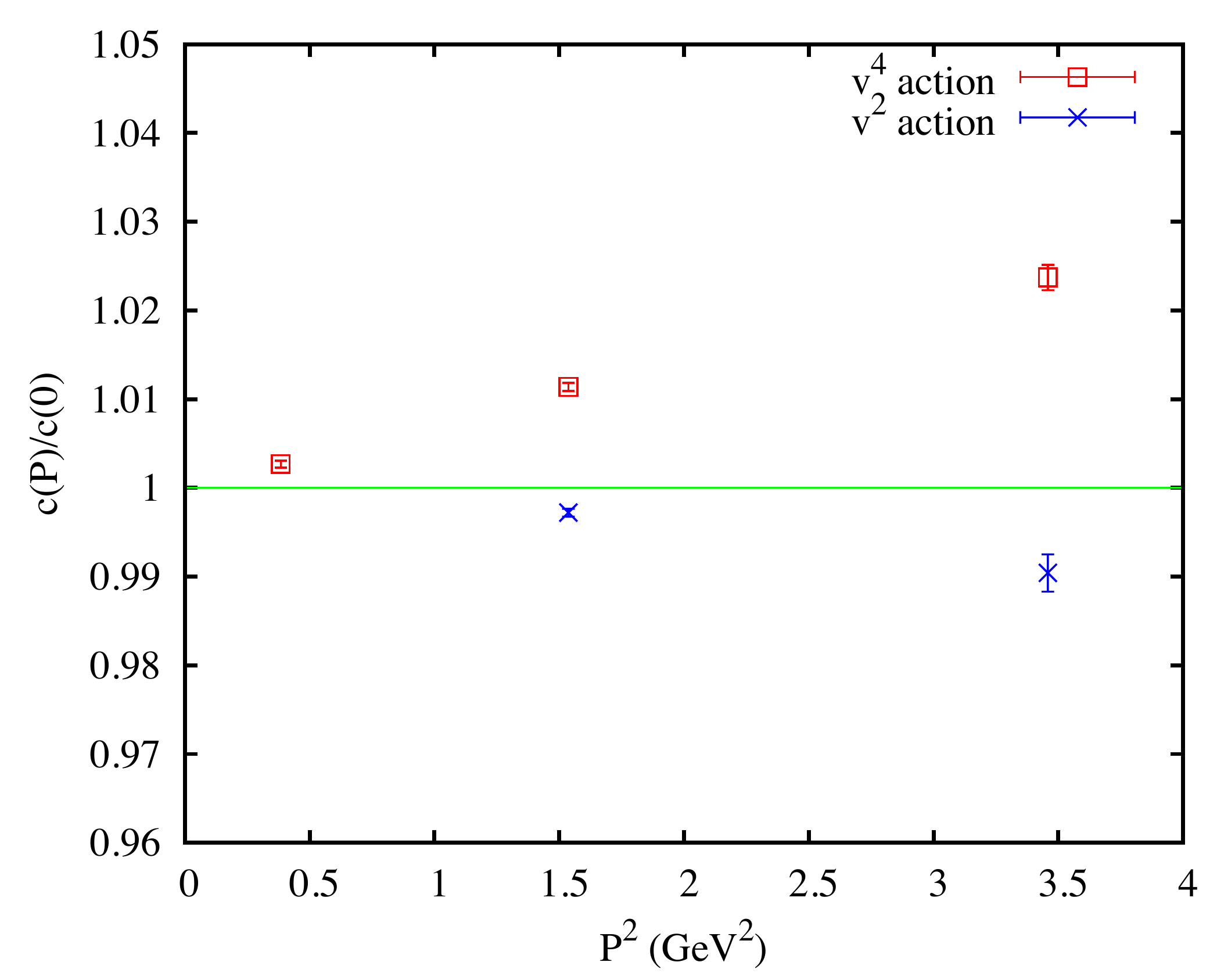}
\end{center}
\caption{$\eta_b$ correlator amplitudes as a function of 
the square of the $\eta_b$ spatial momentum. Amplitudes 
are given as a ratio to the zero momentum amplitude.  
Results are from set 5 fine configurations. Blue crosses indicated 
results from a purely $v^2$ NRQCD action ($H_0$ only) and red squares 
give results from the full $v^4$ action used here (eq.~(\ref{eq:deltaH})). 
}
\label{fig:p2p4}
\end{figure}

\begin{figure}
\begin{center}
\includegraphics[width=0.9\hsize]{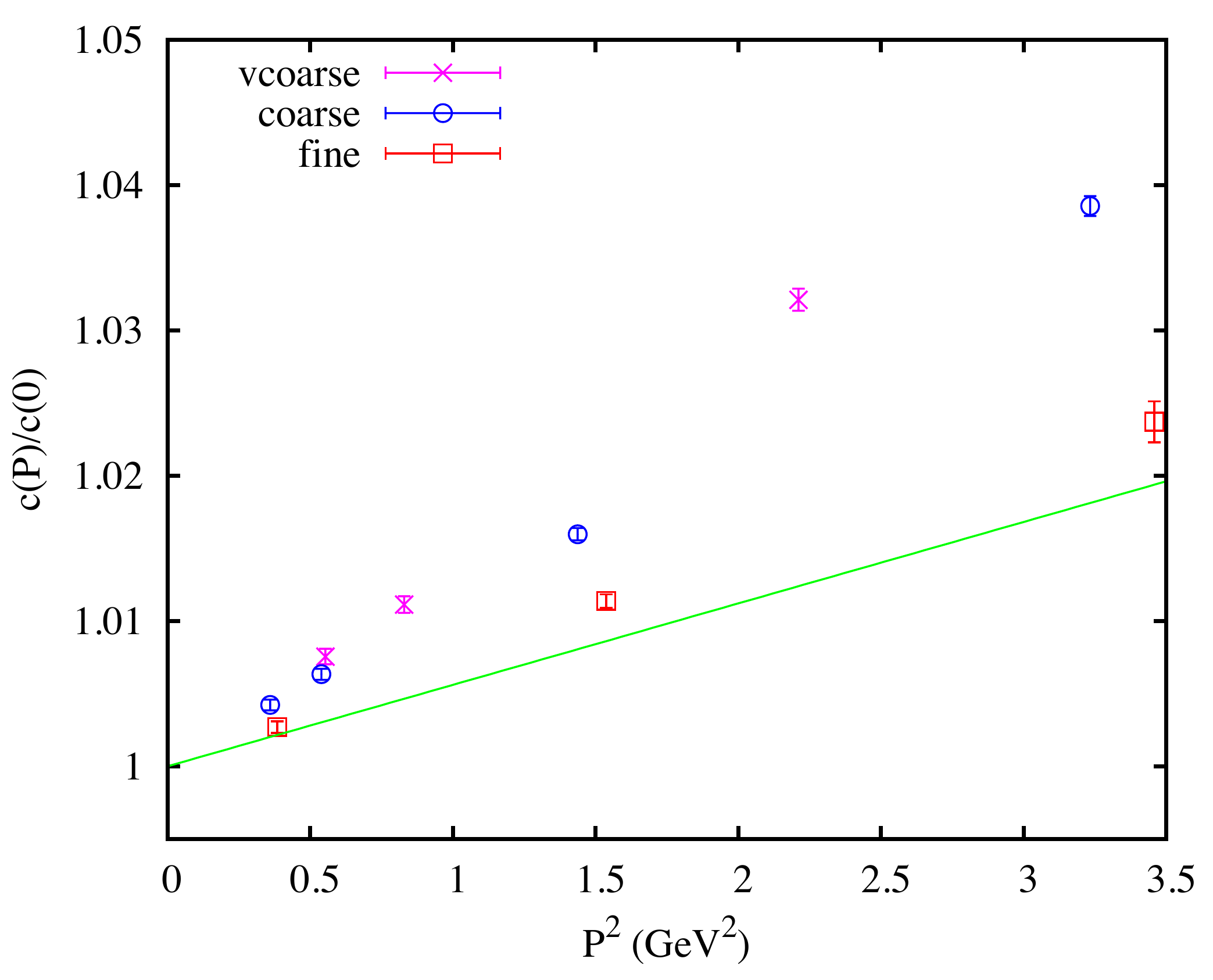}
\end{center}
\caption{Temporal axial current matrix elements derived from 
NRQCD $\eta_b$ correlator amplitudes as a function of 
the square of the $\eta_b$ spatial momentum. Amplitudes 
are given as a ratio to the zero momentum amplitude.  
Results are from very coarse set 1 (pink crosses), coarse set 3 (blue 
circles) and 
fine set 5 (red squares). 
The green line gives the ratio of meson energy to mass, using the 
experimental value for the $\eta_b$ mass.
}
\label{fig:tax}
\end{figure}

\begin{figure}
\begin{center}
\includegraphics[width=0.9\hsize]{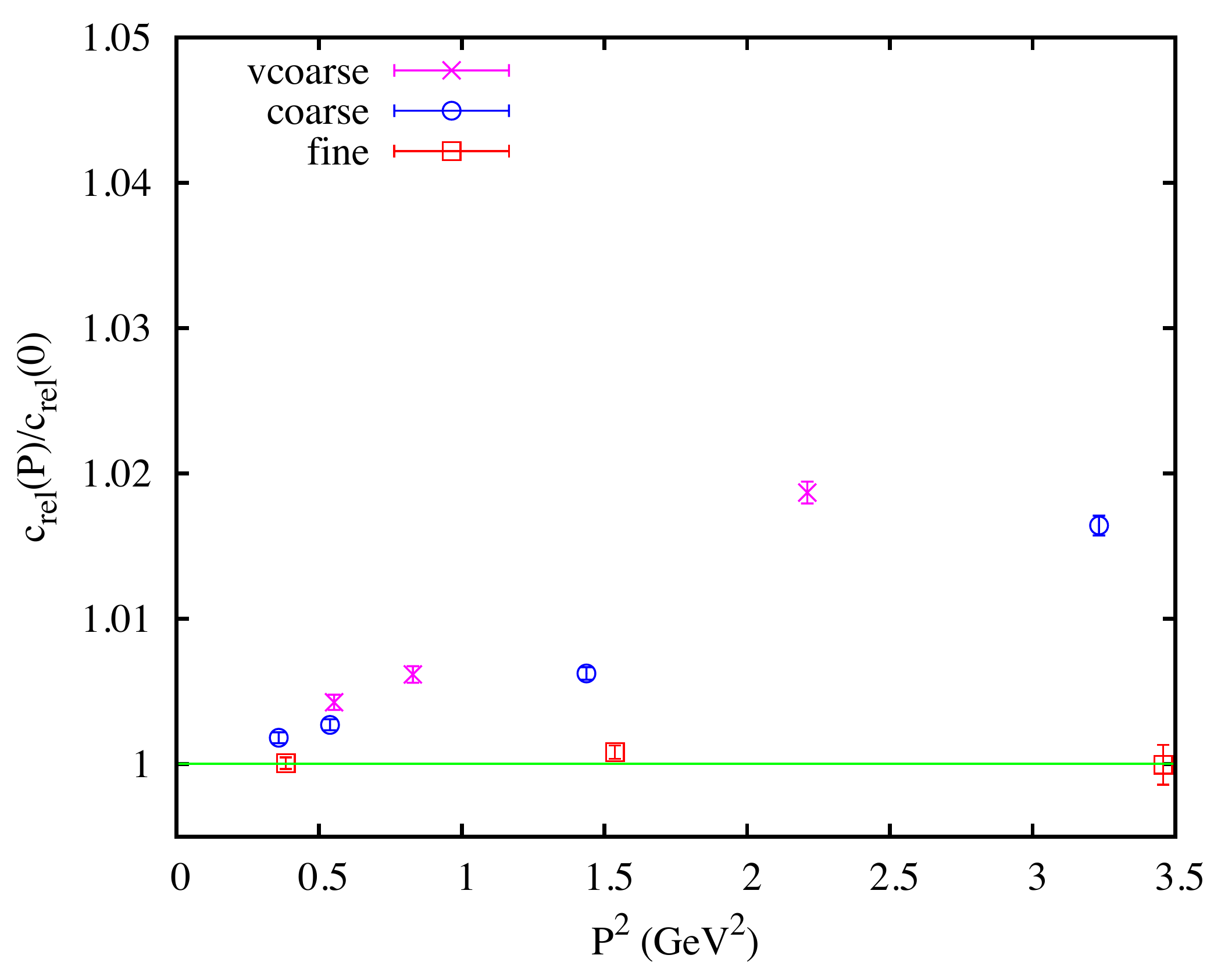}
\end{center}
\caption{Pseudoscalar current matrix elements derived from 
NRQCD $\eta_b$ correlator amplitudes by applying the matching 
factor given in eq.~\ref{eq:crelps} as a function of 
the square of the $\eta_b$ spatial momentum. Amplitudes 
are given as a ratio to the zero momentum amplitude.  
Results are from very coarse set 1 (pink crosses), coarse set 3 (blue 
circles) and 
fine set 5 (red squares). 
The green line shows $P$-independent behaviour. 
}
\label{fig:ps}
\end{figure}

\begin{figure}
\begin{center}
\includegraphics[width=0.9\hsize]{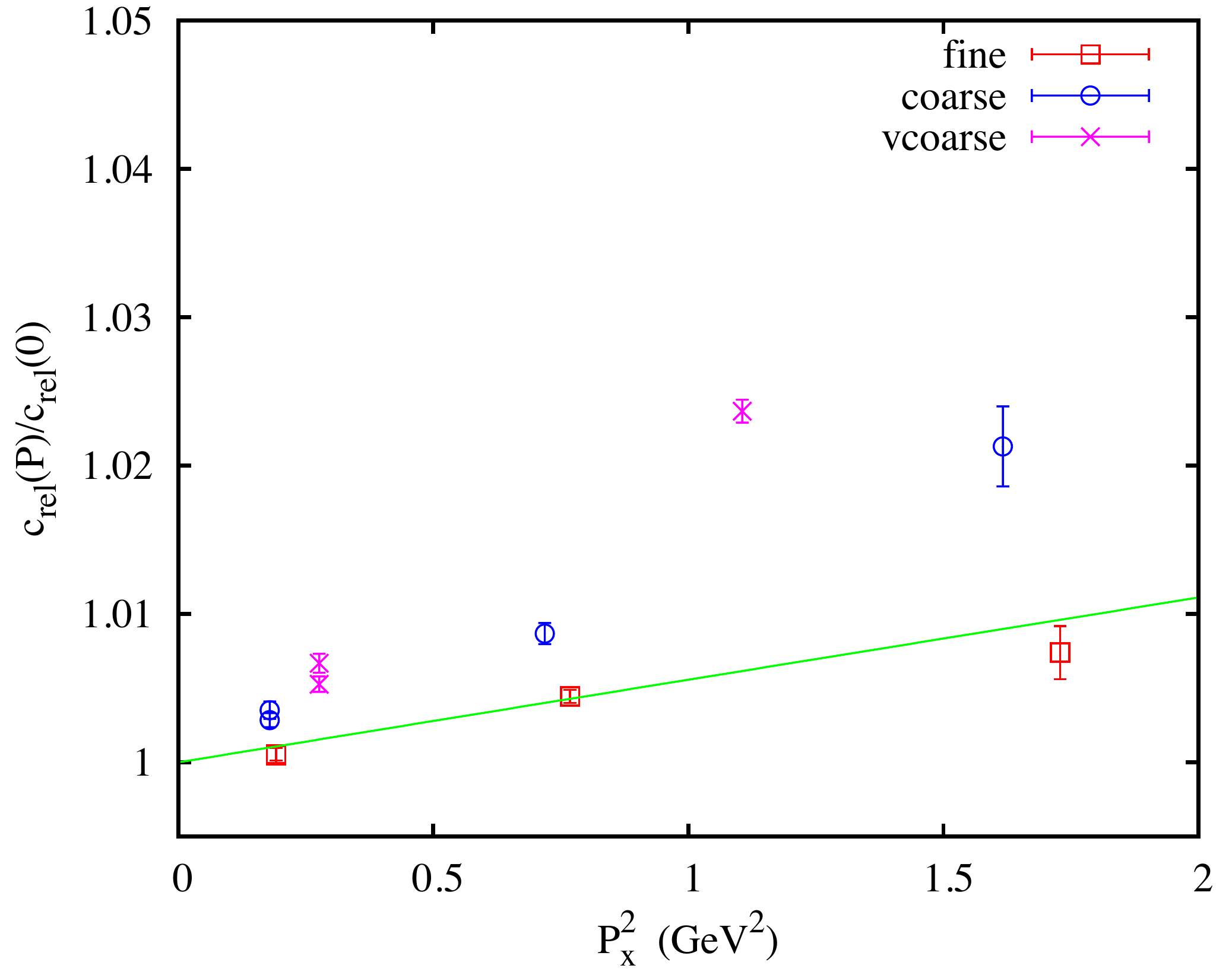}
\end{center}
\caption{Vector current matrix elements derived from 
NRQCD $\Upsilon$ correlator amplitudes for the case 
of a current with polarisation $x$ at both source and sink 
and meson momentum ${\bf P}$ of the form $(p,p,0)$ or 
$(p,p,p)$. 
$c_{rel}$ is obtained from $c$ by applying the matching 
factor given in eq.~\ref{eq:crelvec}. 
Amplitudes 
are given as a ratio to the zero momentum amplitude.  
Results are from very coarse set 1 (pink crosses), coarse set 3 (blue 
circles) and 
fine set 5 (red squares). 
The green line shows the expected behaviour as $\sqrt{1+P_x^2/M_{\Upsilon}^2}$, 
using the experimental value of the $\Upsilon$ mass. 
On all 3 lattices the leftmost points correspond to momentum 
(in units of $2\pi/L_s$) of (1,1,0) and (1,1,1). On the fine 
lattices these two points are on top of each other. 
}
\label{fig:vec}
\end{figure}

Here we discuss the behaviour of the ground-state 
amplitudes/overlaps for mesons made from quark 
propagators from a delta function source, 
$\phi(x) = \delta(x)$ in eq.~(\ref{eq:nrqcdstart}).
This corresponds to the leading order vector current,
$J^{(0)}_{V,\mathrm{NRQCD}}$ (eq.~(\ref{eq:j0})),
in the case of vector mesons and the leading order 
current $J=\chi^{\dag}\psi$ for pseudoscalar mesons. 
We study the amplitudes for these operators
(eq.~(\ref{eq:corrfit})), as a function of spatial 
momentum and show how the correct relativistically covariant behaviour 
develops for moving mesons once $v^4$ terms are added to the NRQCD action.  

For local current operators we expect the following relativistically covariant behaviour:
\begin{eqnarray}
\label{eq:relexpect}
\langle 0 | \overline{\psi} \gamma_5 \psi | \eta_b({\bf p})\rangle &=& \mathrm{constant} \\
\langle 0 | \overline{\psi} \gamma_5 \gamma_0 \psi | \eta_b({\bf p})\rangle &\propto& E_{\eta_b}({\bf p}) \nonumber \\
\langle 0 | \overline{\psi} \gamma_i \psi | \Upsilon({\bf p}, \lambda)\rangle &\propto& \epsilon({\bf p}, \lambda) \nonumber 
\end{eqnarray}
where $\epsilon({\bf p}, \lambda)$ is the $\Upsilon$  polarisation vector. 

To test this for NRQCD we must match NRQCD current operators 
to continuum ones for a quark-antiquark 
pair with net spatial momentum. We work simply at tree-level and perform a nonrelativistic expansion 
of the Dirac bilinear $\overline{v}({\bf p}) \Gamma u({\bf p})$ in terms of Pauli spinors 
using 
\begin{eqnarray}
\label{eq:pauli}
u({\bf p}) &=& \left( \begin{array}{c} \psi \\ \frac{\bf \sigma \cdot p }{E+m} \psi \end{array} \right) \sqrt{\frac{E+m}{2E}}\\ 
v({\bf p}) &=& \left( \begin{array}{c} \frac{\bf \sigma \cdot p }{E+m} \chi \\ \chi \end{array} \right) \sqrt{\frac{E+m}{2E}}\nonumber 
\end{eqnarray}
where $m$ is the quark mass, $E$, its energy and 
we have chosen a nonrelativistic normalisation for the states ($u^{\dag}u$ = $\psi^{\dag}\psi$ = 1). 
When NRQCD to continuum matching is done for mesons at rest~\cite{Hart:2006ij} 
we expand $\overline{v}(-{\bf k})\Gamma u({\bf k})$ in powers of ${\bf k}$ 
where ${\bf k}$ is an internal 
momentum ($\ll m$) for the quarks inside the meson. 
Higher order terms in ${\bf k}$ become relativistic 
corrections to the leading order NRQCD current and implemented via derivative 
operators on the fields. This is the approach taken for the NRQCD vector 
current in section~\ref{subsec:vec} (see eq.~(\ref{eq:j1})). 
In the case where
the meson has momentum ${\bf P}$ we must expand  
$\overline{v}({\bf P/2-k})\Gamma u({\bf P/2+k})$ in powers of ${\bf P}$ 
to identify current correction terms coming from this momentum. 
For simplicity we work to lowest order in ${\bf k}$ i.e. we set
${\bf k}$ to zero. This is sufficient here because the terms 
in ${\bf P}$ dominate those in ${\bf k}$ when we take a ratio 
of results between mesons at rest and moving mesons. The terms 
in ${\bf k}$ will largely cancel because the internal momenta 
change little, whereas effects from the external momentum, ${\bf P}$,
are only present for moving mesons and are highlighted in such a ratio.  

Then to this order : 
\begin{eqnarray}
\label{eq:extcurrcorr}
&&\overline{v}({\bf P}) \gamma_5 u({\bf P}) = \frac{m}{E}\chi^{\dag}\psi \\
&&\overline{v}({\bf P}) \gamma_0 \gamma_5 u({\bf P}) = \chi^{\dag}\psi \nonumber \\
&&\overline{v}({\bf P}) \gamma_i u({\bf P}) = \frac{E+m}{2E}\chi^{\dag}\left[ \sigma_i + \frac{\bf \sigma \cdot P/2}{E+m} \sigma_i \frac{\bf \sigma \cdot P/2}{E+m}\right]\psi \nonumber 
\end{eqnarray}
where $E\equiv E({\bf P})$. 
We see that in the first case the matching generates a simple numerical factor which is a 
function of ${\bf P}$ by which to multiply the NRQCD operator. 
In the second case this factor is simply 1. 
The third case is more complicated since there is a interplay between momentum components 
and meson polarisation, but in fact this is exactly what is required to generate the correct 
sum over polarisation vectors ($\delta_{ij}+P_iP_j/M^2$) in the meson correlation function
when two such operators are combined~\cite{dudekcharm}. 

To test whether and how the factors given in eq.~(\ref{eq:extcurrcorr}) give the 
results expected in eq.~(\ref{eq:relexpect}) we generate NRQCD $b$ quark propagators 
and meson correlation functions $C_{\mathrm{PS,NRQCD}}$ for $\eta_b$ and 
$C_{\mathrm{V,NRQCD}}$ for $\Upsilon$ using local source and sink operators $\chi^{\dag}\psi$
and $\chi^{\dag}\sigma_x\psi$ respectively. The correlators are 
generated at zero and non-zero spatial momentum using a random wall source as described 
for the determination of the kinetic mass
in Section~\ref{subsec:mass}. 
We fit the correlators averaged over configurations to the multi-exponential form 
given in eq.~\ref{eq:corrfit}, extracting the ground state amplitude for 
different momenta, $c({\bf P})$, so that:  
\begin{equation}
C_{NRQCD}(t) \stackrel{t \rightarrow \infty}{=} c({\bf P})c^*({\bf P}) e^{-E_0t} + \ldots
\label{eq:larget}
\end{equation}

Figure~\ref{fig:p2p4} shows results for $c({\bf P})/c(0)$ for the $\eta_b$ 
comparing the NRQCD action we use here that includes terms at $v^4$, with the 
result from just using $H_0$ along with discretisation corrections to $H_0$ 
(i.e. a purely $v^2$ action) on the fine lattices, set 5. We see very 
different behaviour --  for the full action the amplitude rises linearly 
with $P^2$ (as the energy); for the $v^2$ action it does not. 

We now consider the impact of the correction factors in eq.~(\ref{eq:extcurrcorr}). 
For the temporal axial current, as indicated in eq.~(\ref{eq:extcurrcorr}), 
the factor connecting the NRQCD operator $\chi^{\dag}\psi$ and the 
Dirac operator is 1. Thus the NRQCD amplitude in this case can be directly 
compared with the expectation in eq.~(\ref{eq:relexpect}) since we 
take a nonrelativistic normalisation for the states. We therefore expect 
growth of the amplitude according to the ratio of the 
meson energy to the mass, $E({\bf P})/M_{\eta_b}$.   
Figure~\ref{fig:tax} shows that indeed, for the full NRQCD action, the temporal 
axial current matrix element does increase with the energy as it should. It is 
clear from comparison with Figure~\ref{fig:p2p4} that this would not happen 
for the $v^2$-only NRQCD action. 

Figure~\ref{fig:ps} shows equivalent results for the amplitude that 
can be related to the pseudoscalar current matrix element. Here we have taken
\begin{equation}
c_{\mathrm{rel}}({\bf P}) = c({\bf P}) \frac{m}{\sqrt{P^2/4+m^2}}
\label{eq:crelps}
\end{equation}
following eq.~(\ref{eq:extcurrcorr}) for that case. This then results 
in an amplitude which becomes $P$-independent on the fine lattices 
for the full $v^4$ NRQCD action,
as it should from eq.~\ref{eq:relexpect}. 

For the vector we illustrate the results for the correlator made 
using $\chi^{\dag}\sigma_x \psi$ at both source and sink. Then 
the numerical matching factor to convert the NRQCD amplitudes 
into Dirac amplitudes is given by: 
\begin{equation}
c_{\mathrm{rel}}({\bf P}) = c({\bf P}) \left[\frac{m}{E}+ \frac{P_x^2}{E(E+m)}\right]
\label{eq:crelvec}
\end{equation}
where $E=\sqrt{P^2+m^2}$.
Figure~\ref{fig:vec} shows the result of applying this matching for 
$\Upsilon$ amplitudes as a function of momentum. The results on successively finer 
lattices again move closer to the expectation from eq.~(\ref{eq:relexpect}), which 
in this case is $\sqrt{1+P_x^2/M_{\Upsilon}^2}$. Note again that this works because 
the original NRQCD amplitudes $c({\bf P})$ in the $\Upsilon$ case behave 
in a very similar way to that of the $\eta_b$ amplitudes shown 
in Figure~\ref{fig:p2p4} i.e. with the $v^2$ only action they are 
approximately $P$-independent (in that case they are indistinguishable 
from $\eta_b$ amplitudes since there are no spin-dependent terms in $H_0$) 
and with the full $v^4$ action they increase approximately as $E_{\Upsilon}$. 

\bibliography{heavy}

\begin{thebibliography}{67}
\expandafter\ifx\csname natexlab\endcsname\relax\def\natexlab#1{#1}\fi
\expandafter\ifx\csname bibnamefont\endcsname\relax
  \def\bibnamefont#1{#1}\fi
\expandafter\ifx\csname bibfnamefont\endcsname\relax
  \def\bibfnamefont#1{#1}\fi
\expandafter\ifx\csname citenamefont\endcsname\relax
  \def\citenamefont#1{#1}\fi
\expandafter\ifx\csname url\endcsname\relax
  \def\url#1{\texttt{#1}}\fi
\expandafter\ifx\csname urlprefix\endcsname\relax\def\urlprefix{URL }\fi
\providecommand{\bibinfo}[2]{#2}
\providecommand{\eprint}[2][]{\url{#2}}

\bibitem[{\citenamefont{Davies et~al.}(2004)}]{ourlatqcd}
\bibinfo{author}{\bibfnamefont{C.}~\bibnamefont{Davies}} \bibnamefont{et~al.}
  (\bibinfo{collaboration}{HPQCD, UKQCD, MILC and Fermilab Lattice
  Collaborations}), \bibinfo{journal}{Phys.Rev.Lett.}
  \textbf{\bibinfo{volume}{92}}, \bibinfo{pages}{022001}
  (\bibinfo{year}{2004}), \eprint{hep-lat/0304004}.

\bibitem[{\citenamefont{Davies}(2011)}]{cdlat11}
\bibinfo{author}{\bibfnamefont{C.}~\bibnamefont{Davies}},
  \bibinfo{journal}{PoS} \textbf{\bibinfo{volume}{LATTICE2011}},
  \bibinfo{pages}{019} (\bibinfo{year}{2011}), \eprint{1203.3862}.

\bibitem[{\citenamefont{Laiho et~al.}(2011)\citenamefont{Laiho, Lunghi, and
  Van~de Water}}]{lunghilat11}
\bibinfo{author}{\bibfnamefont{J.}~\bibnamefont{Laiho}},
  \bibinfo{author}{\bibfnamefont{E.}~\bibnamefont{Lunghi}}, \bibnamefont{and}
  \bibinfo{author}{\bibfnamefont{R.}~\bibnamefont{Van~de Water}},
  \bibinfo{journal}{PoS} \textbf{\bibinfo{volume}{LATTICE2011}},
  \bibinfo{pages}{018} (\bibinfo{year}{2011}), \eprint{1204.0791}.

\bibitem[{\citenamefont{Eichten and Quigg}(1995)}]{Eichten:1995ch}
\bibinfo{author}{\bibfnamefont{E.~J.} \bibnamefont{Eichten}} \bibnamefont{and}
  \bibinfo{author}{\bibfnamefont{C.}~\bibnamefont{Quigg}},
  \bibinfo{journal}{Phys.Rev.} \textbf{\bibinfo{volume}{D52}},
  \bibinfo{pages}{1726} (\bibinfo{year}{1995}), \eprint{hep-ph/9503356}.

\bibitem[{\citenamefont{Pineda and Signer}(2007)}]{Pineda:2006ri}
\bibinfo{author}{\bibfnamefont{A.}~\bibnamefont{Pineda}} \bibnamefont{and}
  \bibinfo{author}{\bibfnamefont{A.}~\bibnamefont{Signer}},
  \bibinfo{journal}{Nucl.Phys.} \textbf{\bibinfo{volume}{B762}},
  \bibinfo{pages}{67} (\bibinfo{year}{2007}), \eprint{hep-ph/0607239}.

\bibitem[{\citenamefont{Brambilla et~al.}(2011)\citenamefont{Brambilla,
  Eidelman, Heltsley, Vogt, Bodwin et~al.}}]{qwg10}
\bibinfo{author}{\bibfnamefont{N.}~\bibnamefont{Brambilla}},
  \bibinfo{author}{\bibfnamefont{S.}~\bibnamefont{Eidelman}},
  \bibinfo{author}{\bibfnamefont{B.}~\bibnamefont{Heltsley}},
  \bibinfo{author}{\bibfnamefont{R.}~\bibnamefont{Vogt}},
  \bibinfo{author}{\bibfnamefont{G.}~\bibnamefont{Bodwin}},
  \bibnamefont{et~al.}, \bibinfo{journal}{Eur.Phys.J.}
  \textbf{\bibinfo{volume}{C71}}, \bibinfo{pages}{1534} (\bibinfo{year}{2011}),
  \eprint{1010.5827}.

\bibitem[{\citenamefont{Beneke et~al.}(2014)\citenamefont{Beneke, Kiyo,
  Marquard, Penin, Piclum et~al.}}]{upsleptpert}
\bibinfo{author}{\bibfnamefont{M.}~\bibnamefont{Beneke}},
  \bibinfo{author}{\bibfnamefont{Y.}~\bibnamefont{Kiyo}},
  \bibinfo{author}{\bibfnamefont{P.}~\bibnamefont{Marquard}},
  \bibinfo{author}{\bibfnamefont{A.}~\bibnamefont{Penin}},
  \bibinfo{author}{\bibfnamefont{J.}~\bibnamefont{Piclum}},
  \bibnamefont{et~al.}, \bibinfo{journal}{Phys.Rev.Lett.}
  \textbf{\bibinfo{volume}{112}}, \bibinfo{pages}{151801}
  (\bibinfo{year}{2014}), \eprint{1401.3005}.

\bibitem[{\citenamefont{Donald et~al.}(2012)\citenamefont{Donald, Davies,
  Dowdall, Follana, Hornbostel et~al.}}]{psipaper}
\bibinfo{author}{\bibfnamefont{G.}~\bibnamefont{Donald}},
  \bibinfo{author}{\bibfnamefont{C.}~\bibnamefont{Davies}},
  \bibinfo{author}{\bibfnamefont{R.}~\bibnamefont{Dowdall}},
  \bibinfo{author}{\bibfnamefont{E.}~\bibnamefont{Follana}},
  \bibinfo{author}{\bibfnamefont{K.}~\bibnamefont{Hornbostel}},
  \bibnamefont{et~al.} (\bibinfo{collaboration}{HPQCD collaboration}),
  \bibinfo{journal}{Phys.Rev.} \textbf{\bibinfo{volume}{D86}},
  \bibinfo{pages}{094501} (\bibinfo{year}{2012}), \eprint{1208.2855}.

\bibitem[{\citenamefont{Becirevic and Sanfilippo}(2012)}]{etmccharm}
\bibinfo{author}{\bibfnamefont{D.}~\bibnamefont{Becirevic}} \bibnamefont{and}
  \bibinfo{author}{\bibfnamefont{F.}~\bibnamefont{Sanfilippo}}
  (\bibinfo{year}{2012}), \eprint{1206.1445}.

\bibitem[{\citenamefont{Davies et~al.}(1994{\natexlab{a}})\citenamefont{Davies,
  Hornbostel, Langnau, Lepage, Lidsey et~al.}}]{Daviesorigups}
\bibinfo{author}{\bibfnamefont{C.}~\bibnamefont{Davies}},
  \bibinfo{author}{\bibfnamefont{K.}~\bibnamefont{Hornbostel}},
  \bibinfo{author}{\bibfnamefont{A.}~\bibnamefont{Langnau}},
  \bibinfo{author}{\bibfnamefont{G.}~\bibnamefont{Lepage}},
  \bibinfo{author}{\bibfnamefont{A.}~\bibnamefont{Lidsey}},
  \bibnamefont{et~al.}, \bibinfo{journal}{Phys.Rev.}
  \textbf{\bibinfo{volume}{D50}}, \bibinfo{pages}{6963}
  (\bibinfo{year}{1994}{\natexlab{a}}), \eprint{hep-lat/9406017}.

\bibitem[{\citenamefont{Bodwin et~al.}(2002)\citenamefont{Bodwin, Sinclair, and
  Kim}}]{bodwin}
\bibinfo{author}{\bibfnamefont{G.~T.} \bibnamefont{Bodwin}},
  \bibinfo{author}{\bibfnamefont{D.}~\bibnamefont{Sinclair}}, \bibnamefont{and}
  \bibinfo{author}{\bibfnamefont{S.}~\bibnamefont{Kim}},
  \bibinfo{journal}{Phys.Rev.} \textbf{\bibinfo{volume}{D65}},
  \bibinfo{pages}{054504} (\bibinfo{year}{2002}), \eprint{hep-lat/0107011}.

\bibitem[{\citenamefont{Gray et~al.}(2005)\citenamefont{Gray, Allison, Davies,
  Dalgic, Lepage et~al.}}]{gray}
\bibinfo{author}{\bibfnamefont{A.}~\bibnamefont{Gray}},
  \bibinfo{author}{\bibfnamefont{I.}~\bibnamefont{Allison}},
  \bibinfo{author}{\bibfnamefont{C.}~\bibnamefont{Davies}},
  \bibinfo{author}{\bibfnamefont{E.}~\bibnamefont{Dalgic}},
  \bibinfo{author}{\bibfnamefont{G.}~\bibnamefont{Lepage}},
  \bibnamefont{et~al.} (\bibinfo{collaboration}{HPQCD Collaboration}),
  \bibinfo{journal}{Phys.Rev.} \textbf{\bibinfo{volume}{D72}},
  \bibinfo{pages}{094507} (\bibinfo{year}{2005}), \eprint{hep-lat/0507013}.

\bibitem[{\citenamefont{Allison et~al.}(2008)\citenamefont{Allison, Dalgic,
  Davies, Follana et~al.}}]{firstcurrcurr}
\bibinfo{author}{\bibfnamefont{I.}~\bibnamefont{Allison}},
  \bibinfo{author}{\bibfnamefont{E.}~\bibnamefont{Dalgic}},
  \bibinfo{author}{\bibfnamefont{C.~T.~H.} \bibnamefont{Davies}},
  \bibinfo{author}{\bibfnamefont{E.}~\bibnamefont{Follana}},
  \bibnamefont{et~al.}, \bibinfo{journal}{Phys.Rev.}
  \textbf{\bibinfo{volume}{D78}}, \bibinfo{pages}{054513}
  (\bibinfo{year}{2008}), \eprint{0805.2999}.

\bibitem[{\citenamefont{McNeile et~al.}(2010)\citenamefont{McNeile, Davies,
  Follana, Hornbostel, and Lepage}}]{bcmasses}
\bibinfo{author}{\bibfnamefont{C.}~\bibnamefont{McNeile}},
  \bibinfo{author}{\bibfnamefont{C.}~\bibnamefont{Davies}},
  \bibinfo{author}{\bibfnamefont{E.}~\bibnamefont{Follana}},
  \bibinfo{author}{\bibfnamefont{K.}~\bibnamefont{Hornbostel}},
  \bibnamefont{and} \bibinfo{author}{\bibfnamefont{G.}~\bibnamefont{Lepage}}
  (\bibinfo{collaboration}{HPQCD Collaboration}), \bibinfo{journal}{Phys.Rev.}
  \textbf{\bibinfo{volume}{D82}}, \bibinfo{pages}{034512}
  (\bibinfo{year}{2010}), \eprint{1004.4285}.

\bibitem[{\citenamefont{Lepage et~al.}(1992)\citenamefont{Lepage, Magnea,
  Nakhleh, Magnea, and Hornbostel}}]{nrqcd}
\bibinfo{author}{\bibfnamefont{G.~P.} \bibnamefont{Lepage}},
  \bibinfo{author}{\bibfnamefont{L.}~\bibnamefont{Magnea}},
  \bibinfo{author}{\bibfnamefont{C.}~\bibnamefont{Nakhleh}},
  \bibinfo{author}{\bibfnamefont{U.}~\bibnamefont{Magnea}}, \bibnamefont{and}
  \bibinfo{author}{\bibfnamefont{K.}~\bibnamefont{Hornbostel}},
  \bibinfo{journal}{Phys.Rev.} \textbf{\bibinfo{volume}{D46}},
  \bibinfo{pages}{4052} (\bibinfo{year}{1992}), \eprint{hep-lat/9205007}.

\bibitem[{\citenamefont{Dowdall
  et~al.}(2012{\natexlab{a}})\citenamefont{Dowdall, Colquhoun, Daldrop, Davies
  et~al.}}]{dowdallr1}
\bibinfo{author}{\bibfnamefont{R.}~\bibnamefont{Dowdall}},
  \bibinfo{author}{\bibfnamefont{B.}~\bibnamefont{Colquhoun}},
  \bibinfo{author}{\bibfnamefont{J.~O.} \bibnamefont{Daldrop}},
  \bibinfo{author}{\bibfnamefont{C.~T.~H.} \bibnamefont{Davies}},
  \bibnamefont{et~al.} (\bibinfo{collaboration}{HPQCD Collaboration}),
  \bibinfo{journal}{Phys.Rev.} \textbf{\bibinfo{volume}{D85}},
  \bibinfo{pages}{054509} (\bibinfo{year}{2012}{\natexlab{a}}),
  \eprint{1110.6887}.

\bibitem[{\citenamefont{Lee et~al.}(2013)\citenamefont{Lee, Monahan, Horgan
  et~al.}}]{mbpert}
\bibinfo{author}{\bibfnamefont{A.}~\bibnamefont{Lee}},
  \bibinfo{author}{\bibfnamefont{C.~J.} \bibnamefont{Monahan}},
  \bibinfo{author}{\bibfnamefont{R.~R.} \bibnamefont{Horgan}},
  \bibnamefont{et~al.} (\bibinfo{collaboration}{HPQCD Collaboration}),
  \bibinfo{journal}{Phys.Rev.} \textbf{\bibinfo{volume}{D87}},
  \bibinfo{pages}{074018} (\bibinfo{year}{2013}), \eprint{1302.3739}.

\bibitem[{\citenamefont{Dawson et~al.}(2013)\citenamefont{Dawson, Gritsan,
  Logan, Qian, Tully et~al.}}]{snowmasshiggs}
\bibinfo{author}{\bibfnamefont{S.}~\bibnamefont{Dawson}},
  \bibinfo{author}{\bibfnamefont{A.}~\bibnamefont{Gritsan}},
  \bibinfo{author}{\bibfnamefont{H.}~\bibnamefont{Logan}},
  \bibinfo{author}{\bibfnamefont{J.}~\bibnamefont{Qian}},
  \bibinfo{author}{\bibfnamefont{C.}~\bibnamefont{Tully}}, \bibnamefont{et~al.}
  (\bibinfo{year}{2013}), \eprint{1310.8361}.

\bibitem[{\citenamefont{Lepage et~al.}(2014)\citenamefont{Lepage, Mackenzie,
  and Peskin}}]{gplhiggs}
\bibinfo{author}{\bibfnamefont{G.~P.} \bibnamefont{Lepage}},
  \bibinfo{author}{\bibfnamefont{P.~B.} \bibnamefont{Mackenzie}},
  \bibnamefont{and} \bibinfo{author}{\bibfnamefont{M.~E.} \bibnamefont{Peskin}}
  (\bibinfo{year}{2014}), \eprint{1404.0319}.

\bibitem[{\citenamefont{Bazavov et~al.}(2013)}]{milchisq}
\bibinfo{author}{\bibfnamefont{A.}~\bibnamefont{Bazavov}} \bibnamefont{et~al.}
  (\bibinfo{collaboration}{MILC Collaboration}), \bibinfo{journal}{Phys.Rev.}
  \textbf{\bibinfo{volume}{D87}}, \bibinfo{pages}{054505}
  (\bibinfo{year}{2013}), \eprint{1212.4768}.

\bibitem[{\citenamefont{Follana et~al.}(2007)}]{hisqdef}
\bibinfo{author}{\bibfnamefont{E.}~\bibnamefont{Follana}} \bibnamefont{et~al.}
  (\bibinfo{collaboration}{HPQCD Collaboration}), \bibinfo{journal}{Phys.Rev.}
  \textbf{\bibinfo{volume}{D75}}, \bibinfo{pages}{054502}
  (\bibinfo{year}{2007}), \eprint{hep-lat/0610092}.

\bibitem[{\citenamefont{Hart et~al.}(2009)\citenamefont{Hart, von Hippel, and
  Horgan}}]{Hart:2008sq}
\bibinfo{author}{\bibfnamefont{A.}~\bibnamefont{Hart}},
  \bibinfo{author}{\bibfnamefont{G.}~\bibnamefont{von Hippel}},
  \bibnamefont{and} \bibinfo{author}{\bibfnamefont{R.}~\bibnamefont{Horgan}}
  (\bibinfo{collaboration}{HPQCD Collaboration}), \bibinfo{journal}{Phys.Rev.}
  \textbf{\bibinfo{volume}{D79}}, \bibinfo{pages}{074008}
  (\bibinfo{year}{2009}), \eprint{0812.0503}.

\bibitem[{\citenamefont{Hammant et~al.}(2013)\citenamefont{Hammant, Hart, von
  Hippel, Horgan, and Monahan}}]{nrqcdimp}
\bibinfo{author}{\bibfnamefont{T.}~\bibnamefont{Hammant}},
  \bibinfo{author}{\bibfnamefont{A.}~\bibnamefont{Hart}},
  \bibinfo{author}{\bibfnamefont{G.}~\bibnamefont{von Hippel}},
  \bibinfo{author}{\bibfnamefont{R.}~\bibnamefont{Horgan}}, \bibnamefont{and}
  \bibinfo{author}{\bibfnamefont{C.}~\bibnamefont{Monahan}},
  \bibinfo{journal}{Phys.Rev.} \textbf{\bibinfo{volume}{D88}},
  \bibinfo{pages}{014505} (\bibinfo{year}{2013}), \eprint{1303.3234}.

\bibitem[{\citenamefont{Dowdall
  et~al.}(2013{\natexlab{a}})\citenamefont{Dowdall, Davies, Horgan, Monahan,
  and Shigemitsu}}]{DowdallBdecay}
\bibinfo{author}{\bibfnamefont{R.}~\bibnamefont{Dowdall}},
  \bibinfo{author}{\bibfnamefont{C.}~\bibnamefont{Davies}},
  \bibinfo{author}{\bibfnamefont{R.}~\bibnamefont{Horgan}},
  \bibinfo{author}{\bibfnamefont{C.}~\bibnamefont{Monahan}}, \bibnamefont{and}
  \bibinfo{author}{\bibfnamefont{J.}~\bibnamefont{Shigemitsu}}
  (\bibinfo{collaboration}{HPQCD Collaboration}),
  \bibinfo{journal}{Phys.Rev.Lett.} \textbf{\bibinfo{volume}{110}},
  \bibinfo{pages}{222003} (\bibinfo{year}{2013}{\natexlab{a}}),
  \eprint{1302.2644}.

\bibitem[{\citenamefont{Daldrop et~al.}(2012)\citenamefont{Daldrop, Davies, and
  Dowdall}}]{Daldrop}
\bibinfo{author}{\bibfnamefont{J.}~\bibnamefont{Daldrop}},
  \bibinfo{author}{\bibfnamefont{C.}~\bibnamefont{Davies}}, \bibnamefont{and}
  \bibinfo{author}{\bibfnamefont{R.}~\bibnamefont{Dowdall}}
  (\bibinfo{collaboration}{HPQCD Collaboration}),
  \bibinfo{journal}{Phys.Rev.Lett.} \textbf{\bibinfo{volume}{108}},
  \bibinfo{pages}{102003} (\bibinfo{year}{2012}), \eprint{1112.2590}.

\bibitem[{\citenamefont{Dowdall
  et~al.}(2013{\natexlab{b}})\citenamefont{Dowdall, Davies, Hammant, and
  Horgan}}]{Dowdallhyp}
\bibinfo{author}{\bibfnamefont{R.}~\bibnamefont{Dowdall}},
  \bibinfo{author}{\bibfnamefont{C.}~\bibnamefont{Davies}},
  \bibinfo{author}{\bibfnamefont{T.}~\bibnamefont{Hammant}}, \bibnamefont{and}
  \bibinfo{author}{\bibfnamefont{R.}~\bibnamefont{Horgan}}
  (\bibinfo{year}{2013}{\natexlab{b}}), \eprint{1309.5797}.

\bibitem[{\citenamefont{Dowdall
  et~al.}(2012{\natexlab{b}})\citenamefont{Dowdall, Davies, Hammant, and
  Horgan}}]{rachelBmass}
\bibinfo{author}{\bibfnamefont{R.}~\bibnamefont{Dowdall}},
  \bibinfo{author}{\bibfnamefont{C.}~\bibnamefont{Davies}},
  \bibinfo{author}{\bibfnamefont{T.}~\bibnamefont{Hammant}}, \bibnamefont{and}
  \bibinfo{author}{\bibfnamefont{R.}~\bibnamefont{Horgan}}
  (\bibinfo{collaboration}{HPQCD Collaboration})
  (\bibinfo{year}{2012}{\natexlab{b}}), \eprint{1207.5149}.

\bibitem[{\citenamefont{Davies et~al.}(2010{\natexlab{a}})\citenamefont{Davies,
  Follana, Kendall, Lepage, and McNeile}}]{oldr1paper}
\bibinfo{author}{\bibfnamefont{C.}~\bibnamefont{Davies}},
  \bibinfo{author}{\bibfnamefont{E.}~\bibnamefont{Follana}},
  \bibinfo{author}{\bibfnamefont{I.}~\bibnamefont{Kendall}},
  \bibinfo{author}{\bibfnamefont{G.~P.} \bibnamefont{Lepage}},
  \bibnamefont{and} \bibinfo{author}{\bibfnamefont{C.}~\bibnamefont{McNeile}}
  (\bibinfo{collaboration}{HPQCD Collaboration}), \bibinfo{journal}{Phys.Rev.}
  \textbf{\bibinfo{volume}{D81}}, \bibinfo{pages}{034506}
  (\bibinfo{year}{2010}{\natexlab{a}}), \eprint{0910.1229}.

\bibitem[{\citenamefont{Beringer et~al.}(2012)}]{pdg}
\bibinfo{author}{\bibfnamefont{J.}~\bibnamefont{Beringer}} \bibnamefont{et~al.}
  (\bibinfo{collaboration}{Particle Data Group}), \bibinfo{journal}{Phys. Rev.}
  \textbf{\bibinfo{volume}{D86}}, \bibinfo{pages}{010001}
  (\bibinfo{year}{2012}).

\bibitem[{\citenamefont{Gregory et~al.}(2011)\citenamefont{Gregory, Davies,
  Kendall, Koponen, Wong et~al.}}]{gregory}
\bibinfo{author}{\bibfnamefont{E.~B.} \bibnamefont{Gregory}},
  \bibinfo{author}{\bibfnamefont{C.~T.} \bibnamefont{Davies}},
  \bibinfo{author}{\bibfnamefont{I.~D.} \bibnamefont{Kendall}},
  \bibinfo{author}{\bibfnamefont{J.}~\bibnamefont{Koponen}},
  \bibinfo{author}{\bibfnamefont{K.}~\bibnamefont{Wong}}, \bibnamefont{et~al.}
  (\bibinfo{collaboration}{HPQCD Collaboration}), \bibinfo{journal}{Phys.Rev.}
  \textbf{\bibinfo{volume}{D83}}, \bibinfo{pages}{014506}
  (\bibinfo{year}{2011}), \eprint{1010.3848}.

\bibitem[{\citenamefont{Erler}(1999)}]{alpha-em}
\bibinfo{author}{\bibfnamefont{J.}~\bibnamefont{Erler}},
  \bibinfo{journal}{Phys.Rev.} \textbf{\bibinfo{volume}{D59}},
  \bibinfo{pages}{054008} (\bibinfo{year}{1999}), \eprint{hep-ph/9803453}.

\bibitem[{\citenamefont{Adams et~al.}(2005)}]{cleo1}
\bibinfo{author}{\bibfnamefont{G.}~\bibnamefont{Adams}} \bibnamefont{et~al.}
  (\bibinfo{collaboration}{CLEO Collaboration}),
  \bibinfo{journal}{Phys.Rev.Lett.} \textbf{\bibinfo{volume}{94}},
  \bibinfo{pages}{012001} (\bibinfo{year}{2005}), \eprint{hep-ex/0409027}.

\bibitem[{\citenamefont{Rosner et~al.}(2006)}]{cleo2}
\bibinfo{author}{\bibfnamefont{J.}~\bibnamefont{Rosner}} \bibnamefont{et~al.}
  (\bibinfo{collaboration}{CLEO Collaboration}),
  \bibinfo{journal}{Phys.Rev.Lett.} \textbf{\bibinfo{volume}{96}},
  \bibinfo{pages}{092003} (\bibinfo{year}{2006}), \eprint{hep-ex/0512056}.

\bibitem[{\citenamefont{Besson et~al.}(2007)}]{cleo3}
\bibinfo{author}{\bibfnamefont{D.}~\bibnamefont{Besson}} \bibnamefont{et~al.}
  (\bibinfo{collaboration}{CLEO Collaboration}),
  \bibinfo{journal}{Phys.Rev.Lett.} \textbf{\bibinfo{volume}{98}},
  \bibinfo{pages}{052002} (\bibinfo{year}{2007}), \eprint{hep-ex/0607019}.

\bibitem[{\citenamefont{Hart et~al.}(2007)\citenamefont{Hart, von Hippel, and
  Horgan}}]{Hart:2006ij}
\bibinfo{author}{\bibfnamefont{A.}~\bibnamefont{Hart}},
  \bibinfo{author}{\bibfnamefont{G.}~\bibnamefont{von Hippel}},
  \bibnamefont{and} \bibinfo{author}{\bibfnamefont{R.}~\bibnamefont{Horgan}},
  \bibinfo{journal}{Phys.Rev.} \textbf{\bibinfo{volume}{D75}},
  \bibinfo{pages}{014008} (\bibinfo{year}{2007}), \eprint{hep-lat/0605007}.

\bibitem[{\citenamefont{Jones and Woloshyn}(1999)}]{Jones:1998ub}
\bibinfo{author}{\bibfnamefont{B.}~\bibnamefont{Jones}} \bibnamefont{and}
  \bibinfo{author}{\bibfnamefont{R.}~\bibnamefont{Woloshyn}},
  \bibinfo{journal}{Phys.Rev.} \textbf{\bibinfo{volume}{D60}},
  \bibinfo{pages}{014502} (\bibinfo{year}{1999}), \eprint{hep-lat/9812008}.

\bibitem[{\citenamefont{Davies et~al.}(1998)}]{daviesoldups}
\bibinfo{author}{\bibfnamefont{C.}~\bibnamefont{Davies}} \bibnamefont{et~al.}
  (\bibinfo{collaboration}{UKQCD Collaboration}), \bibinfo{journal}{Phys.Rev.}
  \textbf{\bibinfo{volume}{D58}}, \bibinfo{pages}{054505}
  (\bibinfo{year}{1998}), \eprint{hep-lat/9802024}.

\bibitem[{\citenamefont{Chetyrkin et~al.}(2006)\citenamefont{Chetyrkin, Kuhn,
  and Sturm}}]{qcdpt1}
\bibinfo{author}{\bibfnamefont{K.}~\bibnamefont{Chetyrkin}},
  \bibinfo{author}{\bibfnamefont{J.~H.} \bibnamefont{Kuhn}}, \bibnamefont{and}
  \bibinfo{author}{\bibfnamefont{C.}~\bibnamefont{Sturm}},
  \bibinfo{journal}{Eur.Phys.J.} \textbf{\bibinfo{volume}{C48}},
  \bibinfo{pages}{107} (\bibinfo{year}{2006}), \eprint{hep-ph/0604234}.

\bibitem[{\citenamefont{Boughezal et~al.}(2006)\citenamefont{Boughezal, Czakon,
  and Schutzmeier}}]{qcdpt2}
\bibinfo{author}{\bibfnamefont{R.}~\bibnamefont{Boughezal}},
  \bibinfo{author}{\bibfnamefont{M.}~\bibnamefont{Czakon}}, \bibnamefont{and}
  \bibinfo{author}{\bibfnamefont{T.}~\bibnamefont{Schutzmeier}},
  \bibinfo{journal}{Phys.Rev.} \textbf{\bibinfo{volume}{D74}},
  \bibinfo{pages}{074006} (\bibinfo{year}{2006}), \eprint{hep-ph/0605023}.

\bibitem[{\citenamefont{Maier et~al.}(2008)\citenamefont{Maier, Maierhofer, and
  Marqaurd}}]{qcdpt3}
\bibinfo{author}{\bibfnamefont{A.}~\bibnamefont{Maier}},
  \bibinfo{author}{\bibfnamefont{P.}~\bibnamefont{Maierhofer}},
  \bibnamefont{and} \bibinfo{author}{\bibfnamefont{P.}~\bibnamefont{Marqaurd}},
  \bibinfo{journal}{Phys.Lett.} \textbf{\bibinfo{volume}{B669}},
  \bibinfo{pages}{88} (\bibinfo{year}{2008}), \eprint{0806.3405}.

\bibitem[{\citenamefont{Maier et~al.}(2010)\citenamefont{Maier, Maierhofer,
  Marquard, and Smirnov}}]{qcdpt4}
\bibinfo{author}{\bibfnamefont{A.}~\bibnamefont{Maier}},
  \bibinfo{author}{\bibfnamefont{P.}~\bibnamefont{Maierhofer}},
  \bibinfo{author}{\bibfnamefont{P.}~\bibnamefont{Marquard}}, \bibnamefont{and}
  \bibinfo{author}{\bibfnamefont{A.}~\bibnamefont{Smirnov}},
  \bibinfo{journal}{Nucl.Phys.} \textbf{\bibinfo{volume}{B824}},
  \bibinfo{pages}{1} (\bibinfo{year}{2010}), \eprint{0907.2117}.

\bibitem[{\citenamefont{Kiyo et~al.}(2009)\citenamefont{Kiyo, Maier,
  Maierhofer, and Marquard}}]{qcdpt5}
\bibinfo{author}{\bibfnamefont{Y.}~\bibnamefont{Kiyo}},
  \bibinfo{author}{\bibfnamefont{A.}~\bibnamefont{Maier}},
  \bibinfo{author}{\bibfnamefont{P.}~\bibnamefont{Maierhofer}},
  \bibnamefont{and} \bibinfo{author}{\bibfnamefont{P.}~\bibnamefont{Marquard}},
  \bibinfo{journal}{Nucl.Phys.} \textbf{\bibinfo{volume}{B823}},
  \bibinfo{pages}{269} (\bibinfo{year}{2009}), \eprint{0907.2120}.

\bibitem[{\citenamefont{Czakon and Schutzmeier}(2008)}]{Czakon:2007qi}
\bibinfo{author}{\bibfnamefont{M.}~\bibnamefont{Czakon}} \bibnamefont{and}
  \bibinfo{author}{\bibfnamefont{T.}~\bibnamefont{Schutzmeier}},
  \bibinfo{journal}{JHEP} \textbf{\bibinfo{volume}{0807}}, \bibinfo{pages}{001}
  (\bibinfo{year}{2008}), \eprint{0712.2762}.

\bibitem[{\citenamefont{Kuhn et~al.}(2007)\citenamefont{Kuhn, Steinhauser, and
  Sturm}}]{kuhnmc07}
\bibinfo{author}{\bibfnamefont{J.~H.} \bibnamefont{Kuhn}},
  \bibinfo{author}{\bibfnamefont{M.}~\bibnamefont{Steinhauser}},
  \bibnamefont{and} \bibinfo{author}{\bibfnamefont{C.}~\bibnamefont{Sturm}},
  \bibinfo{journal}{Nucl.Phys.} \textbf{\bibinfo{volume}{B778}},
  \bibinfo{pages}{192} (\bibinfo{year}{2007}), \eprint{hep-ph/0702103}.

\bibitem[{\citenamefont{Lepage et~al.}(2002)}]{gplbayes}
\bibinfo{author}{\bibfnamefont{G.~P.} \bibnamefont{Lepage}}
  \bibnamefont{et~al.}, \bibinfo{journal}{Nucl. Phys. Proc. Suppl.}
  \textbf{\bibinfo{volume}{106}}, \bibinfo{pages}{12} (\bibinfo{year}{2002}),
  \eprint{hep-lat/0110175}.

\bibitem[{\citenamefont{Davies et~al.}(2010{\natexlab{b}})\citenamefont{Davies,
  McNeile, Follana, Lepage, Na et~al.}}]{fdsupdate}
\bibinfo{author}{\bibfnamefont{C.}~\bibnamefont{Davies}},
  \bibinfo{author}{\bibfnamefont{C.}~\bibnamefont{McNeile}},
  \bibinfo{author}{\bibfnamefont{E.}~\bibnamefont{Follana}},
  \bibinfo{author}{\bibfnamefont{G.}~\bibnamefont{Lepage}},
  \bibinfo{author}{\bibfnamefont{H.}~\bibnamefont{Na}}, \bibnamefont{et~al.}
  (\bibinfo{collaboration}{HPQCD Collaboration}), \bibinfo{journal}{Phys.Rev.}
  \textbf{\bibinfo{volume}{D82}}, \bibinfo{pages}{114504}
  (\bibinfo{year}{2010}{\natexlab{b}}), \eprint{1008.4018}.

\bibitem[{\citenamefont{McNeile et~al.}(2012)\citenamefont{McNeile, Davies,
  Follana, Hornbostel, and Lepage}}]{McNeile:2012qf}
\bibinfo{author}{\bibfnamefont{C.}~\bibnamefont{McNeile}},
  \bibinfo{author}{\bibfnamefont{C.}~\bibnamefont{Davies}},
  \bibinfo{author}{\bibfnamefont{E.}~\bibnamefont{Follana}},
  \bibinfo{author}{\bibfnamefont{K.}~\bibnamefont{Hornbostel}},
  \bibnamefont{and} \bibinfo{author}{\bibfnamefont{G.}~\bibnamefont{Lepage}},
  \bibinfo{journal}{Phys.Rev.} \textbf{\bibinfo{volume}{D86}},
  \bibinfo{pages}{074503} (\bibinfo{year}{2012}), \eprint{1207.0994}.

\bibitem[{\citenamefont{Chetyrkin et~al.}(2009)\citenamefont{Chetyrkin, Kuhn,
  Maier, Maierhofer, Marquard et~al.}}]{kuhn09update}
\bibinfo{author}{\bibfnamefont{K.}~\bibnamefont{Chetyrkin}},
  \bibinfo{author}{\bibfnamefont{J.}~\bibnamefont{Kuhn}},
  \bibinfo{author}{\bibfnamefont{A.}~\bibnamefont{Maier}},
  \bibinfo{author}{\bibfnamefont{P.}~\bibnamefont{Maierhofer}},
  \bibinfo{author}{\bibfnamefont{P.}~\bibnamefont{Marquard}},
  \bibnamefont{et~al.}, \bibinfo{journal}{Phys.Rev.}
  \textbf{\bibinfo{volume}{D80}}, \bibinfo{pages}{074010}
  (\bibinfo{year}{2009}), \eprint{0907.2110}.

\bibitem[{\citenamefont{Chakraborty
  et~al.}(2014{\natexlab{a}})\citenamefont{Chakraborty, Davies, Donald,
  Dowdall, Koponen et~al.}}]{Chakraborty:2014mwa}
\bibinfo{author}{\bibfnamefont{B.}~\bibnamefont{Chakraborty}},
  \bibinfo{author}{\bibfnamefont{C.}~\bibnamefont{Davies}},
  \bibinfo{author}{\bibfnamefont{G.}~\bibnamefont{Donald}},
  \bibinfo{author}{\bibfnamefont{R.}~\bibnamefont{Dowdall}},
  \bibinfo{author}{\bibfnamefont{J.}~\bibnamefont{Koponen}},
  \bibnamefont{et~al.} (\bibinfo{year}{2014}{\natexlab{a}}),
  \eprint{1403.1778}.

\bibitem[{\citenamefont{Bodenstein et~al.}(2012)\citenamefont{Bodenstein,
  Dominguez, and Schilcher}}]{Bodenstein:2011qy}
\bibinfo{author}{\bibfnamefont{S.}~\bibnamefont{Bodenstein}},
  \bibinfo{author}{\bibfnamefont{C.}~\bibnamefont{Dominguez}},
  \bibnamefont{and}
  \bibinfo{author}{\bibfnamefont{K.}~\bibnamefont{Schilcher}},
  \bibinfo{journal}{Phys.Rev.} \textbf{\bibinfo{volume}{D85}},
  \bibinfo{pages}{014029} (\bibinfo{year}{2012}), \eprint{1106.0427}.

\bibitem[{\citenamefont{Davies et~al.}(2013)\citenamefont{Davies, Colquhoun,
  Galloway, Donald, Koponen et~al.}}]{Davies:2013dem}
\bibinfo{author}{\bibfnamefont{C.}~\bibnamefont{Davies}},
  \bibinfo{author}{\bibfnamefont{B.}~\bibnamefont{Colquhoun}},
  \bibinfo{author}{\bibfnamefont{B.}~\bibnamefont{Galloway}},
  \bibinfo{author}{\bibfnamefont{G.}~\bibnamefont{Donald}},
  \bibinfo{author}{\bibfnamefont{J.}~\bibnamefont{Koponen}},
  \bibnamefont{et~al.}, \bibinfo{journal}{PoS}
  \textbf{\bibinfo{volume}{LATTICE2013}}, \bibinfo{pages}{438}
  (\bibinfo{year}{2013}), \eprint{1312.5874}.

\bibitem[{\citenamefont{Chetyrkin et~al.}(1998)\citenamefont{Chetyrkin, Kniehl,
  and Steinhauser}}]{Chetyrkin:1997un}
\bibinfo{author}{\bibfnamefont{K.}~\bibnamefont{Chetyrkin}},
  \bibinfo{author}{\bibfnamefont{B.~A.} \bibnamefont{Kniehl}},
  \bibnamefont{and}
  \bibinfo{author}{\bibfnamefont{M.}~\bibnamefont{Steinhauser}},
  \bibinfo{journal}{Nucl.Phys.} \textbf{\bibinfo{volume}{B510}},
  \bibinfo{pages}{61} (\bibinfo{year}{1998}), \eprint{hep-ph/9708255}.

\bibitem[{\citenamefont{Adachi et~al.}(2013)}]{belle1}
\bibinfo{author}{\bibfnamefont{I.}~\bibnamefont{Adachi}} \bibnamefont{et~al.}
  (\bibinfo{collaboration}{Belle Collaboration}),
  \bibinfo{journal}{Phys.Rev.Lett.} \textbf{\bibinfo{volume}{110}},
  \bibinfo{pages}{131801} (\bibinfo{year}{2013}), \eprint{1208.4678}.

\bibitem[{\citenamefont{Hara et~al.}(2010)}]{belle2}
\bibinfo{author}{\bibfnamefont{K.}~\bibnamefont{Hara}} \bibnamefont{et~al.}
  (\bibinfo{collaboration}{Belle collaboration}), \bibinfo{journal}{Phys.Rev.}
  \textbf{\bibinfo{volume}{D82}}, \bibinfo{pages}{071101}
  (\bibinfo{year}{2010}), \eprint{1006.4201}.

\bibitem[{\citenamefont{Aubert et~al.}(2010)}]{babar1}
\bibinfo{author}{\bibfnamefont{B.}~\bibnamefont{Aubert}} \bibnamefont{et~al.}
  (\bibinfo{collaboration}{BaBar Collaboration}), \bibinfo{journal}{Phys.Rev.}
  \textbf{\bibinfo{volume}{D81}}, \bibinfo{pages}{051101}
  (\bibinfo{year}{2010}), \eprint{0912.2453}.

\bibitem[{\citenamefont{Aubert et~al.}(2008)}]{babar2}
\bibinfo{author}{\bibfnamefont{B.}~\bibnamefont{Aubert}} \bibnamefont{et~al.}
  (\bibinfo{collaboration}{BaBar Collaboration}), \bibinfo{journal}{Phys.Rev.}
  \textbf{\bibinfo{volume}{D77}}, \bibinfo{pages}{011107}
  (\bibinfo{year}{2008}), \eprint{0708.2260}.

\bibitem[{\citenamefont{Donald et~al.}(2013)\citenamefont{Donald, Davies,
  Koponen, and Lepage}}]{Donald:2013pea}
\bibinfo{author}{\bibfnamefont{G.}~\bibnamefont{Donald}},
  \bibinfo{author}{\bibfnamefont{C.}~\bibnamefont{Davies}},
  \bibinfo{author}{\bibfnamefont{J.}~\bibnamefont{Koponen}}, \bibnamefont{and}
  \bibinfo{author}{\bibfnamefont{G.}~\bibnamefont{Lepage}}
  (\bibinfo{collaboration}{HPQCD Collaboration}) (\bibinfo{year}{2013}),
  \eprint{1311.6669}.

\bibitem[{\citenamefont{Donald et~al.}(2014)\citenamefont{Donald, Davies,
  Koponen, and Lepage}}]{dsstar}
\bibinfo{author}{\bibfnamefont{G.}~\bibnamefont{Donald}},
  \bibinfo{author}{\bibfnamefont{C.}~\bibnamefont{Davies}},
  \bibinfo{author}{\bibfnamefont{J.}~\bibnamefont{Koponen}}, \bibnamefont{and}
  \bibinfo{author}{\bibfnamefont{G.}~\bibnamefont{Lepage}}
  (\bibinfo{collaboration}{HPQCD Collaboration}),
  \bibinfo{journal}{Phys.Rev.Lett.} \textbf{\bibinfo{volume}{112}},
  \bibinfo{pages}{212002} (\bibinfo{year}{2014}), \eprint{1312.5264}.

\bibitem[{\citenamefont{Dowdall
  et~al.}(2013{\natexlab{c}})\citenamefont{Dowdall, Davies, Lepage, and
  McNeile}}]{Dowdallfkpi}
\bibinfo{author}{\bibfnamefont{R.}~\bibnamefont{Dowdall}},
  \bibinfo{author}{\bibfnamefont{C.}~\bibnamefont{Davies}},
  \bibinfo{author}{\bibfnamefont{G.}~\bibnamefont{Lepage}}, \bibnamefont{and}
  \bibinfo{author}{\bibfnamefont{C.}~\bibnamefont{McNeile}},
  \bibinfo{journal}{Phys.Rev.} \textbf{\bibinfo{volume}{D88}},
  \bibinfo{pages}{074504} (\bibinfo{year}{2013}{\natexlab{c}}),
  \eprint{1303.1670}.

\bibitem[{\citenamefont{Bazavov et~al.}(2014)}]{milc14}
\bibinfo{author}{\bibfnamefont{A.}~\bibnamefont{Bazavov}} \bibnamefont{et~al.}
  (\bibinfo{collaboration}{Fermilab Lattice and MILC Collaborations})
  (\bibinfo{year}{2014}), \eprint{1407.3772}.

\bibitem[{\citenamefont{Follana et~al.}(2008)\citenamefont{Follana, Davies,
  Lepage, and Shigemitsu}}]{fdsorig}
\bibinfo{author}{\bibfnamefont{E.}~\bibnamefont{Follana}},
  \bibinfo{author}{\bibfnamefont{C.}~\bibnamefont{Davies}},
  \bibinfo{author}{\bibfnamefont{G.}~\bibnamefont{Lepage}}, \bibnamefont{and}
  \bibinfo{author}{\bibfnamefont{J.}~\bibnamefont{Shigemitsu}}
  (\bibinfo{collaboration}{HPQCD Collaboration}),
  \bibinfo{journal}{Phys.Rev.Lett.} \textbf{\bibinfo{volume}{100}},
  \bibinfo{pages}{062002} (\bibinfo{year}{2008}), \eprint{0706.1726}.

\bibitem[{\citenamefont{Chakraborty
  et~al.}(2014{\natexlab{b}})\citenamefont{Chakraborty, Davies, Galloway
  et~al.}}]{hisqnew}
\bibinfo{author}{\bibfnamefont{B.}~\bibnamefont{Chakraborty}},
  \bibinfo{author}{\bibfnamefont{C.~T.~H.} \bibnamefont{Davies}},
  \bibinfo{author}{\bibfnamefont{B.}~\bibnamefont{Galloway}},
  \bibnamefont{et~al.} (\bibinfo{collaboration}{HPQCD Collaboration})
  (\bibinfo{year}{2014}{\natexlab{b}}), \eprint{1408.4169}.

\bibitem[{\citenamefont{Carrasco et~al.}(2013)\citenamefont{Carrasco,
  Dimopoulos, Frezzotti, Giménez, Lami et~al.}}]{Carrasco:2013naa}
\bibinfo{author}{\bibfnamefont{N.}~\bibnamefont{Carrasco}},
  \bibinfo{author}{\bibfnamefont{P.}~\bibnamefont{Dimopoulos}},
  \bibinfo{author}{\bibfnamefont{R.}~\bibnamefont{Frezzotti}},
  \bibinfo{author}{\bibfnamefont{V.}~\bibnamefont{Giménez}},
  \bibinfo{author}{\bibfnamefont{P.}~\bibnamefont{Lami}}, \bibnamefont{et~al.}
  (\bibinfo{year}{2013}), \eprint{1311.2837}.

\bibitem[{\citenamefont{Davies et~al.}(1994{\natexlab{b}})\citenamefont{Davies,
  Hornbostel, Langnau, Lepage, Lidsey et~al.}}]{Davies:1994pz}
\bibinfo{author}{\bibfnamefont{C.}~\bibnamefont{Davies}},
  \bibinfo{author}{\bibfnamefont{K.}~\bibnamefont{Hornbostel}},
  \bibinfo{author}{\bibfnamefont{A.}~\bibnamefont{Langnau}},
  \bibinfo{author}{\bibfnamefont{G.}~\bibnamefont{Lepage}},
  \bibinfo{author}{\bibfnamefont{A.}~\bibnamefont{Lidsey}},
  \bibnamefont{et~al.}, \bibinfo{journal}{Phys.Rev.Lett.}
  \textbf{\bibinfo{volume}{73}}, \bibinfo{pages}{2654}
  (\bibinfo{year}{1994}{\natexlab{b}}), \eprint{hep-lat/9404012}.

\bibitem[{\citenamefont{El-Khadra et~al.}(1997)\citenamefont{El-Khadra,
  Kronfeld, and Mackenzie}}]{fermilab}
\bibinfo{author}{\bibfnamefont{A.~X.} \bibnamefont{El-Khadra}},
  \bibinfo{author}{\bibfnamefont{A.~S.} \bibnamefont{Kronfeld}},
  \bibnamefont{and} \bibinfo{author}{\bibfnamefont{P.~B.}
  \bibnamefont{Mackenzie}}, \bibinfo{journal}{Phys.Rev.}
  \textbf{\bibinfo{volume}{D55}}, \bibinfo{pages}{3933} (\bibinfo{year}{1997}),
  \eprint{hep-lat/9604004}.

\bibitem[{\citenamefont{Bernard et~al.}(2011)}]{Bernard:2010fr}
\bibinfo{author}{\bibfnamefont{C.}~\bibnamefont{Bernard}} \bibnamefont{et~al.}
  (\bibinfo{collaboration}{Fermilab Lattice Collaboration, MILC
  Collaboration}), \bibinfo{journal}{Phys.Rev.} \textbf{\bibinfo{volume}{D83}},
  \bibinfo{pages}{034503} (\bibinfo{year}{2011}), \eprint{1003.1937}.

\bibitem[{\citenamefont{Dudek et~al.}(2006)\citenamefont{Dudek, Edwards, and
  Richards}}]{dudekcharm}
\bibinfo{author}{\bibfnamefont{J.~J.} \bibnamefont{Dudek}},
  \bibinfo{author}{\bibfnamefont{R.~G.} \bibnamefont{Edwards}},
  \bibnamefont{and} \bibinfo{author}{\bibfnamefont{D.~G.}
  \bibnamefont{Richards}}, \bibinfo{journal}{Phys.Rev.}
  \textbf{\bibinfo{volume}{D73}}, \bibinfo{pages}{074507}
  (\bibinfo{year}{2006}), \eprint{hep-ph/0601137}.

\end{thebibliography}

\end{document}